\documentclass{amsart}
\usepackage{amsmath}
\usepackage{amssymb}
\usepackage{amsthm} 
\usepackage{pxfonts}
\usepackage{times}
\usepackage{bbm}
\usepackage{graphicx}
\usepackage{algorithm}
\usepackage{algorithmic}
\usepackage{clrscode}
\usepackage{fullpage}
\usepackage{url}
\usepackage{caption}
\usepackage{subcaption}

\newtheorem{prop}{Proposition}

\newtheorem{thm}{Theorem}
\newtheorem{lem}{Lemma}

\newtheorem{Def}{Definition}

\DeclareMathOperator*{\argmin}{arg\,min}

\DeclareGraphicsExtensions{.eps}
\graphicspath{{Figures/}}

\begin{document}
\pagestyle{plain}
\newenvironment{frcseries}{\fontfamily{frc} \selectfont}{}
\newcommand{\textfrc}[1]{{\frcseries #1}}
\newcommand{\mathfrc}[1]{\text{\textfrc{#1}}}

\title{Approximation of Points on Low-Dimensional Manifolds Via Random Linear Projections}\thanks{This research was partially supported by ONR N00014-07-1-0625, The Alfred P. Sloan Foundation, DARPA N66001-11-1-4002, NSF DMS-1045153, and NSF DMS-0847388.}
%
\author{Mark A. Iwen \hspace{0.15in} Mauro Maggioni\\
Duke University, Box 90320 \\
Durham, NC 27708-0320}

\begin{abstract}
This paper considers the approximate reconstruction of points, $\vec{x} \in \mathbbm{R}^D$, which are close to a given compact $d$-dimensional submanifold, $\mathcal{M}$, of $\mathbbm{R}^D$ using a small number of linear measurements of 
$\vec{x}$.  In particular, it is shown that a number of measurements of $\vec{x}$ which is independent of the extrinsic dimension $D$ suffices for highly accurate reconstruction of a given $\vec{x}$ with high probability.  
Furthermore, it is also proven that all vectors, $\vec{x}$, which are sufficiently close to $\mathcal{M}$ can be reconstructed with uniform approximation guarantees when the number of linear measurements of $\vec{x}$ depends logarithmically on $D$.  Finally, the proofs of these facts are constructive:  A practical algorithm for manifold-based signal recovery is presented in the process of proving the two main results mentioned above.
\end{abstract}

\maketitle
\thispagestyle{empty}

\section{Introduction}
\label{sec:Intro}

In this paper we present a simple reconstruction technique which facilitates compressive sensing for general classes of high-dimensional signals with low intrinsic dimension.  
Two types of models are often considered: sparse models and low-dimensional/manifold models.
The former type of model assumes that each data point has a sparse representation in terms of a (typically known) dictionary $\Phi$, which geometrically means that data points lie on unions of a small number of planes spanned by the elements of the dictionary \cite{Candes:CSredundantdictionaries}.
The latter type of model assumes that data possesses an intrinsically low-dimensional geometrical structure, for example that of a manifold (see e.g. \cite{isomap,belkin:nc,DiffusionPNAS,DG_HessianEigenmaps}, among many others) or a union of planes (see e.g. \cite{HLMsurvey-Vidal-IEEESPM11,CM:CVPR2011,GPCA-MYDF-SiamRev08,TPMM-HRS-IJCV08}), motivated by many applications, for example in image processing \cite{Multiscale-HWHM-ICCV05}, computer vision \cite{MSL-SK-IEICE04}, and pattern recognition \cite{GPCA-MYDF-SiamRev08}.

Given the low-intrinsic dimension of these models, it is natural to ask whether a small number of linear projections (``measurements'') of a data point, together with knowledge of the low-dimensional model, suffices to encode and reconstruct a data point. In the setting of sparsity, compressed sensing \cite{CS3,rauhut2010compressive} not only says that, under suitable assumptions \cite{Candes:CSredundantdictionaries}, this is indeed possible, but a convex optimization problem leads to the stable recovery of the original data point. 
In the setting where data lies on a low-dimensional manifold, the work of Wakin et al. \cite{RandProjSMan,WakinManRecov} on manifold-based signal recovery shows that low-dimensional (random) projections provide small distortion embeddings for manifolds, but leave open the question of reconstructing a data point. 

Standard compressed sensing \cite{CS3,rauhut2010compressive} deals with the approximation of vectors, $\vec{x} \in \mathbbm{R}^D$, which can be sparsely represented in terms of a given $D \times n$ dictionary matrix, $\Phi$.  Note that such $\Phi$-sparse vectors can be compactly stored in a compressed form which is easy to transmit and store.  Moreover, they can be recovered from their compressed representations when necessary.  This compression/recovery problem has been well studied when $\vec{x}$ is available in its entirety before compression (see, e.g., \cite{Mallat}).  However, in situations where $\vec{x}$ is costly to observe one may only have the ability to collect a very small set of measurements of $\vec{x}$ to begin with, thus making standard compression techniques inapplicable (see, e.g., \cite{Apps1,Apps2} and references therein).  This is the \textit{compressed sensing regime}, where lossless compression must occur \textit{before} one determines which vector components or transform coefficients are actually important.  Hence, the goal of standard compressed sensing becomes to design an $m \times D$ measurement matrix $M$, with $m$ as small as absolutely possible, subject to the constraint that a computationally efficient reconstruction algorithm, $\mathcal{A}: \mathbbm{R}^m \rightarrow \mathbbm{R}^D$, exists such that $\mathcal{A} \left( M\vec{x} \right)  \approx \vec{x}$ anytime $\vec{x} \in \mathbbm{R}^D$ is sufficiently compressible with respect to a given dictionary matrix $\Phi$.

More precisely, given an integer $d \ll n$ suppose that
$$\vec{x} ~=~ \Phi \left( \vec{f}_d ~+~ \vec{\epsilon} \right)$$
where $\vec{\epsilon} \in \mathbbm{R}^n$ is in the row space of $\Phi \in \mathbbm{R}^{D \times n}$ and
$$\vec{f}_d ~=~ \argmin_{\vec{y} \in \mathbbm{R}^n ~\textrm{with}~\| \vec{y} \|_0 \leq d} \left\| \vec{x} - \Phi \vec{y} \right\|.$$
The goal of a compressed sensing method is to approximate $\vec{x}$ as well as possible by approximating the at most $d$ nonzero elements of $\vec{f}_d \in \mathbbm{R}^n$.  Furthermore, compressed sensing techniques aim to accomplish this task using as few linear measurements of $\vec{x} \in \mathbbm{R}^D$, 
\begin{equation}
\left \langle \vec{m}_j, \vec{x} \right \rangle = \left \langle \vec{m}_j, \Phi \vec{f}_d ~\right \rangle + \left \langle \vec{m}_j, \Phi \vec{\epsilon} \right \rangle,
\label{Equ:LinMeas}
\end{equation}
as absolutely possible.

Let $M \in \mathbbm{R}^{m \times D}$ be the matrix whose $j^{\rm th}$-row is the measurement vector $\vec{m}_j \in \mathbbm{R}^D$ from Equation~\ref{Equ:LinMeas} above.  A compressed sensing method consists of both a choice of $M \in \mathbbm{R}^{m \times D}$, and a recovery algorithm, $\mathcal{A}: \mathbbm{R}^m \rightarrow \mathbbm{R}^D$,
such that
\begin{equation}
\left\| \vec{x} - \mathcal{A} \left( M \vec{x} \right) \right\|_p ~=~ \left\| \vec{x} - \mathcal{A} \left( M \Phi \vec{f}_d + M \Phi \vec{\epsilon} \right) \right\|_p ~\leq~ C_{p,q} \cdot d^{\frac{1}{p} - \frac{1}{q}} \left \| ~\vec{\epsilon} ~\right \|_q
\label{Equ:Error}
\end{equation}
in fixed $\ell_p$,$\ell_q$ norms, $1 \leq q \leq p \leq 2$, for an absolute constant $C_{p,q} \in \mathbbm{R}$.  Note that $M \in \mathbbm{R}^{m \times D}$ forms a compressed representation of $\vec{x} \in \mathbbm{R}^D$ whenever $m < D$, which is then stably inverted by $\mathcal{A}$.  Many recovery algorithms, $\mathcal{A}$, have been developed for solving this problem when $\Phi$ is a square $D \times (n = D)$ orthonormal matrix, and $M \Phi$ has either restricted isometry \cite{NearOpt} or incoherence \cite{Incoherent1,Incoherent2} properties (e.g., see \cite{candes2005decoding,CS1,CS4,CS2,ROMP,ROMPstable,needell2009cosamp,HardThreshforCS}).  Perhaps the best such results are achieved by $\left( m = O(d \log (D/d)) \right) \times D$ measurement matrices, $M$, whose entries are independent and identically distributed standard Gaussian random variables.  These Gaussian matrices allow for near optimal compression (i.e., a near minimal size for $m$) while still allowing for the existence of recovery algorithms, $\mathcal{A}$, which achieve Equation~\ref{Equ:Error} for an arbitrarily given square orthonormal matrix $\Phi$.  Furthermore, if $M$ is Gaussian then $\Phi$ need not be known when the measurements, $M \vec{x}$, are computed:  It suffices to know $\Phi$ only during reconstruction with $\mathcal{A}$.

One strand of work in compressed sensing has dealt with extending the results mentioned above concerning square orthonormal matrices to include settings where $\Phi$ is a more general (i.e., rectangular) $D \times n$ matrix.  The first of these results extended compressed sensing to include $D \times n$ dictionaries, $\Phi$, whose columns are all nearly pairwise orthogonal \cite{Incoherent3}.  This work shares all of the advantages of the aforementioned results concerning compressed sensing when $\Phi$ is square orthonormal matrix (e.g., nearly orthogonal $\Phi$ also do not need to be known until reconstruction via $\mathcal{A}$) when $M$ is a random matrix exhibiting concentration of measure properties (e.g., if $M$ is Gaussian as above).  These results were later generalized further to allow recovery along the lines of Equation~\ref{Equ:Error} when $\Phi$ has columns with less limited forms of coherence and redundancy \cite{RedundantDict_CS} (e.g., if $\Phi$ is a tight frame).  

In this paper we consider a geometric generalization of standard compressed sensing results for signals which are sparsely representable with respect to a square orthonormal matrix, $\Phi$, by focussing instead on signals which are well represented by manifold models.   More specifically, herein the $D \times n$ dictionary matrix $\Phi$ utilized in standard compressed sensing models will be replaced by a piecewise linear approximation to a given submanifold of $\mathbbm{R}^D$.  To understand why this represents a generalization, note that the set of all vectors which are at least $d$-sparse with respect to an orthogonal matrix $\Phi$ defines a form of Grassmannian manifold consisting of $O \left( D^d \right)$ at most $d$-dimensional linear subspaces of $\mathbbm{R}^D$.  Hence, standard compressed sensing methods concerning square orthonormal matrices, $\Phi$, can be viewed as dealing with a limited class of Grassmannian manifolds.  In contrast, this paper allows for the approximation of signals which belong to much more general types of submanifolds of $\mathbbm{R}^D$.

The work herein utilizes ideas introduced by Baranuik and Wakin which demonstrate the existence of simple linear operators capable of (nearly) isometrically embedding a given compact $d$-dimensional submanifold of $\mathbbm{R}^D$ into $\mathbbm{R}^{O(d \log D)}$ without utilizing detailed knowledge regarding the submanifold's structure
\cite{RandProjSMan}.  In some sense, this work immediately yields measurement matrices, $M \in \mathbbm{R}^{m \times D}$, for manifold-based compressed sensing.  However, a complete compressed sensing strategy also requires an associated reconstruction algorithm, $\mathcal{A}: \mathbbm{R}^m \rightarrow \mathbbm{R}^D$, capable of accurately approximating points near the given manifold in a computationally efficient fashion.  Algorithms of this kind were first considered by Wakin in \cite{WakinManRecov}.  Therein, Wakin showed that approximating a given point, $\vec{x}$, near a compact $d$-dimensional submanifold of $\mathbbm{R}^D$ via an $O(d \log D)$ linear measurements (i.e., see Equation~\ref{Equ:LinMeas}) was possible with high probability if the measurements were randomly regenerated for each new $\vec{x}$.  Furthermore, he concluded that achieving strong reconstruction guarantees using one fixed set of linear measurements for all possible points, $\vec{x}$, near a given compact submanifold of $\mathbbm{R}^D$ was difficult.  However, it is important to mention that the results presented in \cite{WakinManRecov} were derived independently of any particular numerical reconstruction algorithm, $\mathcal{A}$.  As a consequence, this line of work did not result an implementable recovery algorithm with accompanying approximation guarantees.

In this paper we propose a computationally efficient reconstruction algorithm for manifold-based compressed sensing and prove accompanying approximation guarantees.  In the process, we prove that a given point, $\vec{x}$, near a compact $d$-dimensional submanifold of $\mathbbm{R}^D$ can be accurately approximated using $O(d \log d)$ linear measurements with high probability when the measurements are randomly regenerated for each new $\vec{x}$.  This improves on previous results \cite{WakinManRecov} by removing all dependence on the extrinsic dimension of the submanifold, $D$, from the number of linear measurements required for accurate approximation.  Furthermore, we provide stability guarantees for the algorithm when one fixed set of $O(d \log D)$ linear measurements are used for all possible points, $\vec{x}$, near a given compact submanifold of $\mathbbm{R}^D$.  Finally, an empirical evaluation of our method indicates that it also works well in practice.

Before moving on to discuss our methods and results in more detail we hasten to add that other techniques have also been proposed for manifold-based compressed sensing since the initial work of Baranuik and Wakin.  Perhaps most notable among these are the statistical methods proposed by Chen et al. \cite{ManifoldCS_Carin}.  Chen et al. use training data from a compact $d$-dimensional submanifold of $\mathbbm{R}^D$ in order to estimate the manifold data's distibution via a Gaussian mixture model composed of Gaussians whose covariance matrices are all rank $O(d)$.  They then use the probability density resulting from their low-rank Gaussian mixture model to approximate points on the manifold, $\vec{x}$, with a maximum likelihood estimator when given only linear measurements, $M \vec{x} \in \mathbbm{R}^m$.  In contrast, we utilize geometric and analytic techniques herein and make no attempt to estimate the statistical properties of any observed manifold data.

\subsection{Methods and Results}
\label{sec:Manifold_CS}

As discussed above, the standard compressed sensing setup assumes that the signal to be approximated has a compressible representation with respect to an orthonormal basis (or frame \cite{RedundantDict_CS}, or incoherent dictionary \cite{Incoherent3}). Although this is certainly a useful setting, there are many applications where signals might be better approximated via more geometrical considerations. For example, consider the setting where the class of potential input signals varies continuously as a function of a small number of parameters (e.g., see \cite{WakinImageMans,RandProjSMan,WakinManRecov}). In this case it makes more sense to consider the approximate reconstruction of signals, $\vec{x} \in \mathbbm{R}^D$, which are close to a given compact $d$-dimensional submanifold, $\mathcal{M}$, of $\mathbbm{R}^D$.  The optimal approximation for $\vec{x} \in \mathbbm{R}^D$ is then defined to be
$$\vec{x}_{\rm opt} = \argmin_{\vec{y} \in \mathcal{M}} \left\| \vec{x} - \vec{y} \right\|_2.$$
In effect, $\vec{x}_{\rm opt}$ is the best approximation to $\vec{x}$ on $\mathcal{M}$.  Our objective is to approximate $\vec{x}_{\rm opt} \in \mathcal{M} \subset \mathbbm{R}^D$ given only a small number of linear measurements, $M \vec{x} \in \mathbbm{R}^m$, where $M$ is an $m \times D$ measurement matrix as above.  Hence, in this paper we seek to design a measurement matrix $M \in \mathbbm{R}^{m \times D}$ with $m$ as small as absolutely possible, together with a computationally efficient reconstruction algorithm $\mathcal{A}: \mathbbm{R}^m \rightarrow \mathbbm{R}^D$, so that $\mathcal{A} \left( M\vec{x} \right)  \approx \vec{x}$ whenever $\vec{x} \in \mathbbm{R}^D$ is sufficiently close to a given compact $d$-dimensional submanifold of $\mathbbm{R}^D$, $\mathcal{M}$.\footnote{Put another way, we require that $\mathcal{A} \left( M\vec{x} \right)  \approx \vec{x}_{\rm opt}$ which implies that $\mathcal{A} \left( M\vec{x} \right)  \approx \vec{x}$ whenever $\vec{x} \approx \vec{x}_{\rm opt}$.}

Note that a manfold, $\mathcal{M}$, is now taking the place of the dictionary matrix, $\Phi \in \mathbbm{R}^{D \times n}$, in the standard compressed sensing setup discussed above.  Of course, it is unreasonable to expect that we can always have an exact representation of the signal manifold at our disposal.  Instead, we assume that we have a set of locally linear approximations to the given manifold which capture the local geometric structure of the manifold's tangent spaces.  In fact, such piecewise linear manifold representations are exactly the type of approximations produced by existing manifold learning algorithms like LTSA \cite{ZhaZha} and Geometric Multi-Resolution Analysis \cite{CM:MGM2}.  Thus, we assume that the signal manifold, $\mathcal{M}$, is approximated by such a method at some point.  However, as in standard compressed sensing methods, the manifold-based compressed sensing strategies developed below do not require that these piecewise linear manifold representations are known when the compressed measurements, $M \vec{x} \in \mathbbm{R}^m$, are collected.  Approximation of the signal manifold can be put off until later when signal reconstruction takes place (i.e., one does not need a piecewise linear manifold approximation until $\mathcal{A} \left( M\vec{x} \right)$ is actually computed).

Although the manifold-based compressed sensing methods developed herein will work with any locally linear approximation to the given signal manifold, $\mathcal{M}$, we will focus on \textit{multiscale} piecewise linear manifold approximations to $\mathcal{M}$ in particular.  As opposed to fixed-scale locally linear approximations, multiscale representatons better approximate non-smooth manifolds, and manifolds contaminated with noise \cite{MultHighDim,CM:MGM2}.  For example, multiscale locally linear approximation is particularly beneficial for signal processing tasks involving image manifolds, which tend to be non-differentiable in many realistic settings \cite{WakinImageMans}.  Hence, we formulate our compressed sensing methods below with respect to general multiscale piecewise linear manifold approximations of the type produced by Geometric Multi-Resolution Analysis (GMRA) \cite{CM:MGM2}.

As mentioned above, the manifold embeddings of Baranuik and Wakin \cite{RandProjSMan,WakinManRecov} can be considered as manifold-based compressed sensing matrices, for which however no associated recovery algorithms were explicitly defined.  Indeed, the measurement matrices, $M \in \mathbbm{R}^{m \times D}$, used in the manifold-based compressed sensing methods developed below are modifications of their embedding matrices.  However, unlike the embedding matrices considered in \cite{RandProjSMan}, the measurement matrices considered herein (nearly) isometrically embed \textit{both} the underlying signal manifold, $\mathcal{M}$, \textit{and} the multiscale piecewise linear approximation to $\mathcal{M}$ into $\mathbbm{R}^m$ in a way which preserves the fidelity of the embedded multiscale locally linear approximation to the embedded image of $\mathcal{M}$.  Accomplishing this requires us to reengineer the arguments from \cite{RandProjSMan} using Johnson-Lindenstrauss embedding \cite{JLoriginal} techniques similar to those utilized in \cite{baraniuk2008simple}.  The resulting measurement matrices, $M$, ultimately justify this complication by allowing us to develop reconstruction algorithms which work exclusively with locally linear approximations to $\mathcal{M}$ while still preserving approximation accuracy with respect to the true manifold, $\mathcal{M}$.

The reconstruction algorithm, $\mathcal{A}: \mathbbm{R}^m \rightarrow \mathbbm{R}^D$, proposed below consists of two well-studied computational subroutines:  a method for solving approximate nearest neighbor problems  (e.g.,  \cite{indyk1998approximate,beygelzimer2006cover,NNsearch}) in a space of dimension comparable to the intrinsic dimension of the data, and a method for solving an overdetermined least squares problem (e.g., via the singular value decomposition of the associated matrix).   The algorithm works by first using the compressed measurements, $M \vec{x}$, of $\vec{x}$ to locate the best local linear approximation to $\mathcal{M}$ at $\vec{x}$.  This is accomplished by running a nearest neighbor algorithm on a set of ``center points'' from near the manifold, each of which represents a particular linear approximation to $\mathcal{M}$ in a neighborhood of the center point.  Because $\mathcal{M}$ has low intrinsic dimension, and the center points are arranged in a multiscale hierarchy as per \cite{CM:MGM2}, this search can be carried out relatively quickly.  To finish, the algorithm then approximates $\vec{x}_{\rm opt}$, the best approximation to $\vec{x}$ on $\mathcal{M}$, by solving an overdetermined least squares problem using the linear approximation to the manifold located in the first step.

In this paper we prove two compressed sensing results for the proposed reconstruction algorithm, each of which utilizes randomly generated measurement matrices, $M \in \mathbbm{R}^{m \times D}$, satisfying a different set of properties.  Roughly speaking, the first result indicates that $m = O \left( d \log (d / \delta) \right)$ linear measurements of a given $\vec{x} \in \mathbbm{R}^D$ suffice to create a compact representation, $M \vec{x} \in \mathbbm{R}^m$, from which the reconstruction algorithm, $\mathcal{A}$, discussed above will recover an approximation to $\vec{x}_{\rm opt} \in \mathcal{M}$ satisfying 
$$\left\| \vec{x} - \mathcal{A}(M \vec{x}) \right\| ~<~ C \left\| \vec{x} - \vec{x}_{\rm opt} \right\| + \delta.$$  
Here $C \in \mathbbm{R}^+$ represents a fixed universal constant, $\delta \in \mathbbm{R}^+$ can be freely chosen, and $\mathcal{M}$ is the given $d$-dimensional submanifold of $\mathbbm{R}^D$.  This result provides what is commonly referred to as a \textit{nonuniform recovery result}, by which we mean that the upper bound on $\left\| \vec{x} - \mathcal{A}(M \vec{x}) \right\|$ holds with high probability \textit{for each} $\vec{x} \in \mathbbm{R}^D$ over the choice of random measurement matrix.

The second theorem proven below provides a type of \textit{uniform recovery result} which holds with high probability \textit{for all} vectors, $\vec{x} \in \mathbbm{R}^D$, of a particular class.  Simply put, it asserts the existence of a $D$-dimensional tube around the given manifold, $T \supset \mathcal{M}$, within which accurate approximation will always take place with high probability over the choice of random measurement matrix $M \in \mathbbm{R}^{m \times D}$.  More specifically, the second theorem says that $m = O \left( d \log (D / \delta) \right)$ linear measurements of any $\vec{x} \in T \subset \mathbbm{R}^D$ suffice to create a compact representation, $M \vec{x} \in \mathbbm{R}^m$, from which the reconstruction algorithm discussed above, $\mathcal{A}$, will recover an approximation to $\vec{x}_{\rm opt} \in \mathcal{M}$ satisfying 
$$\left\| \vec{x} - \mathcal{A}\left(M \vec{x}\right) \right\| ~<~ C \left\| \vec{x} - \vec{x}_{\rm opt} \right\|_2 + \frac{C}{\sqrt{d}} \left\| \vec{x} - \vec{x}_{\rm opt} \right\|_1 + \delta.$$
Here, as above, $C \in \mathbbm{R}^+$ represents a fixed universal constant and $\delta \in \mathbbm{R}^+$ can be freely chosen.  

The reminder of this paper is organized as follows:  In the next section we begin by fixing terminology and reviewing relevant definitions and theorems.  Having established the necessary notation, we then give precise statements of the two main results proven in this paper in Section~\ref{sec:Results}.  Finally, in Section~\ref{sec:MeasMatrix}, we conclude Section~\ref{sec:Setup} with a discussion of the different types of measurement matrices, $M \in \mathbbm{R}^{m \times D}$, associated with each of our two main results.  In Section~\ref{sec:ReconAlg} the recovery algorithm, $\mathcal{A}$, is presented and analyzed.  In particular, the approximation error of $\mathcal{A}$ for a given $\vec{x}$, $\left\| \vec{x} - \mathcal{A} \left( M\vec{x} \right) \right\|$, is bounded for each of the two possible types of measurement matrices, $M$, considered herein.  The runtime complexity of $\mathcal{A}$ is also determined.  Next, in Section~\ref{sec:UpperBounds}, the number of rows, $m$, required for each type of measurement matrix defined in Section~\ref{sec:MeasMatrix} is upper bounded.  This formally establishes the amount of compression possible in our manifold-based compressed sensing schemes.  To finish, the compressed sensing methods developed herein are evaluated empirically in Section~\ref{sec:Empirical}.

\section{Notation and Setup}
\label{sec:Setup}

Given $n \in \mathbbm{N}$ we will define $[n]$ to be the set $\{ 0, 1, 2, \dots, n \} \subset \mathbbm{Z}$.  All norms, $\| \cdot \|$, will refer to the standard Euclidean norm unless otherwise stated.
We will denote an open ball of radius $\delta \in \mathbbm{R}^+$ centered at $\vec{y} \in \mathbbm{R}^D$ by $\mathcal{B}_{\delta} \left( \vec{y} \right)$.  Our real valued $m \times D$ measurement matrix will always be denoted by 
$M$.  Furthermore, $M$ will always be linear Johnson-Lindenstrauss embedding \cite{JLoriginal,frankl1988johnson,dasgupta2003elementary,krahmer2010new} of a finite set $S \subset \mathbbm{R}^D$ into $\mathbbm{R}^m$.
\begin{Def}
Let $\epsilon \in (0, 1/2)$, and $S \subset \mathbbm{R}^D$ be finite.  An $m \times D$ matrix $M$ is a linear Johnson-Lindenstrauss embedding of $S$ into $\mathbbm{R}^m$ if
$$(1 - \epsilon) \| \vec{u} - \vec{v} \|^2 \leq \| M\vec{u} - M\vec{v} \|^2 \leq (1 + \epsilon) \| \vec{u} - \vec{v} \|^2$$
for all $\vec{u},\vec{v} \in S$.  In this case we will say that $M$ embeds $S$ into $\mathbbm{R}^m$ with $\epsilon$-distortion.
\label{Def:JL}
\end{Def}
The following theorem is proven by showing that an $m \times D$ matrix with randomized entries will satisfy Definition~\ref{Def:JL} for a given set $S \subset \mathbbm{R}^D$ with high probability whenever $m$ is sufficiently large 
(e.g., see \cite{dasgupta2003elementary}).  
\begin{thm} \textrm{\textbf{(See \cite{JLoriginal,dasgupta2003elementary}.)}}
Let $\epsilon \in (0, 1/2)$, and $S \subset \mathbbm{R}^D$ be finite.  Let $m = O( \epsilon^{-2} \log |S|)$ be a natural number. Then, there exists an $m \times D$ linear Johnson-Lindenstrauss embedding of $S$ into $\mathbbm{R}^m$ 
with $\epsilon$-distortion.
\label{thm:JLexists}
\end{thm}

For the remainder of this paper $\mathcal{M}$ will denote a compact $d$-dimensional submanifold of $\mathbbm{R}^D$ with $d$-dimensional volume $V$.  We will characterize results concerning any such manifold $\mathcal{M}$ 
via its reach \cite{FedererCurvMeas}, denoted ${\rm reach}\left( \mathcal{M} \right)$, which is defined as follows:  Let
$$D \left( \mathcal{M} \right) = \left\{ \vec{x} \in \mathbbm{R}^D ~\big|~ \exists~ \textrm{a unique} ~\vec{y} \in \mathcal{M} \textrm{ with } \| \vec{x} - \vec{y} \| = d \left( \vec{x}, \mathcal{M} \right) \right\}$$
and
$${\rm tube}_r\left( \mathcal{M} \right) = \left\{ \vec{x} \in \mathbbm{R}^D ~\big|~ d \left( \vec{x}, \mathcal{M} \right) < r \right\},$$
where $d(\vec{x}, \mathcal{M})$ is the standard Hausdorff distance.  We then define 
\begin{equation}
{\rm reach}\left( \mathcal{M} \right) = \sup \{ r \geq 0 ~|~ {\rm tube}_r\left( \mathcal{M} \right) \subset D \left( \mathcal{M} \right)\}.
\label{eqn:ReachDef}
\end{equation}
Intuitively ${\rm reach}\left( \mathcal{M} \right)$ is the radius of the largest possible non-self-intersecting tube around $\mathcal{M}$.  For example, if $\mathcal{M}$ is a $d$-sphere of radius $r$, then ${\rm reach}\left( \mathcal{M} \right) = r$.  The reach of a manifold is particularly useful because it allows the development of concise bounds for many manifold properties of interest (e.g., curvature, self-avoidance, packing numbers, etcetera).  See \cite{FedererCurvMeas, FindHomofSubMan, RandProjSMan, ImprovedManBounds} for more details.

Given a compact set $\mathcal{S} \subset \mathbbm{R}^D$ we define a \textit{$\delta$-cover of $\mathcal{S}$} to be any finite set $S \subset \mathbbm{R}^D$ with the following property:
$$\forall \vec{x} \in \mathcal{S}, ~ \exists \vec{y} \in S \textrm{ such that } \vec{x} \in \mathcal{B}_{\delta} \left( \vec{y} \right).$$
We will refer to a $\delta$-cover of $\mathcal{S}$, $S$, as \textit{minimal} if $|S| ~\leq~ |\tilde{S}|$ for all other $\delta$-covers of $\mathcal{S}$, $\tilde{S}$.  Hereafter, 
$C_{\delta} \left( \mathcal{S} \right)$ will denote a minimal $\delta$-cover of a given compact set $\mathcal{S}$ in $\mathbbm{R}^D$.  The following lemma, easily proven using results from \cite{FindHomofSubMan}, bounds $\left| C_{\delta} \left( \mathcal{M} \right) \right|$ for any compact $d$-dimensional Riemannian manifold, $\mathcal{M}$, in terms of $\delta$ and ${\rm reach}\left( \mathcal{M} \right)$.

\begin{lem} \textrm{\textbf{(See \cite{FindHomofSubMan}.)}}
Let $\mathcal{M} \subset \mathbbm{R}^D$ be a compact $d$-dimensional Riemannian manifold with $d$-dimensional volume $V$, and suppose that $\delta \in \mathbbm{R}^+$ is less than ${\rm reach}\left( \mathcal{M} \right)$.  Then, any minimal $\delta$-cover of $\mathcal{M}$, $C_{\delta} \left( \mathcal{M} \right)$, will have
$$\big| C_{\delta} \left( \mathcal{M} \right) \big| ~<~ \frac{V \left( \frac{d}{2} + 1 \right)^{\frac{d}{2} + 1}}{2^{\frac{d}{2}} \delta^d}.$$
\label{lem:CovBound}
\end{lem} 

In order to help us develop a practical recovery algorithm we will assume we have a multiscale piecewise linear approximation of $\mathcal{M}$ of the type yielded by GMRA \cite{CM:MGM2}.  Let $J \in \mathbbm{N}$ and $K_0, K_1, \dots, K_J \in \mathbbm{N}$.  For each $j \in [J]$ we assume that we have a set of affine projectors, 
$$\mathbbm{P}_j = \left\{ \mathbbm{P}_{j,k}:  \mathbbm{R}^D \rightarrow \mathbbm{R}^D ~\big|~ k \in [K_j] \right\},$$ 
which approximate $\mathcal{M}$ at scale $j$.  More precisely, these affine projectors will collectively satisfy the three following properties:
\begin{enumerate}
\item \textbf{Affine Projections}:  Every $\mathbbm{P}_{j,k}$ has both an associated vector, $\vec{c}_{j,k} \in \mathbbm{R}^D$, and an associated orthogonal $d \times D$ matrix, $\Phi_{j,k}$, so that 
$$\mathbbm{P}_{j,k} \left( \vec{x} \right) = \Phi^{\rm T}_{j,k}\Phi_{j,k} \left( \vec{x} - \vec{c}_{j,k} \right) + \vec{c}_{j,k}.$$ 
\label{item:ProjectorDef}
\item \textbf{Dyadic Structure}:  There exist two universal constants, $C_1 \in \mathbbm{R}^+$ and $C_2 \in (0,1]$, so that the following conditions are satisfied:  
\begin{enumerate}
\item $K_{j} \leq K_{j+1}$ for all $j \in [J-1]$.
\item $\| \vec{c}_{j,k_1} - \vec{c}_{j,k_2} \| > C_1 \cdot 2^{-j}$ for all $j \in [J]$ and $k_1, k_2 \in [K_j]$ with $k_1 \neq k_2$.  In other words, the $\vec{c}_{j,k}$-vectors at each scale $j \in [J]$ are well separated from one another.  
\label{item:WellSep}
\item For each $j \in [J] - \{ 0 \}$ there is exactly one well defined parent function, $p_j:  [K_j] \rightarrow [K_{j-1}]$, with the property that 
$$\left\| \vec{c}_{j,k} - \vec{c}_{j-1,p_j \left( k \right)} \right\| < C_2 \min_{k' \in [K_{j-1}] - \{ p_j \left( k \right) \}} \left\| \vec{c}_{j,k} - \vec{c}_{j-1,k'} \right\|.$$
\label{item:ParentFunc}
Together these $J$ parent functions collectively define a tree structure on the $\vec{c}_{j,k}$-vectors.  In particular, each $\vec{c}_{0,k}$ with $k \in [K_0]$ is a root node while each $\vec{c}_{J,k}$ with $k \in [K_J]$ is a leaf.
\end{enumerate}
\item \textbf{Multiscale Approximation}:  When $\mathcal{M}$ is sufficiently smooth the affine projectors at each scale $j \in [J]$, $\left\{ \mathbbm{P}_{j,k} ~\big|~ k \in [K_j] \right\}$, approximate $\mathcal{M}$ pointwise with error $O \left(2^{-2j} \right)$.  
\begin{enumerate}
\item There exists a constant $j_0 \in [J-1]$ so that $\vec{c}_{j,k} \in {\rm tube}_{C_1 \cdot 2^{-j-2}}\left( \mathcal{M} \right)$ for all $j \in [J] - [j_0]$ and $k \in [K_j]$.  Note that $j_0$ is a function of the 
constant $C_1$ from Property~\ref{item:WellSep}.  We will generally assume that a $j_0 \in [J-1]$ satisfying this condition exists when $C_1$ is chosen to be as large as possible above.
\label{item:MultiscaleCenters}
\item For each $j \in [J]$ and $\vec{x} \in \mathbbm{R}^D$ let $k_j \left( \vec{x} \right) \in [K_j]$ be such that $\vec{c}_{j,k_j(\vec{x})}$ is one of the nearest neighbors of $\vec{x}$ in the set $\left\{ \vec{c}_{j',k} ~\big|~ j' = j,~k \in [K_j] \right\}$.  That is, for each $j \in [J]$, let
$$k_j \left( \vec{x} \right) = \argmin_{k \in [K_j]} \| \vec{x} - \vec{c}_{j,k} \| .$$
Then, for each $\vec{x} \in \mathcal{M}$ there exists a constant $C \in \mathbbm{R}^+$ such that
$$ \left \| \vec{x} - \mathbbm{P}_{j,k_j \left( \vec{x} \right)} \left( \vec{x} \right) \right\| \leq C \cdot 2^{-2j}$$
for all $j \in [J]$.  In addition, affine projectors associated with $\vec{c}_{j,k}$-vectors that are nearly as close to any $\vec{x} \in \mathcal{M}$ as $\vec{c}_{j,k_j \left( \vec{x} \right)}$ can also accurately 
represent $\vec{x}$.  Hence, for each $\vec{x} \in \mathcal{M}$ their exists a constant $\tilde{C} \in \mathbbm{R}^+$ such that 
$$ \left \| \vec{x} - \mathbbm{P}_{j,k'} \left( \vec{x} \right) \right\| \leq \tilde{C} \cdot 2^{-j}$$
for all $j \in [J]$ and $k' \in [K_j]$ satisfying
$$\left\| \vec{x} - \vec{c}_{j,k'} \right\| \leq 16 \cdot \max \left\{ \left\| \vec{x} - \vec{c}_{j,k_j \left( \vec{x} \right)} \right\|, ~C_1 \cdot 2^{-j-1} \right\}.$$
\label{item:MultiscaleError}
\end{enumerate}
Note that the affine projectors approximate $\mathcal{M}$ more accurately as the scale $j \in [J]$ increases.  The finest scale resolution is obtained when $j = J$.  See \cite{CM:MGM2} for details.
\end{enumerate}

The remainder of this paper is devoted to analyzing the number of measurements required in order to approximately reconstruct an arbitrary point $\vec{x} \in \mathbbm{R}^D$ which is nearly on a compact $d$-dimensional submanifold $\mathcal{M} \subset \mathbbm{R}^D$.  In order to yield substantive progress we must first assume some knowledge of $\mathcal{M}$ (i.e., our manifold-based signal dictionary).  Thus, we will assume below that we have a set of affine projectors, $\left\{ \mathbbm{P}_{j,k} ~\big|~ j \in [J], k \in [K_j] \right\}$, for $\mathcal{M}$ as discussed above, and will primarily focus our analysis on bounding the number of measurements, $m$, sufficient to accurately compute $\mathbbm{P}_{j,k_j \left( \vec{x} \right)} \left( \vec{x} \right)$ for any given input vector $\vec{x} \in \mathbbm{R}^D$ and scale $j \in [J]$.

\subsection{The Goal:  Approximating Manifold Data via Compressive Measurements}

Let $\mathbbm{P} = \left\{ \mathbbm{P}_j ~\big|~ j \in [J] \right\}$ be a multiscale piecewise linear approximation to $\mathcal{M}$ as discussed above.  Given such a $\mathbbm{P}$ we can accurately approximate any 
$\vec{x} \in \mathcal{M} \subset \mathbbm{R}^D$ (e.g., see Property~\ref{item:MultiscaleError}).  However, herein we are primarily interested in approximating arbitrary vectors, $\vec{x} \in \mathbbm{R}^D - \mathcal{M}$, as well as 
they can be approximated by a nearest neighbor on the manifold, $\vec{x}_{\rm opt} \in \mathcal{M}$.  As we shall see, $\mathbbm{P}$ can be utilized for this task.  The following lemma demonstrates that 
$\mathbbm{P}_{j,k_j \left( \vec{x} \right)} \left( \vec{x} \right)$ approximates any vector $\vec{x} \in \mathbbm{R}^D$ nearly as well as $\vec{x}_{\rm opt} \in \mathcal{M}$ does.

\begin{lem}
Let $\mathcal{M} \subset \mathbbm{R}^D$ be a compact $d$-dimensional Riemannian submanifold of $\mathbbm{R}^D$, and $\vec{x} \in \mathbbm{R}^D$.  Furthermore, let $\mathbbm{P}_j = \left\{ \mathbbm{P}_{j,k} ~|~ k \in [K_j] \right\}$ 
be a scale $j \in [J]$ GWRA approximation to $\mathcal{M}$.  Then,
$$\left\| \vec{x} - \mathbbm{P}_{j,k'} \left( \vec{x} \right) \right\| ~\leq~ 17 \left\| \vec{x} - \vec{x}_{\rm opt} \right\| + O \left( 2^{-j} \right)$$
for all $k' \in [K_j]$ satisfying 
$$\left\| \vec{x} - \vec{c}_{j,k'} \right\| \leq 8 \cdot \max \left\{ \left\| \vec{x} - \vec{c}_{j,k_j \left( \vec{x} \right)} \right\|, ~C_1 \cdot 2^{-j-1} \right\}.$$
\label{lem:ErrorXvsPX}
\end{lem}
\noindent \textit{Proof:}  Let $\delta = \max \left\{ \left\| \vec{x} - \vec{c}_{j,k_j \left( \vec{x} \right)} \right\|, ~C_1 \cdot 2^{-j-1} \right\}$, where $C_1 \in \mathbbm{R}$ is defined as in Property~\ref{item:WellSep} above.
Furthermore, let $k' \in [K_j]$ be such that $\left\| \vec{x} - \vec{c}_{j,k'} \right\| \leq 8 \delta$.  To begin, suppose that $\left\| \vec{x} - \vec{c}_{j,k'} \right\| ~\leq~ 17 \left\| \vec{x} - \vec{x}_{\rm opt} \right\|$.  
In this case we are essentially finished since 
$$\left\| \vec{x} - \mathbbm{P}_{j,k'} \left( \vec{x} \right) \right\| ~=~ \left\| \left[ I - \Phi^{\rm T}_{j,k'}\Phi_{j,k'} \right] \left( \vec{x} - \vec{c}_{j,k'} \right) \right\| ~\leq~ \left\| \vec{x} - \vec{c}_{j,k'} 
\right\| ~\leq~ 17 \left\| \vec{x} - \vec{x}_{\rm opt} \right\|.$$
Thus, we will hereafter assume that $\left\| \vec{x} - \vec{c}_{j,k'} \right\| ~>~ 17 \left\| \vec{x} - \vec{x}_{\rm opt} \right\|$ without loss of generality.

Repeatedly applying the triangle inequality we see that $\left\| \vec{x} - \mathbbm{P}_{j,k'} \left( \vec{x} \right) \right\|$ is bounded above by
$$\left\| \vec{x} - \vec{x}_{\rm opt} \right\| + \left\| \vec{x}_{\rm opt} - \mathbbm{P}_{j,k'} \left( \vec{x}_{\rm opt} \right) \right\| + 
\left\| \mathbbm{P}_{j,k'} \left( \vec{x}_{\rm opt} \right) - \mathbbm{P}_{j,k'} \left( \vec{x} \right) \right\|.$$
The third term in the sum immediately above can be bounded by
$$\left\| \mathbbm{P}_{j,k'} \left( \vec{x}_{\rm opt} \right) - \mathbbm{P}_{j,k'} \left( \vec{x} \right) \right\| ~=~ 
\left\| \Phi^T_{j,k'} \Phi_{j,k'} \left( \vec{x} - \vec{x}_{\rm opt} \right) \right\| ~\leq~ \left\| \vec{x} - \vec{x}_{\rm opt} \right\|.$$  
To bound the second term we note that $\left\| \vec{x} - \vec{c}_{j,k'} \right\| > 17 \left\| \vec{x} - \vec{x}_{\rm opt} \right\|$ implies that 
$\left\| \vec{x}_{\rm opt} - \vec{c}_{j,k_j \left( \vec{x}_{\rm opt} \right)} \right\| > 9 \left\| \vec{x} - \vec{x}_{\rm opt} \right\| / 8$.  Therefore,
\begin{align}
\left\| \vec{x}_{\rm opt} - \vec{c}_{j,k'} \right\| & ~\leq~ \left\| \vec{x} - \vec{x}_{\rm opt} \right\| + \left\| \vec{x} - \vec{c}_{j,k'} \right\| ~\leq~ \left\| \vec{x} - \vec{x}_{\rm opt} \right\| + 
8 \cdot \max \left\{ \left\| \vec{x} - \vec{c}_{j,k_j \left( \vec{x}_{\rm opt} \right)} \right\|,~ C_1 \cdot 2^{-j-1} \right\} \nonumber \\
& ~\leq~ 9\left\| \vec{x} - \vec{x}_{\rm opt} \right\| + 8 \cdot \max \left\{ \left\| \vec{x}_{\rm opt} - \vec{c}_{j,k_j \left( \vec{x}_{\rm opt} \right)} \right\|,~ C_1 \cdot 2^{-j-1} \right\} ~<~ 
16 \cdot \max \left\{ \left\| \vec{x}_{\rm opt} - \vec{c}_{j,k_j \left( \vec{x}_{\rm opt} \right)} \right\|,~ C_1 \cdot 2^{-j-1}\right\}. \nonumber 
\end{align}
Property~\ref{item:MultiscaleError} now guarantees that $\left\| \vec{x}_{\rm opt} - \mathbbm{P}_{j,k'} \left( \vec{x}_{\rm opt} \right) \right\| ~\leq~ \tilde{C} \cdot 2^{-j}$.  Hence, we now have
$$\left\| \vec{x} - \mathbbm{P}_{j,k'} \left( \vec{x} \right) \right\| ~\leq~ 2\left\| \vec{x} - \vec{x}_{\rm opt} \right\| + \tilde{C} \cdot 2^{-j}.$$
The result follows.~~$\Box$\\

In this paper we are primarily concerned with achieving approximation results akin to Lemma~\ref{lem:ErrorXvsPX} utilizing \textit{compressive measurements}.  This will allow us to extend the successful sparse approximation 
techniques and results of compressive sensing to the recovery of signals which belong to low dimensional submanifolds of $\mathbbm{R}^D$.  In order to accomplish this goal we must first propose and then subsequently analyze both a 
measurement operator and an associated recovery algorithm.  Furthermore, in order for it to be of practical value, we must demonstrate that the proposed recovery algorithm is computationally efficient, easy to implement, and 
provably accurate.  We begin this process by considering our measurement matrices in Section~\ref{sec:MeasMatrix}.  We then develop a practical reconstruction algorithm in Section~\ref{sec:ReconAlg}.  Before we begin, however, we 
will first state the main results proven herein.

\subsection{Main Results}
\label{sec:Results}

In the statements of the two propositions below, $C \in \mathbbm{R}^{+}$ is an absolute universal constant which is independent of $\vec{x}$, $\mathcal{M}$, $\mathcal{M}$'s GMRA approximation, etcetera.  Note that the upper bounds 
provided for this constant in Section~\ref{sec:ReconAlg} are almost surely quite loose.  We state our first result.

\begin{prop}
Fix precision parameter $\delta \in \mathbbm{R}^{+}$ and let $\vec{x} \in \mathbbm{R}^D$.  In addition, let $\mathbbm{P}_J$, $J = O \left(\log \left[ 1 / (\delta\, {\rm reach}\left( \mathcal{M}) \right) \right] \right)$, be a GMRA approximation to a given compact $d$-dimensional Riemannian manifold, $\mathcal{M} \subset \mathbbm{R}^D$, with volume $V$.  Finally, let 
$$m = O \left( d \log \left( \frac{d}{{\delta\,\rm reach}\left( \mathcal{M} \right)} \right) + \log V \right)$$ 
be a natural number, and define $\mathcal{A}: \mathbbm{R}^m \rightarrow \mathbbm{R}^D$ to be Algorithm~\ref{alg:RecovAlg} from Section~\ref{sec:ReconAlg} below.  
Then, there exists an $m \times D$ matrix, $M$, such that 
$$\left\| \vec{x} - \mathcal{A}\left(M \vec{x}\right) \right\| ~<~ C \cdot \left\| \vec{x} - \vec{x}_{\rm opt} \right\| + \delta$$
with arbitrarily high probability.  Furthermore, $\mathcal{A}\left(M \vec{x}\right)$ can be evaluated in $\left(m^{O(1)} + O(dD)\right)$-time.
\label{MainRes1}
\end{prop}

\noindent \textit{Proof:}  The result follows from Theorem~\ref{thm:Assump1rowbound}, the first part of Theorem~\ref{thm:StableRecov}, and the discussion in Section~\ref{sec:PracticalAlg}.~~$\Box$\\

Proposition~\ref{MainRes1} provides a nonuniform recovery guarantee for each given $\vec{x} \in \mathbbm{R}^D$.  If desired, bounds could be altered to depend on the desired probability of success, $p \in (0,1)$, by including an 
additional multiplicative factor of $O \left( \log(1 / 1 - p) \right)$ in both the runtime of the algorithm and the upper bound for $m$.  The measurement matrices, $M$, referred to by the proposition can be any standard 
Johnson-Lindenstrauss embedding matrix (e.g., a Gaussian random matrix, a random orthogonal projection, etc.).  Hence, they are well understood.  The worst case theoretical runtime complexity of the recovery algorithm is polynomial 
in $m$.  We refer the reader to Section~\ref{sec:Empirical} for an empirical evaluation of the recovery algorithm's computational efficiently in practice.  Finally, we note that the number of required measurements, $m$, is entirely 
independent of the extrinsic dimension, $D$.  Next, we state a uniform approximation guarantee for Algorithm~\ref{alg:RecovAlg}.

\begin{prop}
Fix precision parameter $\delta \in \mathbbm{R}^{+}$.  In addition, let $\mathbbm{P}_J$, $J = O \left(\log \left[ 1 / (\delta\,{\rm reach}\left( \mathcal{M}) \right) \right] \right)$, be a GMRA approximation to a given compact 
$d$-dimensional Riemannian manifold, $\mathcal{M} \subset \mathbbm{R}^D$, with volume $V$.  Finally, let 
$$m =  O \left( d \log \left( \frac{ D}{{\delta\,\rm reach}\left( \mathcal{M} \right)} \right) + \log V \right)$$ 
be a natural number, and define $\mathcal{A}: \mathbbm{R}^m \rightarrow \mathbbm{R}^D$ to be Algorithm~\ref{alg:RecovAlg} from Section~\ref{sec:ReconAlg} below.  
Then, there exists an $m \times D$ matrix, $M$, such that
$$\left\| \vec{x} - \mathcal{A}\left(M \vec{x}\right) \right\| ~<~ C \left\| \vec{x} - \vec{x}_{\rm opt} \right\|_2 + \frac{C}{\sqrt{d}} \left\| \vec{x} - \vec{x}_{\rm opt} \right\|_1 + \delta$$
for all $\vec{x} \in \mathbbm{R}^D$ with
$$2\left\| \vec{x} - \vec{x}_{\rm opt} \right\|_2 ~+~ \frac{6}{5 \sqrt{d}}\left\| \vec{x} - \vec{x}_{\rm opt} \right\|_1 ~\leq~ \max \left\{ \left\| \vec{x} - \vec{c}_{J,k_J \left( \vec{x} \right)} \right\|,~\delta\right\}.$$  
Furthermore, $\mathcal{A}\left(M \vec{x}\right)$ can be evaluated in worst case $\left( 2^{O(d)} \log V + O \left(md^2 + dD \right) \right)$-time.
\label{MainRes2}
\end{prop}

\noindent \textit{Proof:}  The result follows from Theorem~\ref{thm:Assump2rowbound}, the second part of Theorem~\ref{thm:StableRecov}, and the discussion in Section~\ref{sec:PracticalAlg}.~~$\Box$\\

Proposition~\ref{MainRes2} is best interpreted as a general stability result.  It guarantees that Algorithm~\ref{alg:RecovAlg} will uniformly approximate all points which are sufficiently close to the manifold $\mathcal{M}$ (i.e., the points
need not be exactly on $\mathcal{M}$).  Thus, Algorithm~\ref{alg:RecovAlg} has some limited robustness to arbitrary additive input noise.  The examples in the experimental section suggest that the constants involved are very mild.

\subsection{The Measurement Matrix}
\label{sec:MeasMatrix}

In the process of developing an algorithm to approximate $\mathbbm{P}_{j,k_j \left( \vec{x} \right)} \left( \vec{x} \right)$, and subsequently demonstrating its accuracy, we will require some knowledge regarding our $m \times D$ measurement matrix $M$.  We shall consider two sets of assumptions regarding $M$'s interaction with both the manifold $\mathcal{M}$ and our given set of affine projectors for $\mathcal{M}$ at each scale $j \in [J]$.  Each set of assumptions will ultimately result in both different approximation guarantees for our reconstruction algorithm, and different measurements bounds (i.e., sufficient upper bounds on $m$) for $M$.  We will postpone discussion of how to create $M$ and how to bound the number of rows it must have in order to satisfy each set of assumptions below until Section~\ref{sec:UpperBounds}.  In Section~\ref{sec:ReconAlg} below we will begin by presenting our reconstruction algorithm together with approximation error bounds under each set of assumptions regarding $M$.

Let $\vec{x} \in \mathbbm{R}^D$ and $\mathbbm{P} = \left\{ \mathbbm{P}_j ~\big|~ j \in [J] \right\}$ be a fixed set of affine projectors for $\mathcal{M}$ for each scale $j \in [J]$.  Fix $\epsilon \in \left(0, \frac{1}{2} \right)$.  In Sections~\ref{sec:ReconAlg} and~\ref{sec:UpperBounds} we will assume that our $m \times D$ measurement matrix $M$ satisfies each of these sets of assumptions in turn.  
\begin{enumerate}
\item \textbf{Assumption Set 1:  Required for Nonuniform Recovery of a Given $\vec{x} \in \mathbbm{R}^D$ (see Proposition~\ref{MainRes1})}
\begin{enumerate}
\item Let $S_1  \subset \mathbbm{R}^D$ be 
$$S_1 ~=~ \left\{ \Phi^{\rm T}_{j,k}\Phi_{j,k} \left( \vec{x} - \vec{c}_{j,k} \right) ~\big|~ j \in [J],~k \in [K_j] \right\} \bigcup \left\{ \vec{x} - \vec{c}_{j,k} ~\big|~ j \in [J],~k \in [K_j] \right\} \bigcup \left\{\vec{0} \right\}.$$  
We will assume that 
$$(1-\epsilon) \left\| \vec{y} - \vec{z} \right\|^2 ~\leq~ \left\| M \vec{y} - M \vec{z} \right\|^2 ~\leq~(1+\epsilon) \left\| \vec{y} - \vec{z} \right\|^2$$
for all $\vec{y}, \vec{z} \in S_1$.
\item Furthermore, we will assume that
$$(1-\epsilon) \left\| \Phi^{\rm T}_{j,k}\Phi_{j,k} \vec{y} \right\| ~\leq~ \left\| M \Phi^{\rm T}_{j,k}\Phi_{j,k} \vec{y} \right\| ~\leq~(1+\epsilon) \left\| \Phi^{\rm T}_{j,k}\Phi_{j,k} \vec{y} \right\|$$
for all $j \in [J],~k \in [K_j]$, and $\vec{y} \in \mathbbm{R}^D$.
\end{enumerate}

\item \textbf{Assumption Set 2:  Required for General Stability (see Proposition~\ref{MainRes2})}
\begin{enumerate}
\item Let $S_2 = \mathcal{M} \bigcup \left\{ \vec{c}_{j,k} ~\big|~ j \in [J],~k \in [K_j] \right\} \subset \mathbbm{R}^D$.  We will assume that 
$$(1-\epsilon) \left\| \vec{y} - \vec{z} \right\|^2 ~\leq~ \left\| M \vec{y} - M \vec{z} \right\|^2 ~\leq~(1+\epsilon) \left\| \vec{y} - \vec{z} \right\|^2$$
for all $\vec{y}, \vec{z} \in S_2$.
\item Furthermore, we will assume that $\left\| M \vec{y} \right\|$ is bounded above by $E_{M} \left( \vec{y} \right)$ for all $\vec{y} \in \mathbbm{R}^D$, where $E_{M}: \mathbbm{R}^D \rightarrow \mathbbm{R}^+$ is a continuous function with $E_{M} \left( \vec{0} \right) = 0$. $E_{M}$ is discussed in detail in Section~\ref{sec:AssumpSet2Rows}.
\item As before, we will assume that
$$(1-\epsilon) \left\| \Phi^{\rm T}_{j,k}\Phi_{j,k} \vec{y} \right\| ~\leq~ \left\| M \Phi^{\rm T}_{j,k}\Phi_{j,k} \vec{y} \right\| ~\leq~(1+\epsilon) \left\| \Phi^{\rm T}_{j,k}\Phi_{j,k} \vec{y} \right\|$$
for all $j \in [J],~k \in [K_j]$, and $\vec{y} \in \mathbbm{R}^D$.
\item Finally, we will also assume that 
$$(1-\epsilon) \left\| \vec{y} - \mathbbm{P}_{j,k} \left( \vec{y} \right) \right\| - 2^{-J} ~\leq~ \left\| M \vec{y} - M \mathbbm{P}_{j,k} \left( \vec{y} \right) \right\| ~\leq~(1+\epsilon) \left\| \vec{y} - \mathbbm{P}_{j,k} \left( \vec{y} \right) \right\| + 2^{-J}$$
for all $j \in [J],~k \in [K_j]$, and $\vec{y} \in \mathcal{M}$.
\end{enumerate}
\end{enumerate}

Note that the critical difference between the two sets of assumptions above concerns the treatment of $\vec{x} \in \mathbbm{R}^D$ and $\vec{x}_{\rm opt} \in \mathcal{M} \subset \mathbbm{R}^D$.  If possible we would like to obtain measurement bounds which are independent of the ambient dimension, $D$.  Since an arbitrary vector $\vec{x}$ may contain a substantial portion of its energy in the subspace orthogonal the tangent space to $\mathcal{M}$ at $\vec{x}_{\rm opt}$, results which are entirely independent of $D$ generally appear to be unattainable unless our measurement matrix happens to successfully preserve information in the direction of $\vec{x} - \vec{x}_{\rm opt}$.  We assume that $M$ preserves lengths of vectors in the general direction of $\vec{x} - \vec{x}_{\rm opt}$ as part of our first set of assumptions.  In the second set of assumptions we do not.  It is primarily this difference which leads to different measurement bounds and error guarantees in each case.

\section{The Reconstruction Algorithm} 
\label{sec:ReconAlg}

We will ultimately upper bound the number of measurements required in order to approximate a given $\vec{x} \in \mathbbm{R}^D$ which is close to $\mathcal{M} \subset \mathbbm{R}^D$ via the simple reconstruction technique presented in this section.  In doing so we will require that the reconstruction algorithm approximates $\vec{x}$ nearly as well as the vector on $\mathcal{M}$ closest to $\vec{x}$,
$$\vec{x}_{\rm opt} = \argmin_{\vec{y} \in \mathcal{M}} \| \vec{x} - \vec{y} \|,$$ 
approximates $\vec{x}$.  Our first order of business, therefore, will be to derive explicit error guarantees for the reconstruction technique considered herein which demonstrate that it is indeed ``near-optimal'' in the sense discussed in Section~\ref{sec:Intro} above.  Let $\mathcal{A}\left(M \vec{x}\right) \in \mathbbm{R}^D$ denote the output of our reconstruction procedure for a given input $\vec{x} \in \mathbbm{R}^D$.  We wish to bound the approximation error
$$\left\| \vec{x} - \mathcal{A}\left(M \vec{x}\right) \right\|$$
in terms of the optimal approximation error, $\| \vec{x} - \vec{x}_{\rm opt} \|$, and an additive error term of size $O \left( 2^{-j} \right)$ whenever possible.  Before this task can be accomplished, however, we must first describe the recovery algorithm we will use to calculate $\mathcal{A}\left(M \vec{x}\right)$.

\begin{algorithm}[tb]
\begin{algorithmic}[1]
\caption{$\proc{Approximate }\mathbbm{P}_{j,k_j \left( \vec{x} \right)} \left( \vec{x} \right)$} \label{alg:RecovAlg}
\STATE \textbf{Input: Measurements $M \vec{x} \in \mathbbm{R}^m$, Scale $j \in [J]$, Approximation $\mathbbm{P}_j = \left\{ \mathbbm{P}_{j,k} ~\big|~ k \in [K_j] \right\}$ to manifold $\mathcal{M} \subset \mathbbm{R}^D$} 
\STATE \textbf{Output: $\mathcal{A}\left(M \vec{x}\right)$, an approximation to $\mathbbm{P}_{j,k_j \left( \vec{x} \right)} \left( \vec{x} \right) \approx \vec{x}$}
\vspace{0.1in}
\STATE $k' ~\longleftarrow~ \argmin_{k \in [K_j]} \left\| M \vec{x} - M \vec{c}_{j,k} \right\|$ 
\vspace{0.1in}
\STATE $\vec{u}~' ~\longleftarrow~ \argmin_{\vec{u} \in \mathbbm{R}^d } \left\|~ M \Phi^{\rm T}_{j,k'} \vec{u} - M \vec{x} + M \vec{c}_{j,k'}~\right\|$
\vspace{0.1in}
\STATE $\mathcal{A}\left(M \vec{x}\right) ~\longleftarrow~ \Phi^{\rm T}_{j,k'} \vec{u}~' + \vec{c}_{j,k'}$
\vspace{0.1in}
\STATE Output $\mathcal{A}\left(M \vec{x}\right)$
\end{algorithmic}
\label{alg:A}
\end{algorithm}

Our reconstruction procedure uses compressive measurements of $\vec{x}$ in order to approximate $\mathbbm{P}_{j,k_j \left( \vec{x} \right)} \left( \vec{x} \right)$ in two steps (see Algorithm~\ref{alg:RecovAlg} above).  First, the compressive measurements of $\vec{x}$ are used to determine a ``center'' vector, $\vec{c}_{j,k'}$, which is nearly as close to $\vec{x}$ as its nearest neighboring center, $\vec{c}_{j,k_j \left( \vec{x} \right)}$, is.  This step is guaranteed to work well as long as our measurement matrix, $M$, preserves appropriate distances between $\vec{x}$ and all the center vectors at scale $j$.  Next, an accurate projection of $\vec{x} - \vec{c}_{j,k'}$ onto the $d$-dimensional subspace associated with $\vec{c}_{j,k'}$ is found by solving an overdetermined least squares problem.  This step will also work well as long as our measurement matrix $M$ is well conditioned on all of the $d$-dimensional subspaces associated with the scale $j$ center vectors.  As we demonstrate below, the two sets of assumptions for $M$ in Section~\ref{sec:MeasMatrix} are sufficient to guarantee that both steps work well.  

The following lemma guarantees that the center found in line 3 of Algorithm~\ref{alg:RecovAlg} is nearly as close to $\vec{x}$ as $\vec{x}$'s true nearest center is.  

\begin{lem}
Fix $\epsilon \in \left(0, \frac{1}{2} \right)$.  Let $\mathcal{M} \subset \mathbbm{R}^D$ be a compact $d$-dimensional Riemannian submanifold of $\mathbbm{R}^D$, and $\vec{x} \in \mathbbm{R}^D$.  Furthermore, let $\mathbbm{P}_j = \left\{ \mathbbm{P}_{j,k} ~|~ k \in [K_j] \right\}$ be a scale $j \in [J]$ GWRA approximation to $\mathcal{M}$.  Then, if our $m \times D$ measurement matrix $M$ satisfies Assumption Set 1 in Section~\ref{sec:MeasMatrix} above, line 3 of Algorithm~\ref{alg:RecovAlg} will select a $k' \in [K_j]$ which has
$$\left\| \vec{x} - \vec{c}_{j,k'} \right\| ~\leq~ \sqrt{\frac{1 + \epsilon}{1 - \epsilon}} \cdot \left\| \vec{x} - \vec{c}_{j,k_j \left( \vec{x} \right)} \right\|.$$
If our $m \times D$ measurement matrix $M$ satisfies Assumption Set 2 in Section~\ref{sec:MeasMatrix} above, then line 3 of Algorithm~\ref{alg:RecovAlg} will select a $k' \in [K_j]$ which has
\begin{equation}
\left\| \vec{x} - \vec{c}_{j,k'} \right\| ~\leq~ \sqrt{\frac{1 + \epsilon}{1 - \epsilon}} \cdot \left\|  \vec{x} - \vec{c}_{j,k_j \left( \vec{x} \right)} \right\| + \left( 1 + \sqrt{\frac{1 + \epsilon}{1 - \epsilon}} \right) \left\| \vec{x} - \vec{x}_{\rm opt} \right\| +\sqrt{\frac{4}{1 - \epsilon}} \cdot E_{M} \left( \vec{x} - \vec{x}_{\rm opt} \right).
\label{equ:Assump2Center}
\end{equation}
\label{lem:Line3}
\end{lem}
\noindent \textit{Proof:}  Using the first set of assumptions for $M$ together with the definition of $k' \in [K_j]$ from Algorithm~\ref{alg:RecovAlg} we can see that
$$\left\| \vec{x} - \vec{c}_{j,k'} \right\| ~\leq~ \sqrt{\frac{1}{1 - \epsilon}}  \cdot  \left\| M \vec{x} - M \vec{c}_{j,k'} \right\| ~\leq~ \sqrt{\frac{1}{1 - \epsilon}}  \cdot  \left\| M \vec{x} - M \vec{c}_{j,k_j \left( \vec{x} \right)} \right\| ~\leq~ \sqrt{\frac{1 + \epsilon}{1 - \epsilon}} \cdot \left\| \vec{x} - \vec{c}_{j,k_j \left( \vec{x} \right)} \right\|.$$
We now turn our attention to the case where $M$ satisfies the second set of assumptions.  We have that
\begin{align}
\left\| \vec{x} - \vec{c}_{j,k'} \right\| &~\leq~ \left\| \vec{x} - \vec{x}_{\rm opt} \right\| + \left\| \vec{x}_{\rm opt} - \vec{c}_{j,k'} \right\| ~\leq~ \left\| \vec{x} - \vec{x}_{\rm opt} \right\| + \sqrt{\frac{1}{1 - \epsilon}}  \cdot \left\| M \vec{x}_{\rm opt} - M \vec{c}_{j,k'} \right\| \nonumber \\
&~\leq~ \left\| \vec{x} - \vec{x}_{\rm opt} \right\| + \sqrt{\frac{1}{1 - \epsilon}}  \left(  \left\| M \left(\vec{x} - \vec{x}_{\rm opt} \right) \right\| + \left\| M \vec{x} - M \vec{c}_{j,k_j \left( \vec{x} \right)} \right\| \right) \nonumber \\
&~\leq~ \left\| \vec{x} - \vec{x}_{\rm opt} \right\| + \sqrt{\frac{4}{1 - \epsilon}} \cdot \left\| M \left(\vec{x} - \vec{x}_{\rm opt} \right) \right\| + \sqrt{\frac{1 + \epsilon}{1 - \epsilon}} \cdot \left\|  \vec{x}_{\rm opt} - \vec{c}_{j,k_j \left( \vec{x} \right)} \right\| \nonumber.
\end{align}
Focusing on the first and third terms in the line immediately above, we note that
$$\left\| \vec{x} - \vec{x}_{\rm opt} \right\| + \sqrt{\frac{1 + \epsilon}{1 - \epsilon}} \cdot \left\|  \vec{x}_{\rm opt} - \vec{c}_{j,k_j \left( \vec{x} \right)} \right\| ~\leq~ \sqrt{\frac{1 + \epsilon}{1 - \epsilon}} \cdot \left\|  \vec{x} - \vec{c}_{j,k_j \left( \vec{x} \right)} \right\| + \left( 1 + \sqrt{\frac{1 + \epsilon}{1 - \epsilon}} \right) \left\| \vec{x} - \vec{x}_{\rm opt} \right\|.$$
The result follows.~~$\Box$\\

Next, we prove a lemma which guarantees the accuracy of the solution of the overdetermined least squares problem produced by line 4 of Algorithm~\ref{alg:RecovAlg}.

\begin{lem}
Let $\mathcal{M} \subset \mathbbm{R}^D$ be a compact $d$-dimensional Riemannian submanifold of $\mathbbm{R}^D$, and $\vec{x} \in \mathbbm{R}^D$.  Furthermore, let $\mathbbm{P}_j = \left\{ \mathbbm{P}_{j,k} ~|~ k \in [K_j] \right\}$ be a scale $j \in [J]$ GWRA approximation to $\mathcal{M}$, and $k' \in [K_j]$ be the value computed by line 3 of Algorithm~\ref{alg:RecovAlg}.  Then, if our $m \times D$ measurement matrix $M$ satisfies either set of assumptions in Section~\ref{sec:MeasMatrix} above, line 5 of Algorithm~\ref{alg:RecovAlg} will produce an $\mathcal{A}\left(M \vec{x}\right) \in \mathbbm{R}^D$ which has
$$\left\| \mathbbm{P}_{j,k'} \left( \vec{x} \right) - \mathcal{A}\left(M \vec{x}\right) \right\| ~\leq~ \frac{2}{1 - \epsilon} \cdot \left\| M \left[ \vec{x} - \mathbbm{P}_{j,k'} \left( \vec{x} \right) \right]  \right\|.$$
\label{lem:Line4}
\end{lem}
\noindent \textit{Proof:}  Let $\vec{u}~' \in \mathbbm{R}^d$ be as defined in line 4 of Algorithm~\ref{alg:RecovAlg}.  Given either set of assumptions for $M$ we will have 
$$\left\| \Phi^{\rm T}_{j,k'} \vec{u}~' - \Phi^{\rm T}_{j,k'}\Phi_{j,k'} \left( \vec{x} - \vec{c}_{j,k'} \right) \right\| ~\leq~ \frac{1}{1 - \epsilon} \left( \left\| M\Phi^{\rm T}_{j,k'} \vec{u}~' - M \left( \vec{x} - \vec{c}_{j,k'} \right) \right\| + \left\| M \left[ I - \Phi^{\rm T}_{j,k'}\Phi_{j,k'} \right] \left( \vec{x} - \vec{c}_{j,k'} \right) \right\| \right),$$
where $I$ is the $D \times D$ identity matrix.  By the definition of $\vec{u}~'$ in Algorithm~\ref{alg:RecovAlg} we can now see that
$$\left\| \Phi^{\rm T}_{j,k'} \vec{u}~' - \Phi^{\rm T}_{j,k'}\Phi_{j,k'} \left( \vec{x} - \vec{c}_{j,k'} \right) \right\| ~\leq~ \frac{2}{1 - \epsilon} \cdot \left\| M \left[ I - \Phi^{\rm T}_{j,k'}\Phi_{j,k'} \right] \left( \vec{x} - \vec{c}_{j,k'} \right) \right\|.$$
The stated result follows.~~$\Box$\\

Finally, we demonstrate the accuracy of the output of Algorithm~\ref{alg:RecovAlg} as an approximation to $\vec{x}$.

\begin{thm}
Fix $\epsilon \in \left(0, \frac{1}{2} \right)$.  Let $\mathcal{M} \subset \mathbbm{R}^D$ be a compact $d$-dimensional Riemannian submanifold of $\mathbbm{R}^D$, and $\vec{x} \in \mathbbm{R}^D$.  Furthermore, let $\mathbbm{P}_j = \left\{ \mathbbm{P}_{j,k} ~|~ k \in [K_j] \right\}$ be a scale $j \in [J]$ GWRA approximation to $\mathcal{M}$.  Then, if our $m \times D$ measurement matrix $M$ satisfies Assumption Set 1 in Section~\ref{sec:MeasMatrix} above, Algorithm~\ref{alg:RecovAlg} will output a point, $\mathcal{A}\left(M \vec{x}\right) \in \mathbbm{R}^D$, which satisfies
$$\left\| \vec{x} - \mathcal{A}\left(M \vec{x}\right) \right\| ~<~ 100.3 \left\| \vec{x} - \vec{x}_{\rm opt} \right\| + O \left( 2^{-j} \right).$$

Now suppose that our $m \times D$ measurement matrix $M$ satisfies Assumption Set 2 in Section~\ref{sec:MeasMatrix} above, and that $\mathbbm{P}_j$ is a scale $j$ GWRA approximation to $\mathcal{M}$ for some $j > j_0$ 
(revisit Properties~\ref{item:MultiscaleCenters} and~\ref{item:MultiscaleError} in Section~\ref{sec:Setup} for the definitions of the constants $j_0$, $C_1$, and $\tilde{C}$).  Furthermore, suppose that 
$\vec{x} \in \mathbbm{R}^D - \mathcal{M}$ has 
\begin{equation}
2 \cdot E_{M} \left( \vec{x} - \vec{x}_{\rm opt} \right) \leq \left( 8\sqrt{1 - \epsilon} - \sqrt{1 + \epsilon} \right) \left\| \vec{x} - \vec{c}_{j,k_j \left( \vec{x} \right)} \right\| - \left( \sqrt{1 - \epsilon} + \sqrt{1 + \epsilon} \right) \left\| \vec{x} - \vec{x}_{\rm opt} \right\|.
\label{equ:EboundThm1}
\end{equation}
Then, Algorithm~\ref{alg:RecovAlg} will output a point, $\mathcal{A}\left(M \vec{x}\right) \in \mathbbm{R}^D$, which satisfies
\begin{equation}
\left\| \vec{x} - \mathcal{A}\left(M \vec{x}\right) \right\| ~<~ 220 \left\| \vec{x} - \vec{x}_{\rm opt} \right\| + 4 \cdot E_{M} \left( \vec{x} - \vec{x}_{\rm opt} \right) + O \left( 2^{-j} \right).
\label{equ:ErboundThm1}
\end{equation}
\label{thm:StableRecov}
\end{thm}
\noindent \textit{Proof:}  To begin we note that
\begin{equation}
\left\| \vec{x} - \mathcal{A}\left(M \vec{x}\right) \right\| ~\leq~ \left\| \vec{x} - \mathbbm{P}_{j,k'} \left( \vec{x} \right) \right\| + \left\| \mathbbm{P}_{j,k'} \left( \vec{x} \right) - \mathcal{A}\left(M \vec{x}\right) \right\|
\label{equ:Thm1pf}
\end{equation}
where $k' \in [K_j]$ is defined as in line 3 of Algorithm~\ref{alg:RecovAlg}.  The first set of assumptions for $M$ together with Lemmas~\ref{lem:ErrorXvsPX} and~\ref{lem:Line3} tells us that
$$\left\| \vec{x} - \mathbbm{P}_{j,k'} \left( \vec{x} \right) \right\| ~\leq~ 17 \left\| \vec{x} - \vec{x}_{\rm opt} \right\| + O \left( 2^{-j} \right)$$
since $\epsilon \in \left(0, \frac{1}{2} \right)$.  Furthermore, the first set of assumptions for $M$ together with Lemma~\ref{lem:Line4} indicates that
$$\left\| \mathbbm{P}_{j,k'} \left( \vec{x} \right) - \mathcal{A}\left(M \vec{x}\right) \right\| ~\leq~ \frac{2}{1 - \epsilon} \cdot \left\| M \left[ \vec{x} - \mathbbm{P}_{j,k'} \left( \vec{x} \right) \right]  \right\| ~\leq~ 2 \frac{\sqrt{1+\epsilon}}{1 - \epsilon} \cdot \left\| \vec{x} - \mathbbm{P}_{j,k'} \left( \vec{x} \right) \right\|.$$
Hence, we obtain the stated bound in the first case.

Now assume that $M$ satisfies Assumption Set 2 in Section~\ref{sec:MeasMatrix}.  
We will begin by bounding the $\left\| \mathbbm{P}_{j,k'} \left( \vec{x} \right) - \mathcal{A}\left(M \vec{x}\right) \right\|$ term in Equation~\ref{equ:Thm1pf}.  Applying Lemma~\ref{lem:Line4} and then utilizing our second set of 
assumptions regarding $M$ we can see that
\begin{align}
\left\| \mathbbm{P}_{j,k'} \left( \vec{x} \right) - \mathcal{A}\left(M \vec{x}\right) \right\| &~\leq~ \frac{2}{1 - \epsilon} \cdot \left\| M \left[ \vec{x} - \mathbbm{P}_{j,k'} \left( \vec{x} \right) \right] \right\| ~\leq~ 
\frac{2}{1 - \epsilon} \left( \left\| M x - M \vec{x}_{\rm opt} \right\| + \left\| M \vec{x}_{\rm opt} - M \mathbbm{P}_{j,k'} \left( \vec{x} \right) \right\| \right) \nonumber \\
&~\leq~ \frac{2}{1 - \epsilon} \left( \left\| M x - M \vec{x}_{\rm opt} \right\| + \left\| M \vec{x}_{\rm opt} - M \mathbbm{P}_{j,k'} \left( \vec{x}_{\rm opt} \right) \right\| + 
\left\| M \mathbbm{P}_{j,k'} \left( \vec{x}_{\rm opt} \right) - M \mathbbm{P}_{j,k'} \left( \vec{x} \right) \right\| \right) \nonumber \\
&~\leq~ 2 \cdot \frac{1 + \epsilon}{1 - \epsilon} \cdot \left\| \vec{x} - \vec{x}_{\rm opt} \right\| + \frac{2}{1 - \epsilon} \left( \left\| M x - M \vec{x}_{\rm opt} \right\| + \left\| M \vec{x}_{\rm opt} - M \mathbbm{P}_{j,k'} 
\left( \vec{x}_{\rm opt} \right) \right\| \right) \label{equ:Thm1pf2}.
\end{align}
In order to bound the last term in Equation~\ref{equ:Thm1pf2} above, we note that $\left\| \vec{x} - \vec{c}_{j,k'} \right\| ~\leq~ 8 \left\| \vec{x} - \vec{c}_{j,k_j \left( \vec{x} \right)} \right\|$ whenever 
$E_{M} \left( \vec{x} - \vec{x}_{\rm opt} \right)$ satisfies Equation~\ref{equ:EboundThm1}.  
Therefore, we will have $\left\| \vec{x}_{\rm opt} - \vec{c}_{j,k'} \right\| ~<~ 16 \left\| \vec{x}_{\rm opt} - \vec{c}_{j,k_j \left( \vec{x}_{\rm opt} \right)} \right\|$ whenever 
$\left\| \vec{x} - \vec{c}_{j,k'} \right\| > 17 \left\| \vec{x} - \vec{x}_{\rm opt} \right\|$ by an argument identical to that presented in the second paragraph of the proof of Lemma~\ref{lem:ErrorXvsPX}.  
Hence, Property~\ref{item:MultiscaleError} in Section~\ref{sec:Setup} guarantees that $\left\| \vec{x}_{\rm opt} - \mathbbm{P}_{j,k'} \left( \vec{x}_{\rm opt} \right) \right\| ~\leq~ \tilde{C} \cdot 2^{-j}$ whenever 
$\left\| \vec{x} - \vec{c}_{j,k'} \right\| > 17 \left\| \vec{x} - \vec{x}_{\rm opt} \right\|$.  Item (d) of Assumption Set 2 in Section~\ref{sec:MeasMatrix} now guarantees that $\left\| M \vec{x}_{\rm opt} - M \mathbbm{P}_{j,k'} 
\left( \vec{x}_{\rm opt} \right) \right\|$ will also be $O \left( 2^{-j} \right)$ whenever $\left\| \vec{x} - \vec{c}_{j,k'} \right\| > 17 \left\| \vec{x} - \vec{x}_{\rm opt} \right\|$.

To finish, suppose that $\left\| \vec{x} - \vec{c}_{j,k'} \right\| \leq 17 \left\| \vec{x} - \vec{x}_{\rm opt} \right\|$.  Continuing to bound the last term of Equation~\ref{equ:Thm1pf2} in this case we obtain
\begin{align}
\frac{2}{1 - \epsilon} \left\| M \vec{x}_{\rm opt} - M \mathbbm{P}_{j,k'} \left( \vec{x}_{\rm opt} \right) \right\| &~\leq~ \frac{2}{1 - \epsilon} \left( \left\| M\vec{x}_{\rm opt} - M\vec{c}_{j,k'} \right\| + \left\| M \Phi^{\rm T}_{j,k'}\Phi_{j,k'} \left( \vec{x}_{\rm opt} - \vec{c}_{j,k'} \right) \right\| \right) \nonumber \\
&~\leq~ 2 \cdot \frac{\sqrt{1+ \epsilon}}{1 - \epsilon} \left\| \vec{x}_{\rm opt} - \vec{c}_{j,k'} \right\| + 2 \cdot \frac{1+ \epsilon}{1 - \epsilon} \left\| \vec{x}_{\rm opt} - \vec{c}_{j,k'} \right\| \nonumber \\
&~\leq~ 36 \cdot \frac{\sqrt{1+ \epsilon}}{1 - \epsilon} \left( 1 + \sqrt{1+ \epsilon} \right) \left\| \vec{x} - \vec{x}_{\rm opt} \right\|. \nonumber 
\end{align}
Combining this bound with the previous paragraph concludes the proof.~~$\Box$\\

Theorem~\ref{thm:StableRecov} demonstrates that Algorithm~\ref{alg:RecovAlg} can stably approximate vectors $\vec{x} \in \mathbbm{R}^D - \mathcal{M}$ as long as the measurement matrix, $M$, satisfies one of the two sets of assumptions detailed in Section~\ref{sec:MeasMatrix}.  However, the strength of the approximation guarantee depends on which set of assumptions $M$ satisfies.  When $M$ possess the attributes listed in Assumption Set 1 (most notably, attribute (a)) the vector returned by Algorithm~\ref{alg:RecovAlg} will always provide an approximation to $\vec{x}$ whose error is a within a constant multiple of the optimal approximation error.  When $M$ satisfies Assumption Set 2, on the other hand, Algorithm~\ref{alg:RecovAlg} is only guaranteed to provide near optimal approximations for vectors, $\vec{x}$, which are relatively close to the manifold $\mathcal{M}$.

\subsection{Practical Implementation of Algorithm~\ref{alg:RecovAlg}}
\label{sec:PracticalAlg}

In line 3 of Algorithm~\ref{alg:RecovAlg} we want to locate the nearest neighbor of $M \vec{x} \in \mathbbm{R}^m$ from the set $\left \{ M \vec{c}_{j,k} ~\big|~ k \in [K_j] \right\} \subset \mathbbm{R}^m$.  This can be accomplished 
naively in $O(m K_j)$-time.  However, $K_j$ is potentially large in the worst case (see Lemma~\ref{lem:CenterBound} below).  Therefore, it is important to note that the runtime's dependence on $K_j$ can be greatly reduced in 
practice with the aid of standard space partitioning techniques (e.g., by building a k-d tree to solve the nearest neighbor problem).  Alternatively, other fast nearest neighbor methods could also be utilized (e.g., 
see \cite{indyk1998approximate,beygelzimer2006cover,NNsearch} and the references therein).  Due to the dyadic structure of our $\vec{c}_{j,k}$-vectors, the worst case theoretical runtime complexity of line 3 can be improved slightly 
to $\left( 2^{O(d)} \log V \right)$-time by using cover trees \cite{beygelzimer2006cover}.\footnote{Here $V$ is the volume of the $d$-dimensional manifold $\mathcal{M} \subset \mathbbm{R}^D$.}  Alternatively, if it suffices to find 
a $(1+\delta)$-nearest neighbor of $M \vec{x}$ with high probability, we can utilize even faster algorithms which run in $m^{O(1)}$-time (see Proposition 3 in \cite{indyk1998approximate} together with the bound for $m$ in 
Theorem~\ref{thm:Assump1rowbound} below).

Line 4 of Algorithm~\ref{alg:RecovAlg} requires the solution of an overdetermined least squares problem.  This can be accomplished in $O(m d^2)$-time via the singular value decomposition of $M \Phi^{\rm T}_{j,k'}$.  Furthermore, the solution can be computed accurately since both sets of assumptions in Section~\ref{sec:MeasMatrix} guarantee that $M \Phi^{\rm T}_{j,k'}$ is well conditioned.   Finally, explicitly forming $\mathcal{A}\left(M \vec{x}\right)$ in line 5 of Algorithm~\ref{alg:RecovAlg} can be accomplished in $O(Dd)$-time.  The total runtime of Algorithm~\ref{alg:RecovAlg} will therefore be
$O\left(d (md + D) + T_{\rm NN}\right)$, where $T_{NN}$ bounds the runtime of the nearest neighbor algorithm used in line 3.

\section{Upper Bounds on the Number of Required Measurements}
\label{sec:UpperBounds}

In this section we will bound the number of rows, $m$, needed in order for our $m \times N$ measurement matrix, $M$, to satisfy each set of assumptions discussed in Section~\ref{sec:MeasMatrix}.  In order to do so, it will suffice to let $M$ be a linear Johnson-Lindenstrauss embedding of a well chosen set of points in $\mathbbm{R}^D$ into $\mathbbm{R}^m$.  Of course, this set of points will vary depending on which set of assumptions from Section~\ref{sec:MeasMatrix} we want $M$ to satisfy.  Below we consider each set of assumptions separately.  However, we will first establish two lemmas which will be useful in both cases.

\begin{lem} 
Let $\epsilon \in \left(0, \frac{1}{2} \right)$.  Furthermore, let $j \in [J]$ and $k \in [K_j]$ denote an affine projector $\mathbbm{P}_{j,k}$ (see Property~\ref{item:ProjectorDef} in Section~\ref{sec:Setup}).  Then, there exists a finite set of vectors, $Q_{j,k} \subset X_{j,k} = \left\{ \Phi^{\rm T}_{j,k}\Phi_{j,k} \vec{y} ~\big|~ \vec{y} \in \mathbbm{R}^D \right\}$ with $\big| Q_{j,k} \big| ~\leq~ \left( 12 / \epsilon \right)^d + 1$, such that 
$$(1-\epsilon) \left\| \Phi^{\rm T}_{j,k}\Phi_{j,k} \vec{y} \right\| ~\leq~ \left\| M \Phi^{\rm T}_{j,k}\Phi_{j,k} \vec{y} \right\| ~\leq~(1+\epsilon) \left\| \Phi^{\rm T}_{j,k}\Phi_{j,k} \vec{y} \right\|$$
for all $\vec{y} \in \mathbbm{R}^D$ whenever $M$ embeds $Q_{j,k}$ into $\mathbbm{R}^m$ with $\epsilon / 2$-distortion.
\label{lem:SubSpaceBound}
\end{lem} 

\noindent \textit{Proof:}  We let $Q'_{j,k}$ be a minimal $\epsilon / 4$-cover of the $d$-dimensional unit ball in $X_{j,k}$ centered at $\vec{0} \in X_{j,k}$.  Now set $Q_{j,k} = Q'_{j,k} \bigcup \left\{ \vec{0} \right\}$.  The stated upper bound of $\big| Q_{j,k} \big|$ follows from existing covering results (see \cite{baraniuk2008simple} for references).  Furthermore, if $M$ embeds $Q_{j,k}$ into $\mathbbm{R}^m$ with $\epsilon / 2$-distortion it is easy to see that
$$\left( 1 - \epsilon / 2 \right) \| q \| ~\leq~ \| Mq \| ~\leq~ \left( 1 + \epsilon / 2 \right) \| q \|$$
for all $q \in Q_{j,k}$.  The remainder of the proof now directly parallels the proof of Lemma 5.1 in \cite{baraniuk2008simple}.~~$\Box$\\

\begin{lem} 
Fix $J \in \mathbbm{N}$ and let $\mathbbm{P}_j$, $j \in [J]$, be a GMRA approximation to a given compact $d$-dimensional Riemannian manifold, $\mathcal{M} \subset \mathbbm{R}^D$, with $d$-dimensional volume $V$.  Furthermore, suppose that $j' \in [J] - \left[ \max \left\{ j_0, \log_2 \left( \frac{C_1}{{\rm reach}\left( \mathcal{M} \right)} \right) - 2 \right\} \right]$, where $j_0$ and $C_1$ are defined as in Property~\ref{item:MultiscaleCenters} of Section~\ref{sec:Setup}.  Then, the number of affine projectors at scale $j'$, $K_{j'}$, is bounded above by 
$\frac{2^{d(j'+ 1.5)}}{C_1^d} \cdot V \cdot \left( \frac{d}{2} + 1 \right)^{\frac{d}{2} + 1}$.
\label{lem:CenterBound}
\end{lem} 

\noindent \textit{Proof:}  We know that $\mathcal{B}_{C_1 \cdot 2^{-j'-2}} \left( \vec{c}_{j',k} \right) \cap \mathcal{M}$ is nonempty for all $k \in [K_{j'}]$ since $j' > j_0$.  Now consider a minimal $C_1 \cdot 2^{-j'-2}$-cover of $\mathcal{M}$, $C_{C_1 \cdot 2^{-j'-2}} \left( \mathcal{M} \right)$.  It is not difficult to see that every $\vec{c}_{j',k}$ will be contained in $\mathcal{B}_{C_1 \cdot 2^{-j'-1}} \left( \vec{y} \right)$ for some $\vec{y} \in C_{C_1 \cdot 2^{-j'-2}} \left( \mathcal{M} \right)$.  Furthermore, there can be no $\vec{y} \in C_{C_1 \cdot 2^{-j'-2}} \left( \mathcal{M} \right)$ such that two distinct $\vec{c}_{j',k}$ are contained in the same ball, $\mathcal{B}_{C_1 \cdot 2^{-j'-1}} \left( \vec{y} \right)$, by Property~\ref{item:WellSep} in Section~\ref{sec:Setup}.  Hence, $K_{j'} \leq \big| C_{C_1 \cdot 2^{-j'-2}} \left( \mathcal{M} \right) \big|$.  Applying Lemma~\ref{lem:CovBound} concludes the proof.~~$\Box$\\

We are now prepared to upper bound the number of rows required by our $m \times N$ measurement matrix, $M$, in order to satisfy each set of assumptions listed in Section~\ref{sec:MeasMatrix}.

\subsection{Bounding the Number of Rows Required to Satisfy Assumption Set 1}
\label{sec:AssumpSet1Rows}

\begin{thm}
Fix $\epsilon \in \left(0, \frac{1}{2} \right)$, $\vec{x} \in \mathbbm{R}^D$, and $J \in \mathbbm{N}$ sufficiently large.  Furthermore, let $\mathbbm{P}_j$, $j \in [J]$, be a GMRA approximation to a given compact $d$-dimensional Riemannian manifold, $\mathcal{M} \subset \mathbbm{R}^D$, with volume $V$.  Then, there exists an $m \times D$ matrix, $M$, which satisfies Assumption Set 1 in Section~\ref{sec:MeasMatrix} with $m = O \left( d \epsilon^{-2} \left( J + \log ( d / \epsilon ) \right) + \epsilon^{-2} \log V \right).$
\label{thm:Assump1rowbound}
\end{thm}

\noindent \textit{Proof:}  The set $S_1 \subset \mathbbm{R}^D$ defined in item (a) of Assumption Set 1 has $|S_1| \leq 2 (J+1) K_J + 1$.  Furthermore, applying Lemma~\ref{lem:SubSpaceBound} to all at most $(J+1) K_J$ affine projectors yields a set of size at most $(J+1) K_J \left(\left( 12 / \epsilon \right)^d + 1 \right)$ for item (b) of Assumption Set 1.  Lemma~\ref{lem:CenterBound} together with Theorem~\ref{thm:JLexists} now finishes the proof.~~$\Box$\\

It is important to recall that Theorem~\ref{thm:JLexists} is proven by showing that a random matrix will (nearly) isometrically embed a given subset of $\mathbbm{R}^D$ into $\mathbbm{R}^m$ with high probability.  In the proof of Theorem~\ref{thm:Assump1rowbound} above, Theorem~\ref{thm:JLexists} is applied to embed a set which depends on the given $\vec{x} \in \mathbbm{R}^D$ we are ultimately interested in approximating (i.e., the set $S_1$ defined in Section~\ref{sec:MeasMatrix} depends on $\vec{x}$).  Thus, Theorem~\ref{thm:Assump1rowbound} provides us with a high probability recovery guarantee for each separate $\vec{x} \in \mathbbm{R}^D$ on which we apply Algorithm~\ref{alg:RecovAlg}.

\subsection{Bounding the Number of Rows Required to Satisfy Assumption Set 2}
\label{sec:AssumpSet2Rows}

We will begin this section by considering item (b) of Assumption Set 2.  Among other things, this will allow us to finally define the function $E_{M}: \mathbbm{R}^D \rightarrow \mathbbm{R}^+$.  However, we must first define the Restricted Isometry Property \cite{candes2005decoding} on which the subsequent discussion relies.

\begin{Def}
Let $D, d \in \mathbbm{N}$, and $\epsilon \in (0, 1)$. An $m \times D$ matrix $M'$ has the Restricted Isometry Property, RIP($D$,$d$,$\epsilon$), if
\begin{equation}
(1 - \epsilon) \left\| \vec{x} \right\|^2 \leq \left\| M' \vec{x} \right\|^2 \leq (1 + \epsilon) \left\| \vec{x} \right\|^2
\label{equ:RIP}
\end{equation}
for all $\vec{x} \in \mathbbm{R}^D$ containing at most $d$ nonzero coordinates.
\end{Def}

We have the following lemma.

\begin{lem}
Let $\epsilon \in \left(0, \frac{1}{2} \right)$.  There exists a finite set of vectors, $Q \subset X = \left\{ \vec{y} ~\big|~ \vec{y} \in \mathbbm{R}^D \textrm{ contains }d \textrm{ nonzero coordinates} \right\}$ with $\big| Q \big| ~\leq~  {D\choose d} \left( \left( 12 / \epsilon \right)^d + 1 \right)$, such that an $m \times D$ matrix $M'$ has the RIP($D$,$d$,$\epsilon$) whenever it embeds $Q$ into $\mathbbm{R}^m$ with $\epsilon / 2$-distortion.  Furthermore, any such matrix $M'$ will have $\left\| M' \vec{y} \right\|_2$ bounded above by 
$$E_{M'} \left( \vec{y} \right) = \sqrt{1 + \epsilon} \cdot \left[ \left\| \vec{y} \right\|_2 + \frac{1}{\sqrt{d}} \left\| \vec{y} \right\|_1 \right]$$ 
for all $\vec{y} \in \mathbbm{R}^D$.
\label{lem:RIPBound}
\end{lem}

\noindent \textit{Proof:}  To prove that $M'$ has the RIP($D$,$d$,$\epsilon$) we employ an argument similar to the proof of Theorem 5.2 in \cite{baraniuk2008simple}.  To begin, we define $\vec{e}_j$, $j \in [D] - \{ 0 \}$, to be the the $j^{\rm th}$ row of the $D \times D$ identity matrix.  Then, for each $d$-element subset $S = \left\{ j_1, \dots, j_d \right\} \subset [D] - \{ 0 \}$, we define $X_S$ to be the $d$-dimensional subspace spanned by $\vec{e}_{j_1}, \dots, \vec{e}_{j_d}$.  Next, we let $Q'_{S}$ be a minimal $\epsilon / 4$-cover of the $d$-dimensional unit ball in $X_{S}$ centered at $\vec{0}$, and define $Q_{S} = Q'_{S} \bigcup \left\{ \vec{0} \right\}$ as per Lemma~\ref{lem:SubSpaceBound}.  Finally, we let
$$Q := \bigcup_{S \subset [D] - \{ 0 \},~\left|S\right|=d} Q_{S}.$$
The upper bound on $|Q|$ follows immediately.

Now suppose that $M'$ embeds $Q$ into $\mathbbm{R}^m$ with $\epsilon / 2$-distortion.  Every $\vec{x} \in \mathbbm{R}^D$ containing at most $d$ nonzero coordinates belongs to some subspace, $X_S$, whose associated set, 
$Q_{S} \subset Q$, is also embedded into $\mathbbm{R}^m$ with $\epsilon / 2$-distortion by $M'$.  Hence, a trivial variant of Lemma~\ref{lem:SubSpaceBound} guarantees that every such $\vec{x}$ will satisfy Equation~\ref{equ:RIP}.  
Therefore, $M'$ will have the RIP($D$,$d$,$\epsilon$) as claimed.  The equation for $E_{M'}$ now follows from Proposition 3.5 in \cite{needell2009cosamp}.~~$\Box$\\

We are now sufficiently equipped to consider item (a) of Assumption Set 2 in Section~\ref{sec:MeasMatrix}.  We have the following lemma.

\begin{lem}
Fix  $\epsilon \in \left(0, \frac{1}{2} \right)$ and $J \in \mathbbm{N} - \left[ \max \left\{ j_0, \log_2 \left( \frac{C_1}{{\rm reach}\left( \mathcal{M} \right)} \right) - 2 \right\} \right]$, where $j_0$ and $C_1$ are defined 
as in Property~\ref{item:MultiscaleCenters} of Section~\ref{sec:Setup}.  In addition, let $\mathbbm{P}_j$, $j \in [J]$, be a GMRA approximation to a given compact $d$-dimensional Riemannian manifold, $\mathcal{M} \subset \mathbbm{R}^D$, with $d$-dimensional volume $V$.  Then, there exist absolute universal constants, $C_3, C_4 \in \mathbbm{R}^{+}$, which are independent of both $\mathcal{M}$ and its GMRA approximation, together with a finite set of vectors, $\tilde{B} \subset \mathbbm{R}^D$, so that any $m \times D$ matrix $M'$ which embeds $\tilde{B}$ into $\mathbbm{R}^m$ with $\left( C_3 \cdot \epsilon \right)$-distortion will satisfy
$$(1-\epsilon) \left\| \vec{y} - \vec{z} \right\|^2 ~\leq~ \left\| M' \vec{y} - M' \vec{z} \right\|^2 ~\leq~(1+\epsilon) \left\| \vec{y} - \vec{z} \right\|^2$$
for all $\vec{y}, \vec{z} \in \mathcal{M} \bigcup \left\{ \vec{c}_{j,k} ~\big|~ j \in [J],~k \in [K_j] \right\} \subset \mathbbm{R}^D$.  Furthermore, $\tilde{B} \subset \mathbbm{R}^D$ will have 
$$\left| \tilde{B} \right| = O \left( 2^{C_4 J \cdot d} V^2 \left( \frac{D}{\epsilon \cdot \min \left\{ 1, {\rm reach}\left( \mathcal{M} \right) \right\} \cdot \min \left\{ 1, C_1 \right\} }  \right)^{C_4 d } \right).$$  
\label{lem:ManCbound}
\end{lem}

\noindent \textit{Proof:}  See Appendix~\ref{app:ManProof}.~~$\Box$\\

Furthermore, a modification of the proof of Lemma~\ref{lem:ManCbound} yeilds our final lemma concerning Assumption Set 2 in Section~\ref{sec:MeasMatrix}.  We have the following result regarding item (d) of Assumption Set 2.

\begin{lem}
Fix  $\epsilon \in \left(0, \frac{1}{2} \right)$ and $J \in \mathbbm{N} - \left[ \max \left\{ j_0, \log_2 \left( \frac{C_1}{{\rm reach}\left( \mathcal{M} \right)} \right) - 2 \right\} \right]$, where $j_0$ and $C_1$ are defined 
as in Property~\ref{item:MultiscaleCenters} of Section~\ref{sec:Setup}.  In addition, let $\mathbbm{P}_j$, $j \in [J]$, be a GMRA approximation to a given compact $d$-dimensional Riemannian manifold, 
$\mathcal{M} \subset \mathbbm{R}^D$, with $d$-dimensional volume $V$.  Then, there exist absolute universal constants, $C_5, C_6 \in \mathbbm{R}^{+}$, which are independent of both $\mathcal{M}$ and its GMRA approximation, 
together with a finite set of vectors, ${B}' \subset \mathbbm{R}^D$, so that any $m \times D$ matrix $M'$ which embeds ${B}'$ into $\mathbbm{R}^m$ with $\left( C_5 \cdot \epsilon \right)$-distortion will satisfy
$$(1-\epsilon) \left\| \vec{y} - \mathbbm{P}_{j,k} \left( \vec{y} \right) \right\| - 2^{-J} ~\leq~ \left\| M' \vec{y} - M' \mathbbm{P}_{j,k} \left( \vec{y} \right) \right\| ~\leq~(1+\epsilon) \left\| \vec{y} - \mathbbm{P}_{j,k} \left( \vec{y} \right) \right\| + 2^{-J}$$
for all $j \in [J],~k \in [K_j]$, and $\vec{y} \in \mathcal{M}$.  Furthermore, ${B}' \subset \mathbbm{R}^D$ will have 
$$\left| {B}' \right| = O \left( 2^{C_6 J \cdot d} V^2 \left( \frac{D}{\epsilon \cdot \min \left\{ 1, {\rm reach}\left( \mathcal{M} \right) \right\} \cdot \min \left\{ 1, C_1 \right\} }  \right)^{C_6 d } \right).$$  
\label{lem:ManApproxbound}
\end{lem}

\noindent \textit{Proof:}  See Appendix~\ref{app:ManApproxProof}.~~$\Box$\\

We are finally ready to provide a useful upper bound for the number of rows required in any measurement matrix satisfying Assumption Set 2 in Section~\ref{sec:MeasMatrix}.  We have the following theorem.

\begin{thm}
Fix  $\epsilon \in \left(0, \frac{1}{2} \right)$ and $J \in \mathbbm{N} - \left[ \max \left\{ j_0, \log_2 \left( \frac{C_1}{{\rm reach}\left( \mathcal{M} \right)} \right) - 2 \right\} \right]$, where $j_0$ and $C_1$ are defined as in Property~\ref{item:MultiscaleCenters} of Section~\ref{sec:Setup}.  In addition, let $\mathbbm{P}_j$, $j \in [J]$, be a GMRA approximation to a given compact $d$-dimensional Riemannian manifold, $\mathcal{M} \subset \mathbbm{R}^D$, with $d$-dimensional volume $V$.  Then, there exists an $m \times D$ matrix, $M$, which satisfies Assumption Set 2 in Section~\ref{sec:MeasMatrix} with 
$$m =  O \left( d \epsilon^{-2} \log \left( \frac{D}{\epsilon \cdot {\rm reach}\left( \mathcal{M} \right)} \right) + d \epsilon^{-2}  J + \epsilon^{-2} \log V \right)$$ 
and 
$$E_{M} \left( \vec{y} \right) = \sqrt{1 + \epsilon} \cdot \left[ \left\| \vec{y} \right\|_2 + \frac{1}{\sqrt{d}} \left\| \vec{y} \right\|_1 \right].$$
\label{thm:Assump2rowbound}
\end{thm}

\noindent \textit{Proof:}  Any $m \times D$ matrix which embeds $\tilde{B} \subset \mathbbm{R}^D$ from Lemma~\ref{lem:ManCbound} into $\mathbbm{R}^m$ with $\left( C_3 \cdot \epsilon \right)$-distortion will satisfy both items (a) 
and (b) of Assumption Set 2 in Section~\ref{sec:MeasMatrix} (see Lemmas~\ref{lem:RIPBound} and \ref{lem:ManCbound}).  Similarly, any given $m \times D$ 
matrix which embeds $B' \subset \mathbbm{R}^D$ from Lemma~\ref{lem:ManApproxbound} into $\mathbbm{R}^m$ with $\left( C_5 \cdot \epsilon \right)$-distortion will satisfy item (d) of Assumption Set 2.  Finally, just as in the proof 
of Theorem~\ref{thm:Assump1rowbound} above, Lemma~\ref{lem:SubSpaceBound} applied to all at most $(J+1) K_J$ affine projectors yields a subset of $\mathbbm{R}^D$ of size at most 
$(J+1) K_J \left(\left( 12 / \epsilon \right)^d + 1 \right)$ for item (c) of Assumption Set 2.  Theorem~\ref{thm:JLexists} 
applied to the union of this subset with $\tilde{B} \cup B'$ guarantees the existence of 
$$O \left( \epsilon^{-2} \log \left( \left| \tilde{B} \right| + \left|B'\right| + (J+1) K_J \left(\left( 12 / \epsilon \right)^d + 1 \right) \right) \right) \times D$$ 
Johnson-Lindenstrauss embedding matrices which satisfy Assumption Set 2 with high probability.  Applying Lemmas~\ref{lem:CenterBound},~\ref{lem:ManCbound}, and~\ref{lem:ManApproxbound} to bound $K_J$, $\left| \tilde{B} \right|$, 
and $\left|B'\right|$, respectively, now finishes the proof.~~$\Box$\\

In the proof of Theorem~\ref{thm:Assump2rowbound} above, Theorem~\ref{thm:JLexists} is applied to embed a set which only depends on the given manifold, $\mathcal{M}$, and its GMRA approximation.  More specifically, no knowledge was 
assumed regarding any point $\vec{x} \in \mathbbm{R}^D - \mathcal{M}$ which we might be interested in approximating via Algorithm~\ref{alg:RecovAlg}.  Thus, Theorem~\ref{thm:Assump2rowbound} provides us with a uniform approximation 
guarantee for all $\vec{x} \in \mathbbm{R}^D$ on which we might apply Algorithm~\ref{alg:RecovAlg}.  However, we pay several penalties for this uniformity.  First, the number of rows in our measurement matrix, $m$, now depends on 
the \textit{extrinsic dimensionality}, $D$, of the given manifold.  Second, the resulting uniform error bounds are only nontrivial for input points, $\vec{x}$, which are close to the given manifold.  Hence, although 
Theorem~\ref{thm:Assump2rowbound} implies that Algorithm~\ref{alg:RecovAlg} enjoys a limited form of stability, it does not provide very robust uniform error guarantees in practice.

\section{Empirical Evaluation}
\label{sec:Empirical}

We implemented Algorithm~\ref{alg:RecovAlg} and present an empirical evaluation of the algorithm in this section.\footnote{All code is freely available at \url{http://www.math.duke.edu/~mauro}}
We consider the following examples:
\begin{itemize}
\item[(i)] $\mathcal{M}_1$: $20,000$ points sampled from a ``swiss roll'', a $2$-dimensional manifold $\mathcal{S}$;
\item[(ii)] $\mathcal{M}_2$: $40,000$ points sampled from a unit $9$-dimensional sphere $\mathbb{S}^9$;
\item[(iii)] $\mathcal{M}_3$:  $5,000$ pictures of the digit `$1$' from the MNIST data base of images, $28\times28$ pixels, of handwritten digits\footnote{Available at \url{http://yann.lecun.com/exdb/mnist/.}}, with each picture having pixel intensity normalized to have unit $L^2$ norm.
\item[(iii)] $\mathcal{M}_4$:  $15,000$ points from the MNIST data base, with $5,000$ points sampled from each of the digits $1,3,5$, with each picture having pixel intensity normalized to have unit $L^2$ norm.
\item[(iv)] $\mathcal{M}_5$: the Science News text document data set, which comprises $1163$ text documents, modeled as vectors in $1153$ dimensions, whose $i$-th entry is the frequency of the $i$-th word in a dictionary (see \cite{CM:MsDataDiffWavelets} for detailed information about this data set), normalized so that every document vector has unit Euclidean norm.
\end{itemize}
We construct the GMRA on these data sets in order to obtain the linear approximations, $\mathbbm{P}_j$ for each scale $j$ considered below in the noiseless setting.

\begin{figure}[t]
\centering
\subcaptionbox{$\mathcal{M}_1$}
{\includegraphics[width=0.19\textwidth]{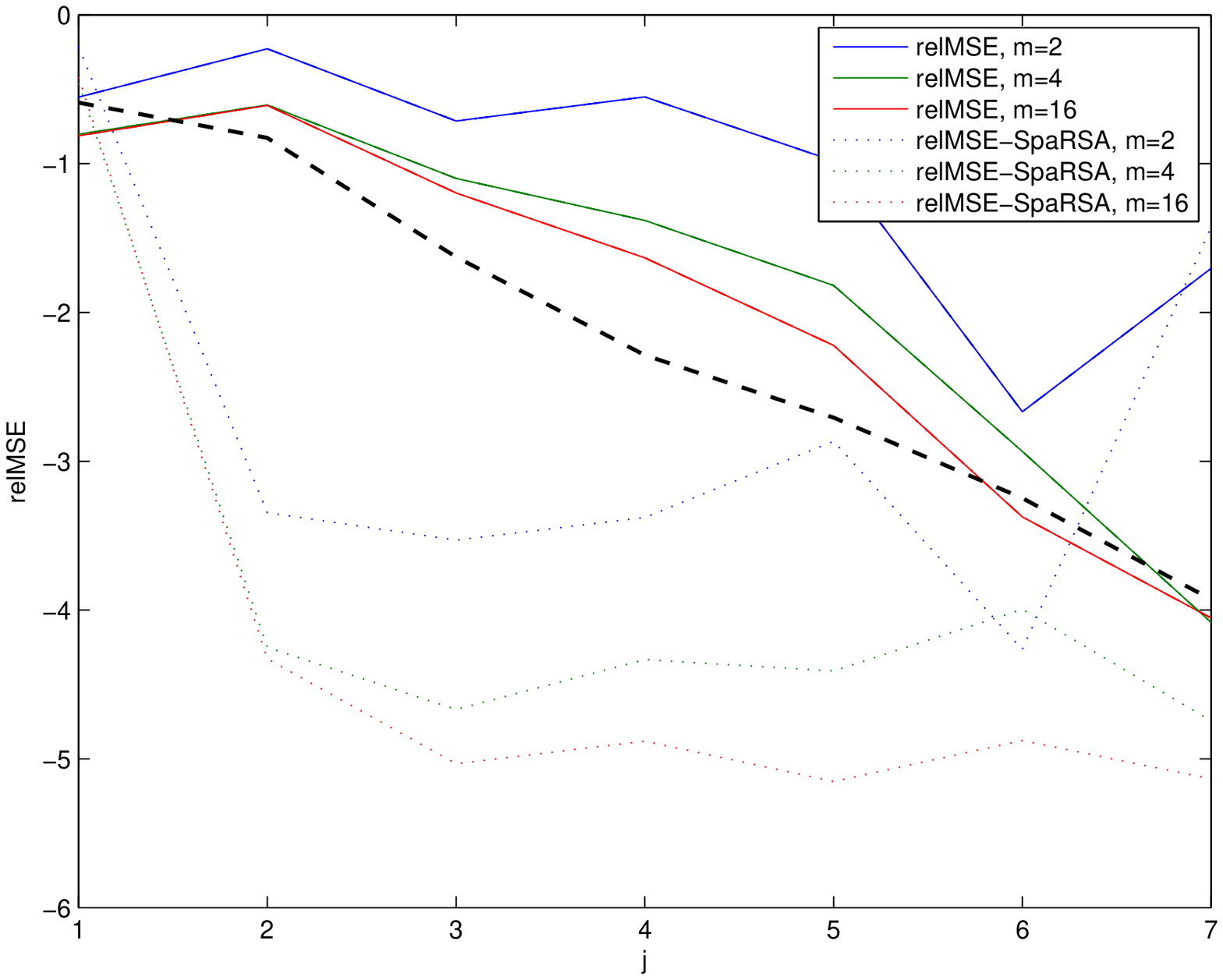}}
\subcaptionbox{$\mathcal{M}_2$}
{\includegraphics[width=0.19\textwidth]{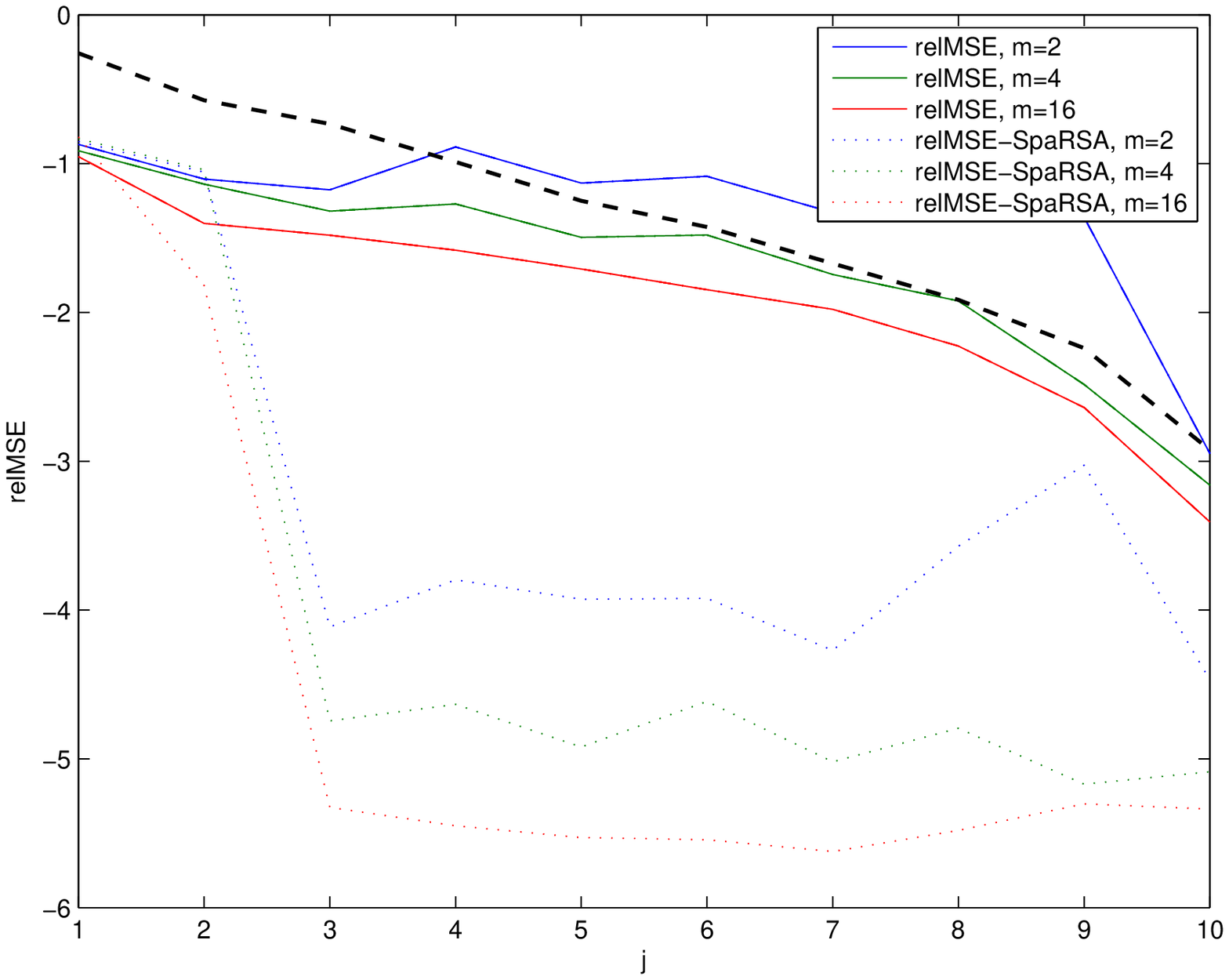}}
\subcaptionbox{$\mathcal{M}_3$}
{\includegraphics[width=0.19\textwidth]{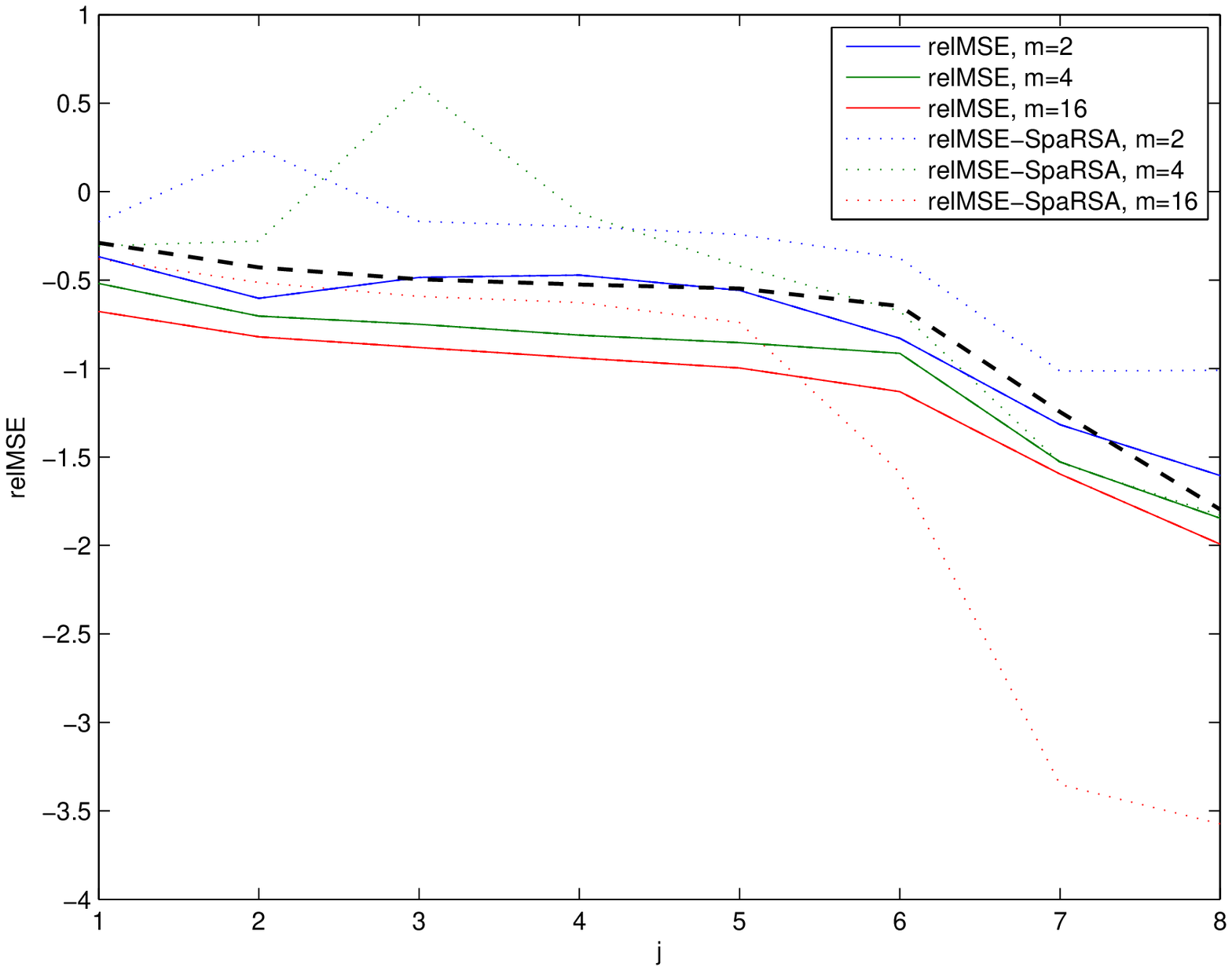}}
\subcaptionbox{$\mathcal{M}_4$}
{\includegraphics[width=0.19\textwidth]{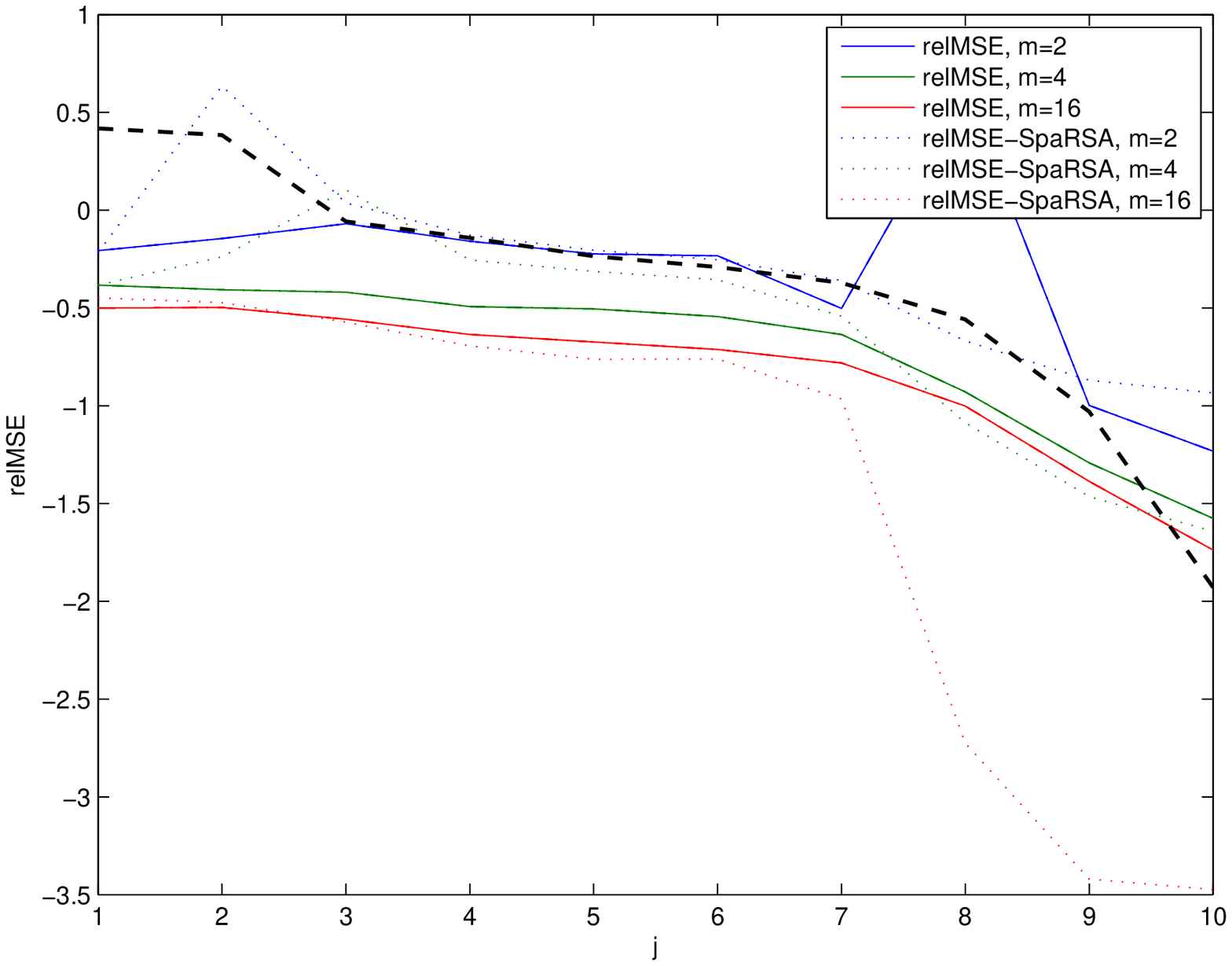}}
\subcaptionbox{$\mathcal{M}_5$}
{\includegraphics[width=0.19\textwidth]{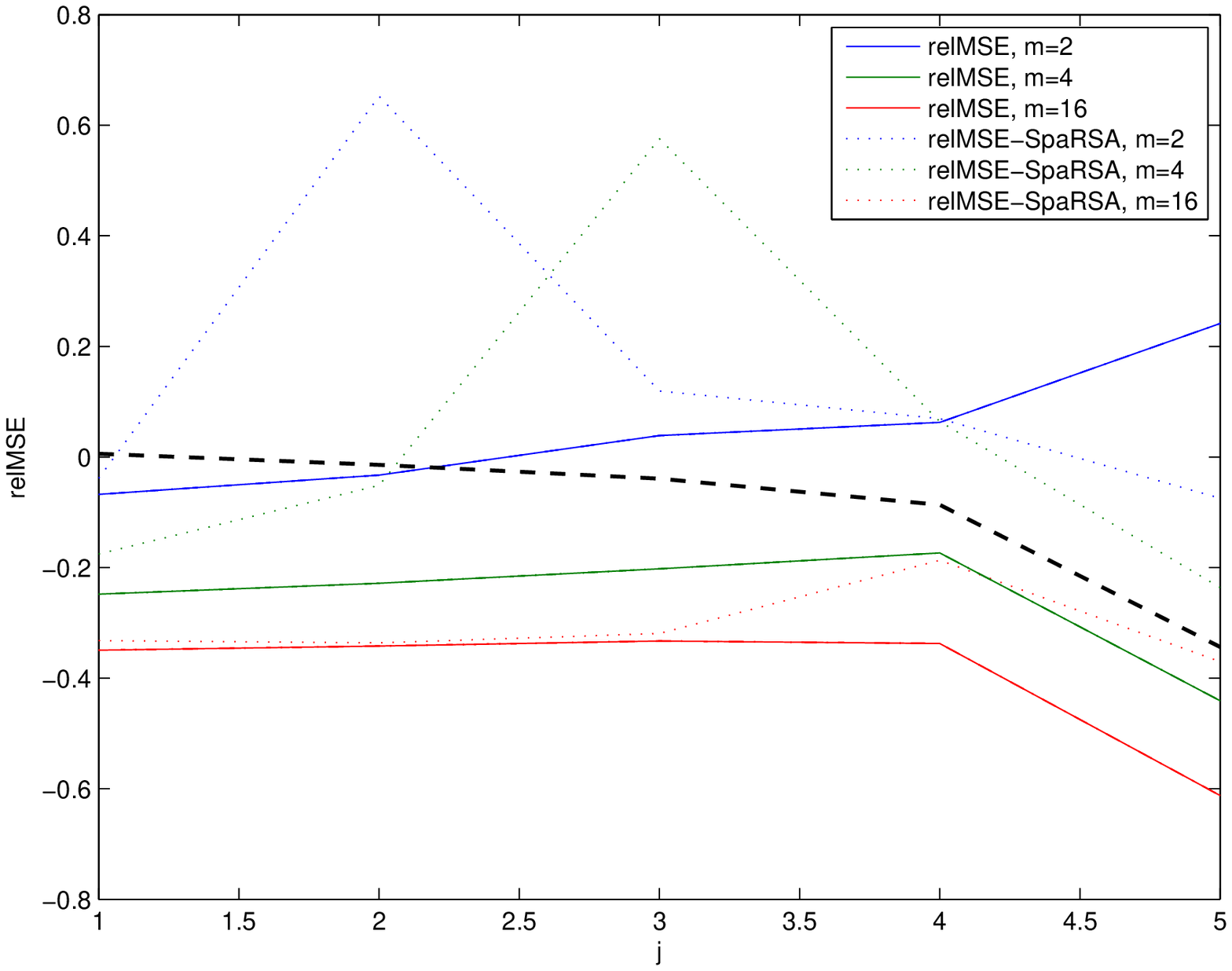}}
\subcaptionbox{$\mathcal{M}_1+\mathcal{N}(0,\frac{0.05^2}{D}I_D)$}
{\includegraphics[width=0.19\textwidth]{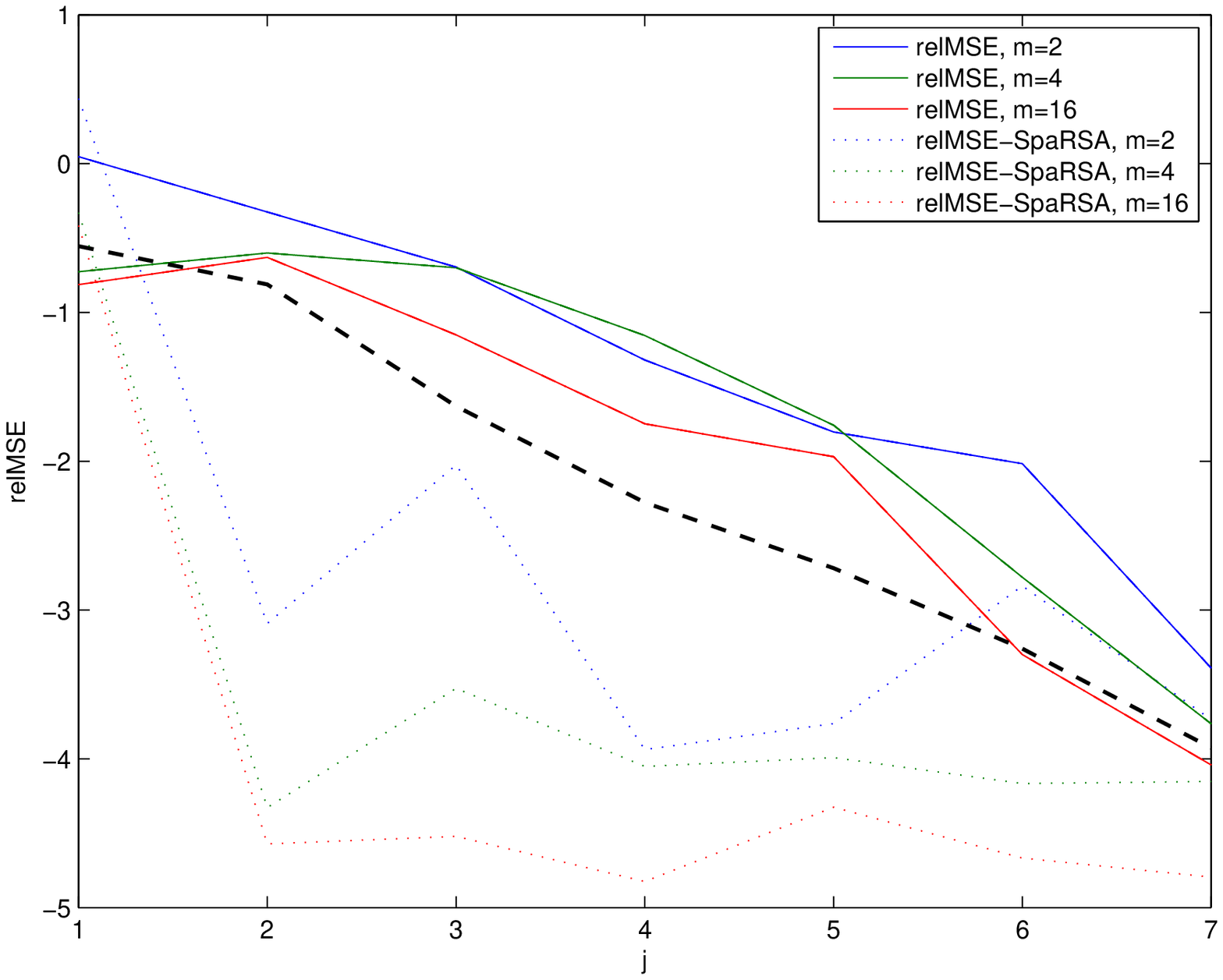}}
\subcaptionbox{$\mathcal{M}_2+\mathcal{N}(0,\frac{0.05^2}{D}I_D)$}
{\includegraphics[width=0.19\textwidth]{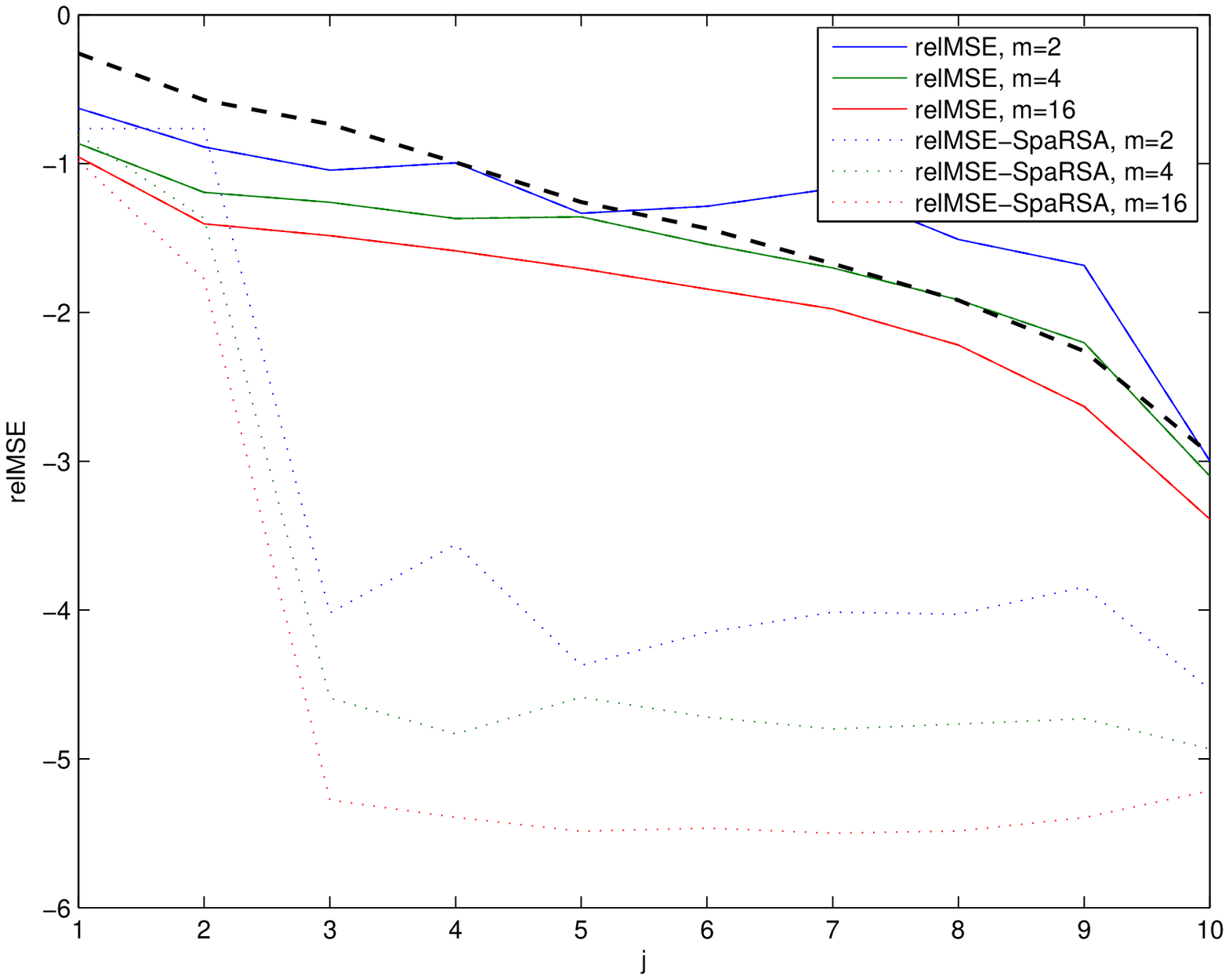}}
\subcaptionbox{$\mathcal{M}_3+\mathcal{N}(0,\frac{0.05^2}{D}I_D)$}
{\includegraphics[width=0.19\textwidth]{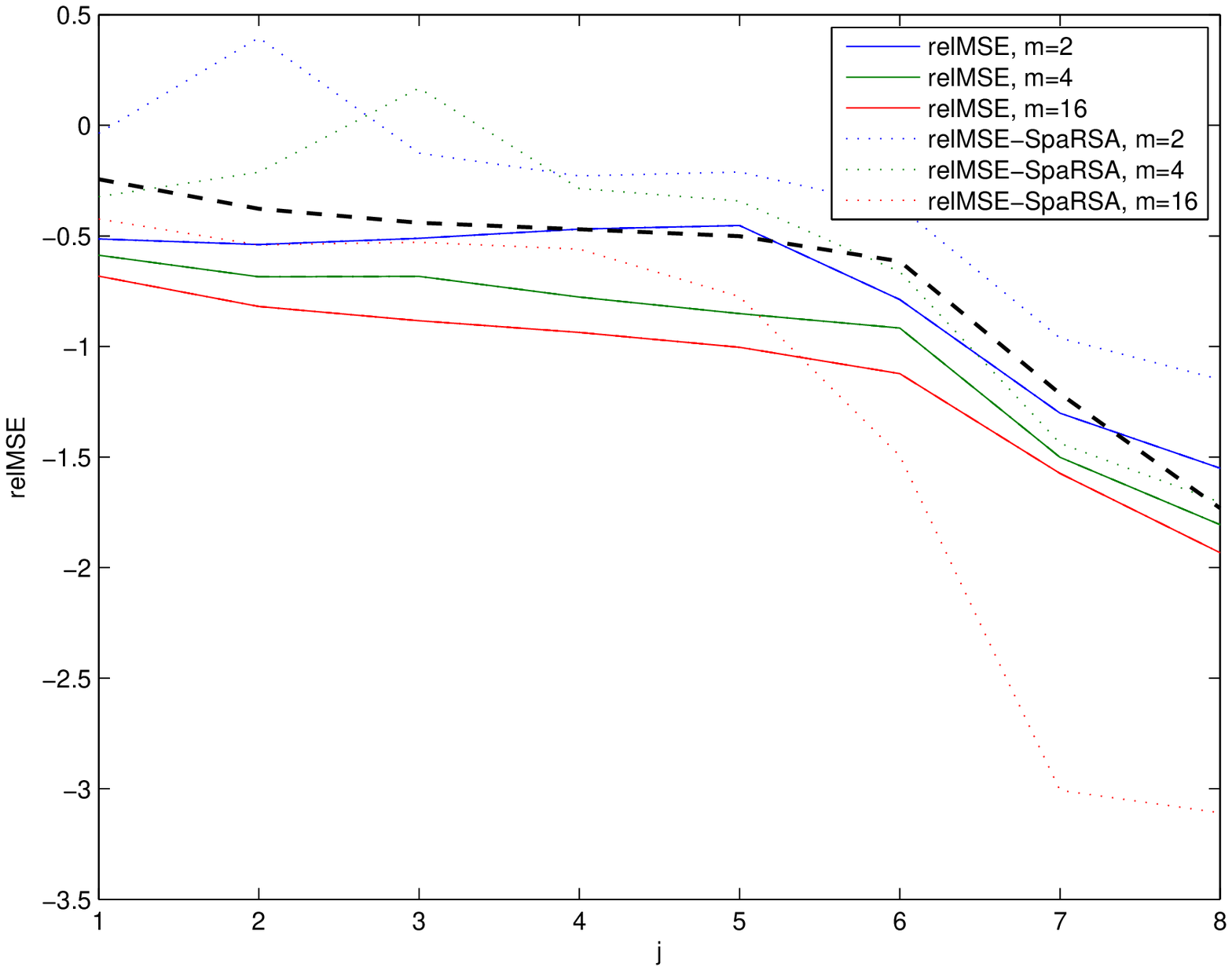}}
\subcaptionbox{$\mathcal{M}_4+\mathcal{N}(0,\frac{0.05^2}{D}I_D)$}
{\includegraphics[width=0.19\textwidth]{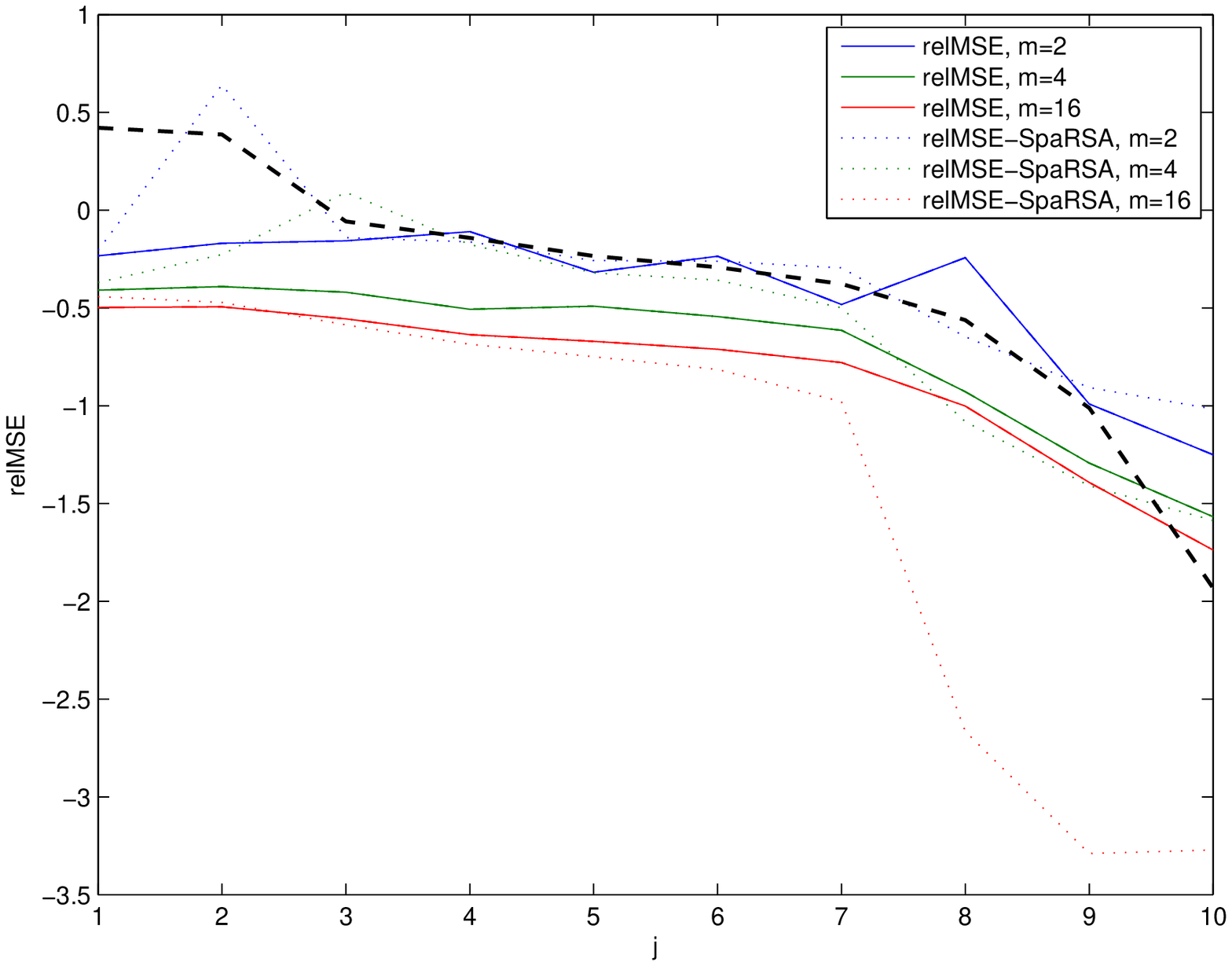}}
\subcaptionbox{$\mathcal{M}_5+\mathcal{N}(0,\frac{0.05^2}{D}I_D)$}
{\includegraphics[width=0.19\textwidth]{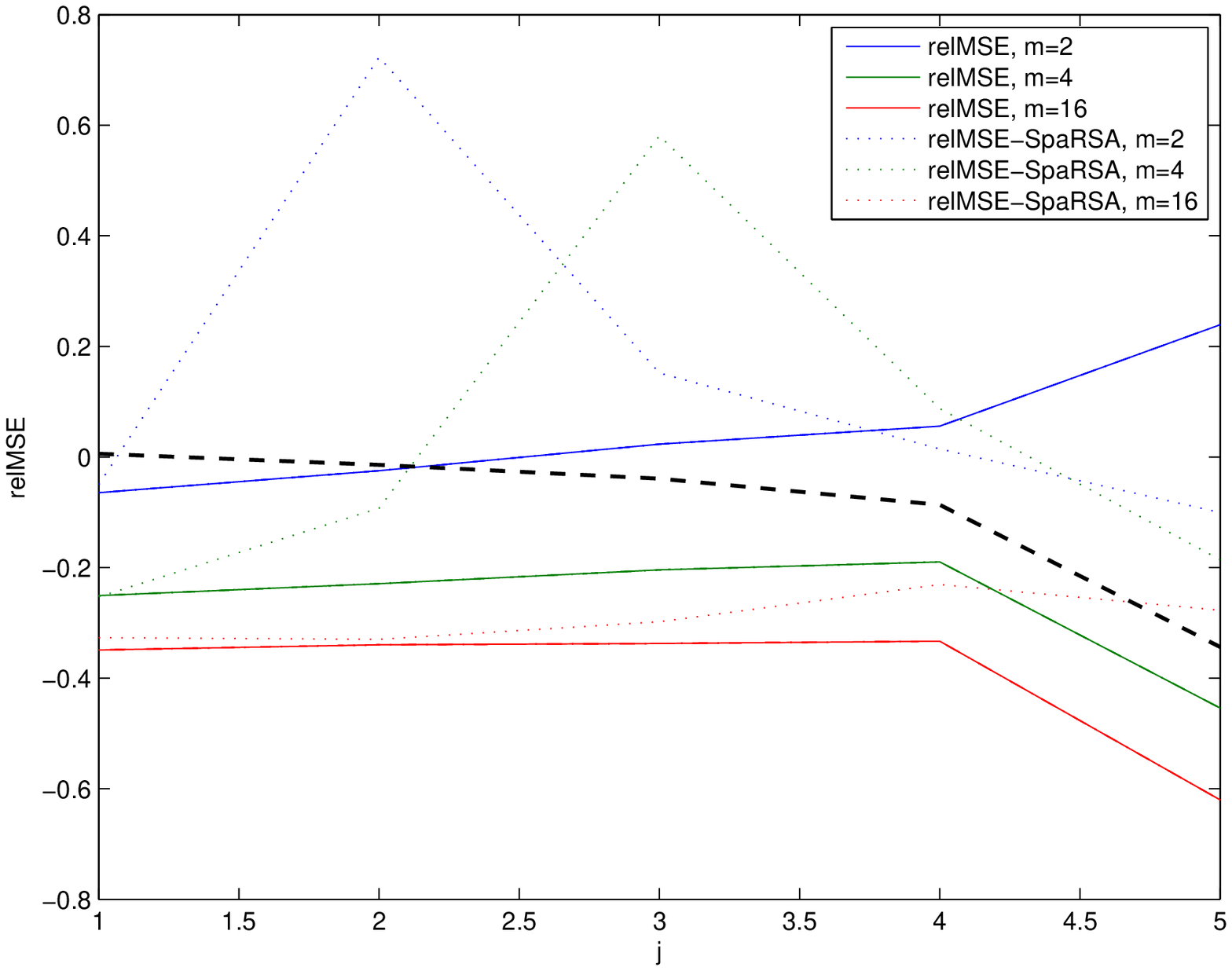}}
\subcaptionbox{$\mathcal{M}_1+\mathcal{N}(0,\frac{0.05^2}{D}I_D)$}
{\includegraphics[width=0.19\textwidth]{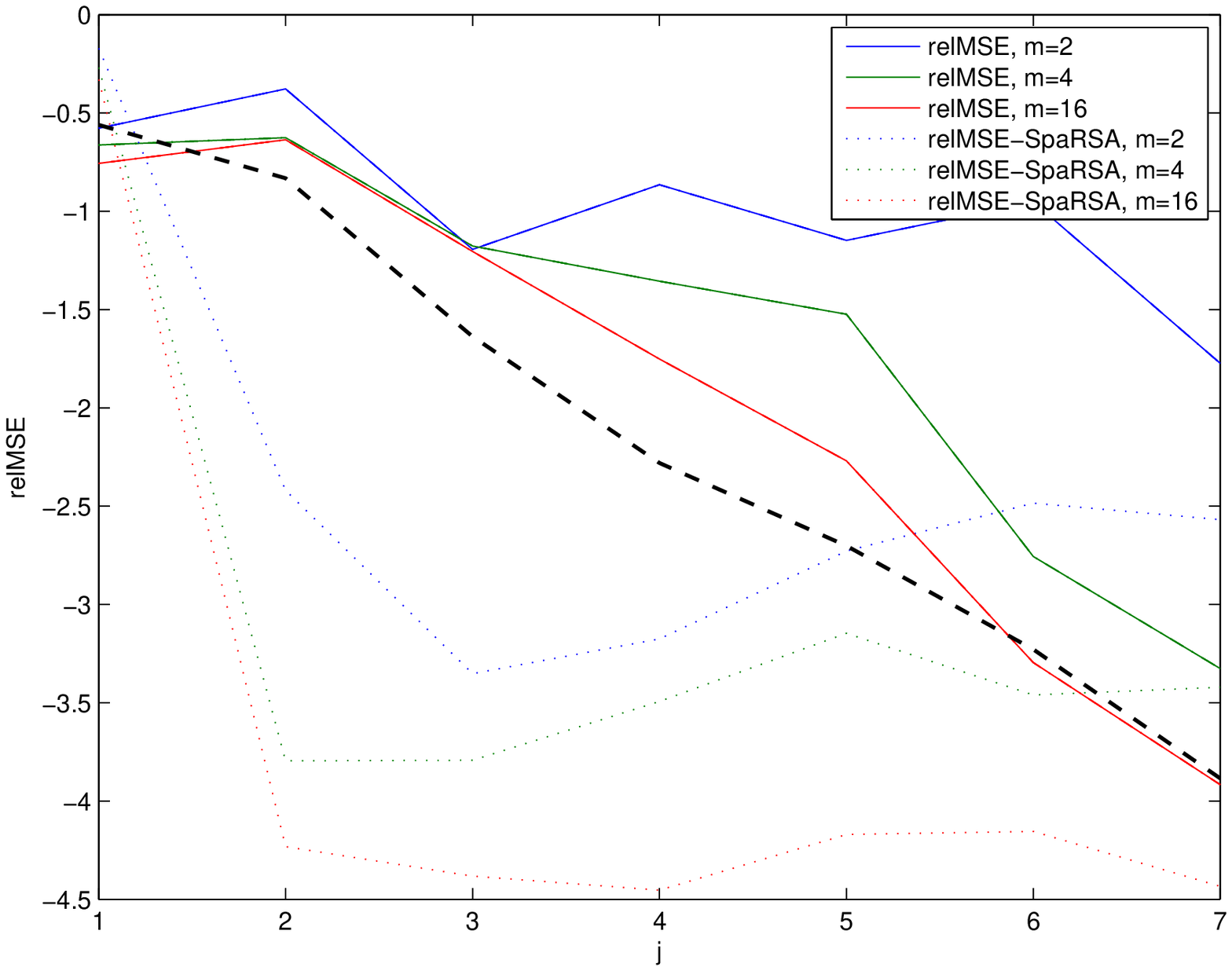}}
\subcaptionbox{$\mathcal{M}_2+\mathcal{N}(0,\frac{0.1^2}{D}I_D)$}
{\includegraphics[width=0.19\textwidth]{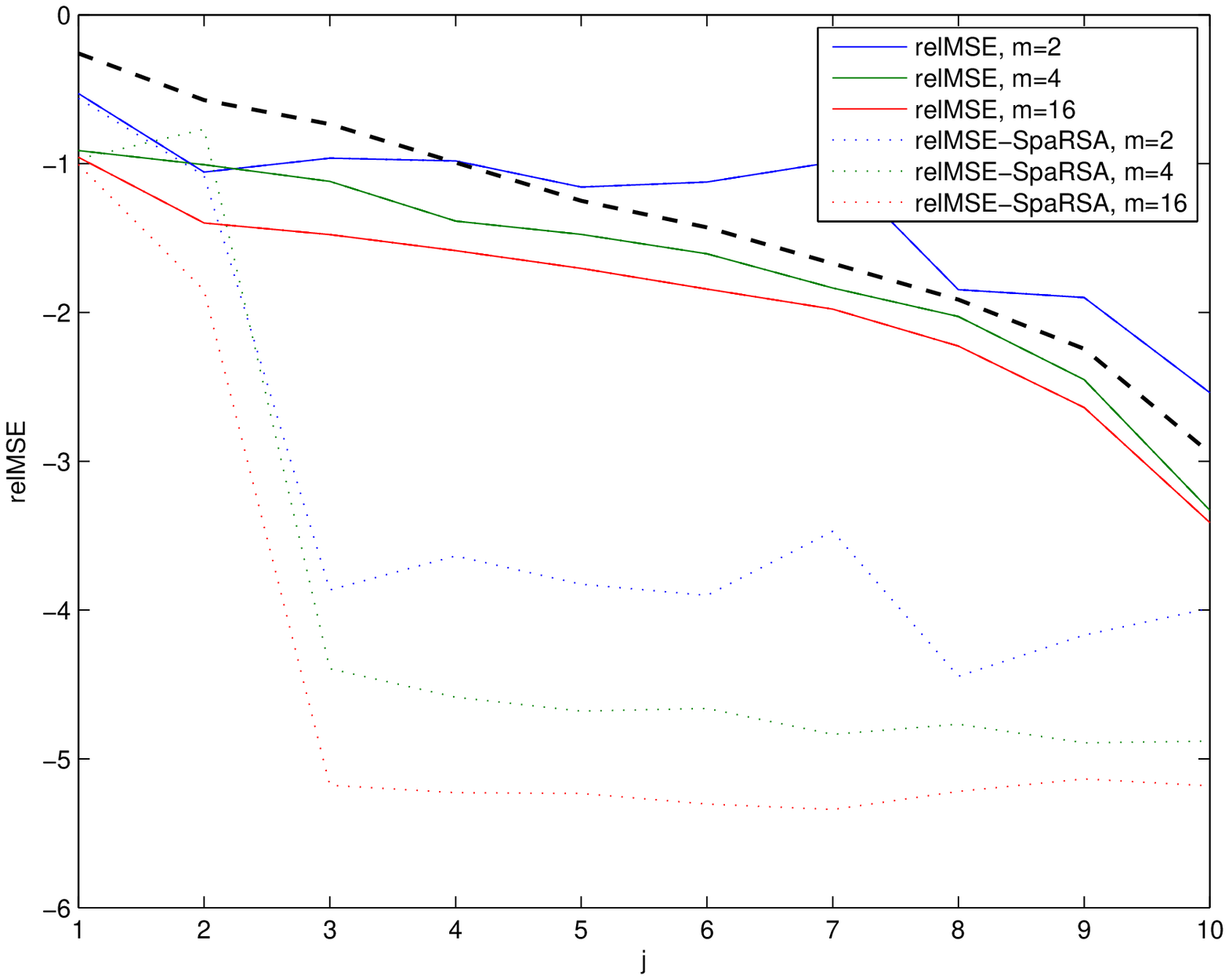}}
\subcaptionbox{$\mathcal{M}_3+\mathcal{N}(0,\frac{0.1^2}{D}I_D)$}
{\includegraphics[width=0.19\textwidth]{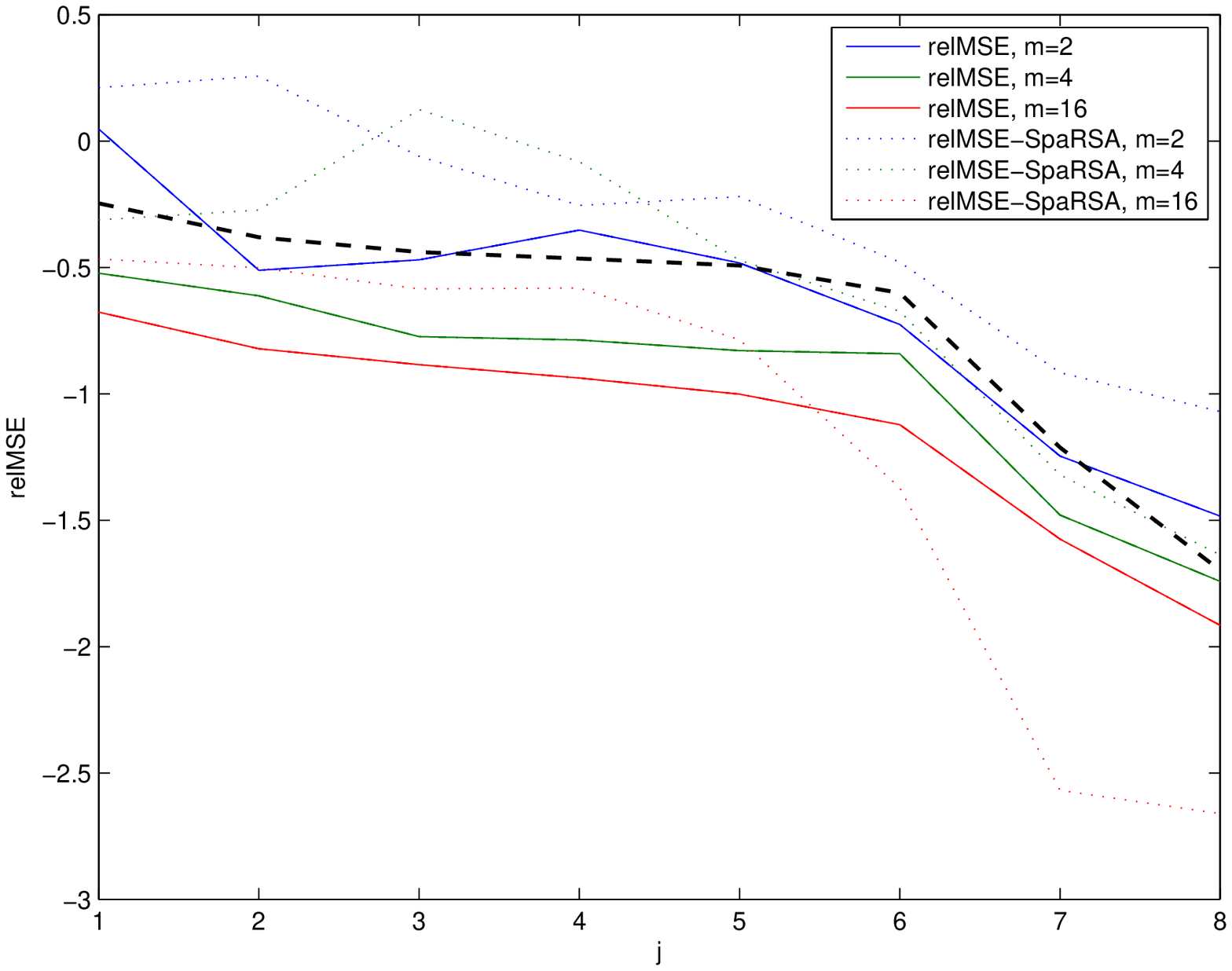}}
\subcaptionbox{$\mathcal{M}_4+\mathcal{N}(0,\frac{0.1^2}{D}I_D)$}
{\includegraphics[width=0.19\textwidth]{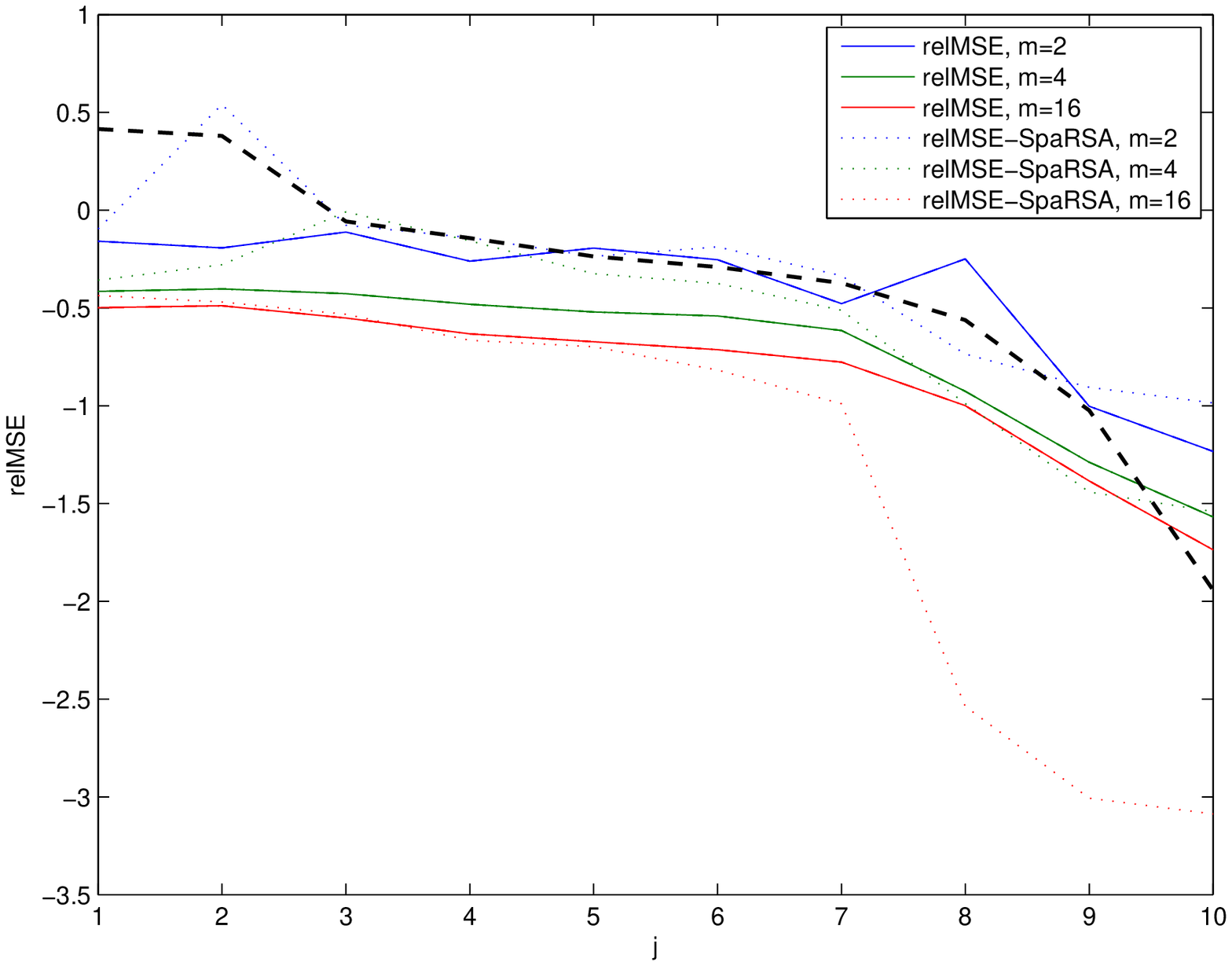}}
\subcaptionbox{$\mathcal{M}_5+\mathcal{N}(0,\frac{0.1^2}{D}I_D)$}
{\includegraphics[width=0.19\textwidth]{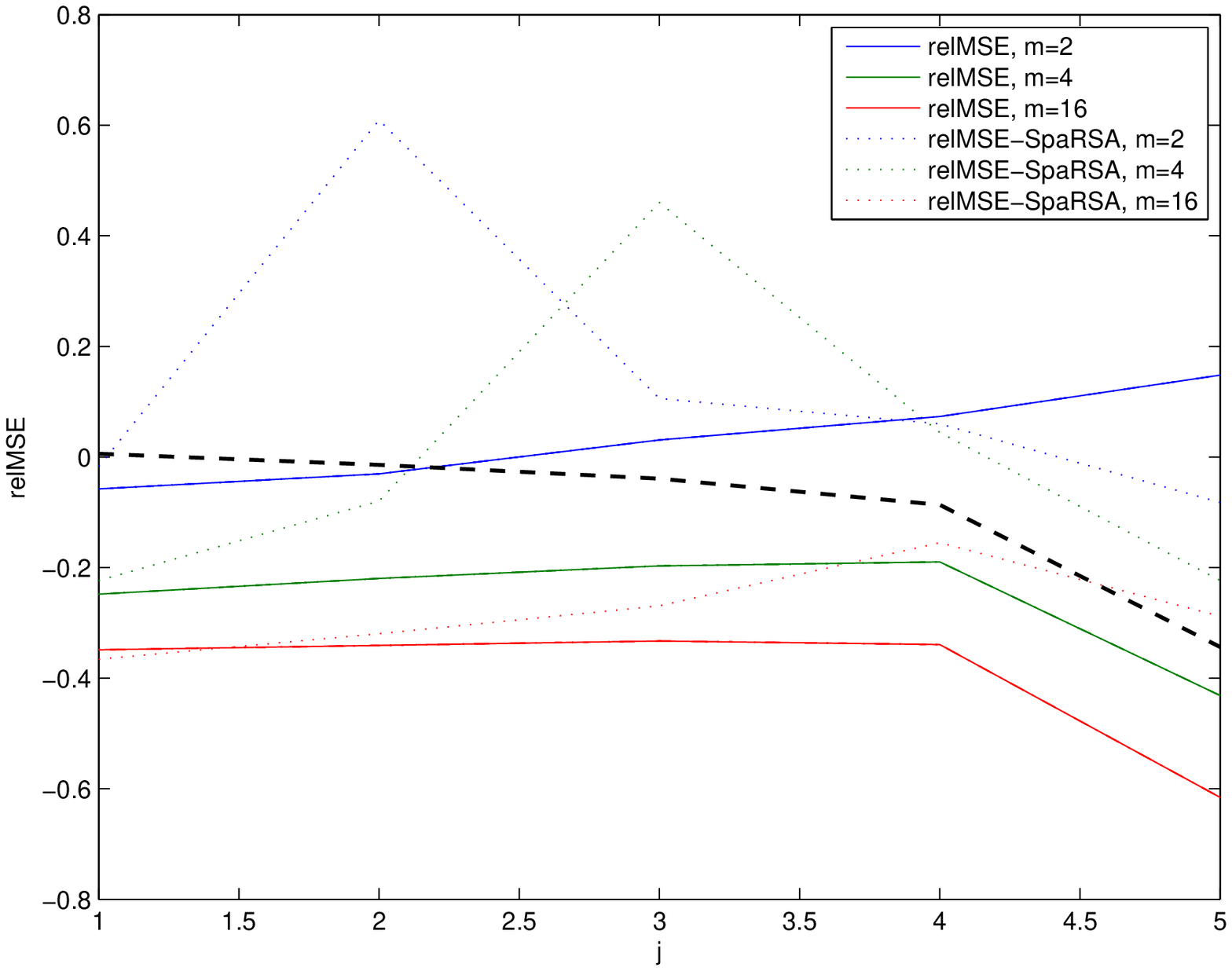}}
\caption{From left to right: data sets $\mathcal{M}_1$ to $\mathcal{M}_5$; from top to bottom: adding noise with increasing variance as above. In each plot, having the horizontal axis representing the scale $j$ and the vertical axis the relative mean square errors $\mathrm{relMSE}(\mathcal{A},M,j)$ as defined in \eqref{e:relerr}, we vary the oversampling parameter $m=2,4,16$, and we also plot (dashed black) the $\mathrm{relMSE_J}$ defined again \eqref{e:relerr}. We report the average result of $10$ draws of the random matrix $M$, and in dashed lines the standard deviation bands around the mean. Note: the maximum square error is about $10$ times larger than the mean square error in all cases (not shown). For $\mathcal{M}_3$ we have $(d_j)_j=(3,3,3,3,9,37,45,45)$ , and for $\mathcal{M}_4$ we have $d_j=(9,7,7,7,6,6,9,36,61,66,66)$, and for $\mathcal{M}_5$ and $d_j=(83,51,33,21,43,50)$. We also run \textrm{SpaRSA} \cite{DBLP:journals/tsp/WrightNF09} (see comments in the text).}
\label{f:exs}
\end{figure}

For the noisy experiments we add Gaussian noise, $\mathcal{N}(0,\frac{\sigma^2}{D} I_D)$ where $D$ is the (ambient) dimension of the data, to each data point for $\sigma=0,0.05,0.1$.  We then use the noisy data to compute the GMRA approximations of the noisy data, as well as the random projections utilized by the proposed reconstruction algorithm $\mathcal{A}$.  We consider the following measures of approximation:
\begin{align}
\mathrm{relMSE(\mathcal{A},M,j)}^2:=\frac1n\sum_{i=1}^n \frac{ ||\vec{x}_i-\mathcal{A}\left(M \vec{x}_i\right)||^2}{||\vec{x}_i||^2}\quad,\quad
\mathrm{relMSE_J}^2:=\frac1n\sum_{i=1}^n \frac{||\vec{x}_i-\mathbb{P}_{J}\left(\vec{x}_i\right)||^2}{||\vec{x}_i||^2}
\label{e:relerr}
\end{align}
where $\{\vec{x}_i\}_{i=1}^n$ are the data points, $j$ is the level in the GMRA, ranging from $0$ to $J$ (dependent on the data set), $\mathcal{A}$ is the proposed Algorithm, and $M$ is a fixed random (with respect to Haar measure) orthogonal projection with range of dimension $(d_j\cdot m)\wedge D$, where the ``oversampling factor'' $m=1,2,4,16$, and the ``intrinsic dimension'' $d_j=\max_k\mathrm{dim}(\mathrm{range}(\mathbb{P}_{j,k}))$. Therefore, $d_j$ is the dimension of the manifold ($2$ and $9$, respectively) for $\mathcal{M}_1$ and $\mathcal{M}_2$.  The dimension parameter, $d_j$, is adaptively chosen in a scale-dependent way for $\mathcal{M}_3,\mathcal{M}_4,\mathcal{M}_5$ as described in \cite{CM:MGM2}, with actual values used in these examples reported in Figure \ref{f:exs}.

There we also run \textrm{SpaRSA} \cite{DBLP:journals/tsp/WrightNF09} (for reasonable choices of the several parameters involved), one of the leading algorithms, among many, for sparse reconstructions. We notice that: (a) for general real world data sets it achieves comparable precision to our algorithm, suggesting that the GMRA dictionaries may be used in the context of standard sparse approximation; (b) for low-dimensional manifold synthetic data sets, which do not curve in many dimensions, it achieves higher accuracy, since the directions of a few tangent planes are sufficient to span a subspace containing the whole manifold.

\begin{figure}[t]
\centering
\subcaptionbox{$\mathcal{M}_1$}
{\includegraphics[width=0.19\textwidth]{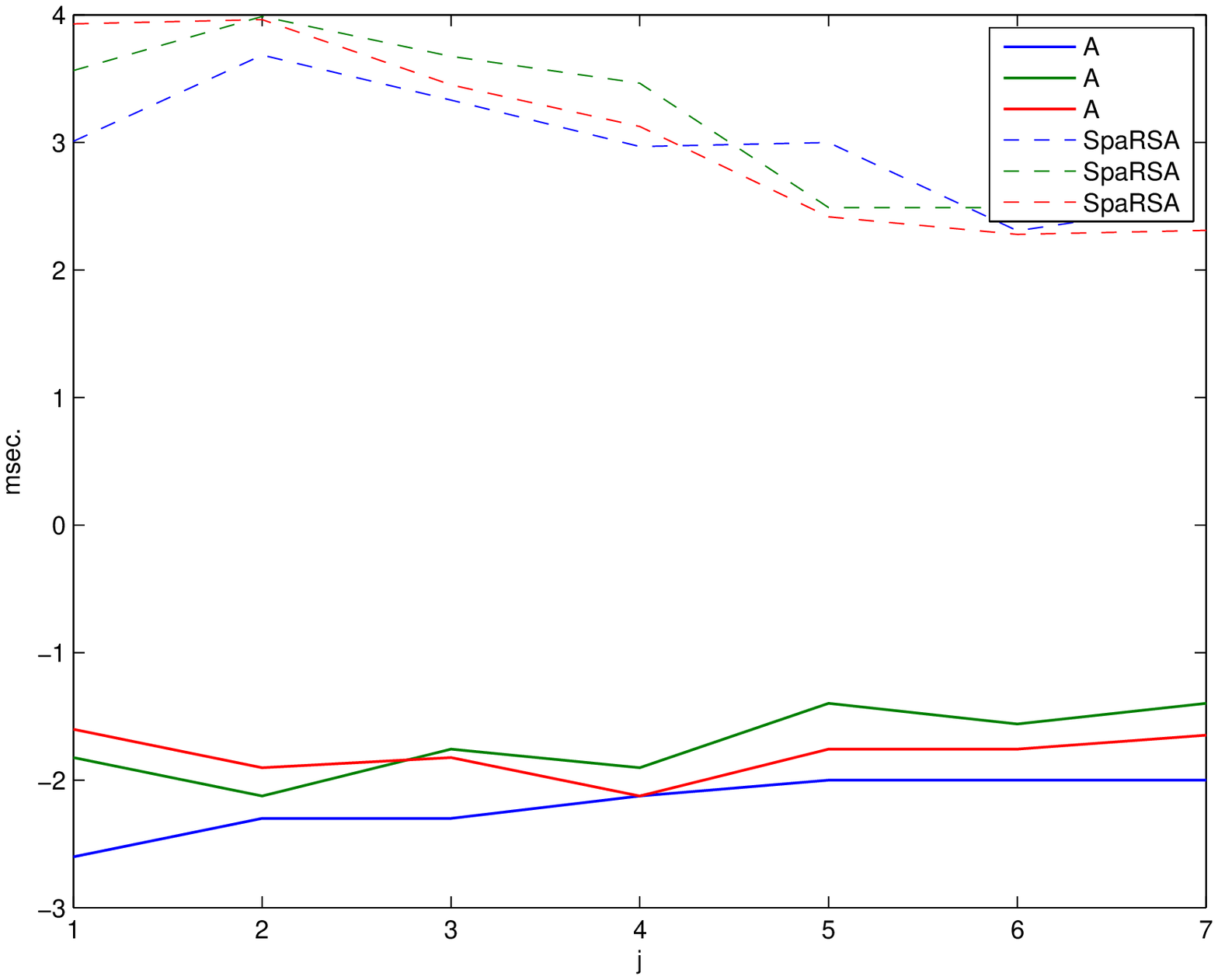}}
\subcaptionbox{$\mathcal{M}_2$}
{\includegraphics[width=0.19\textwidth]{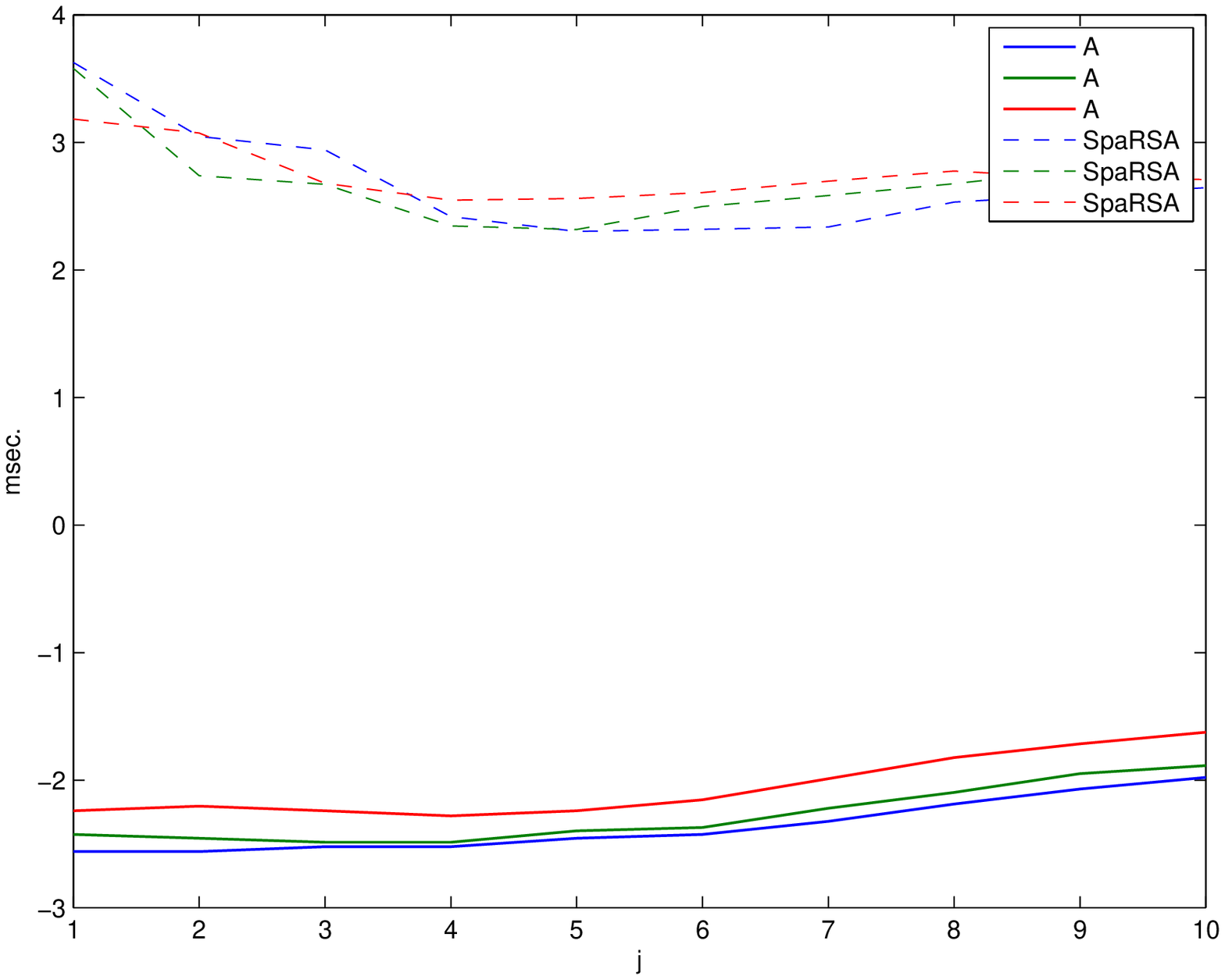}}
\subcaptionbox{$\mathcal{M}_3$}
{\includegraphics[width=0.19\textwidth]{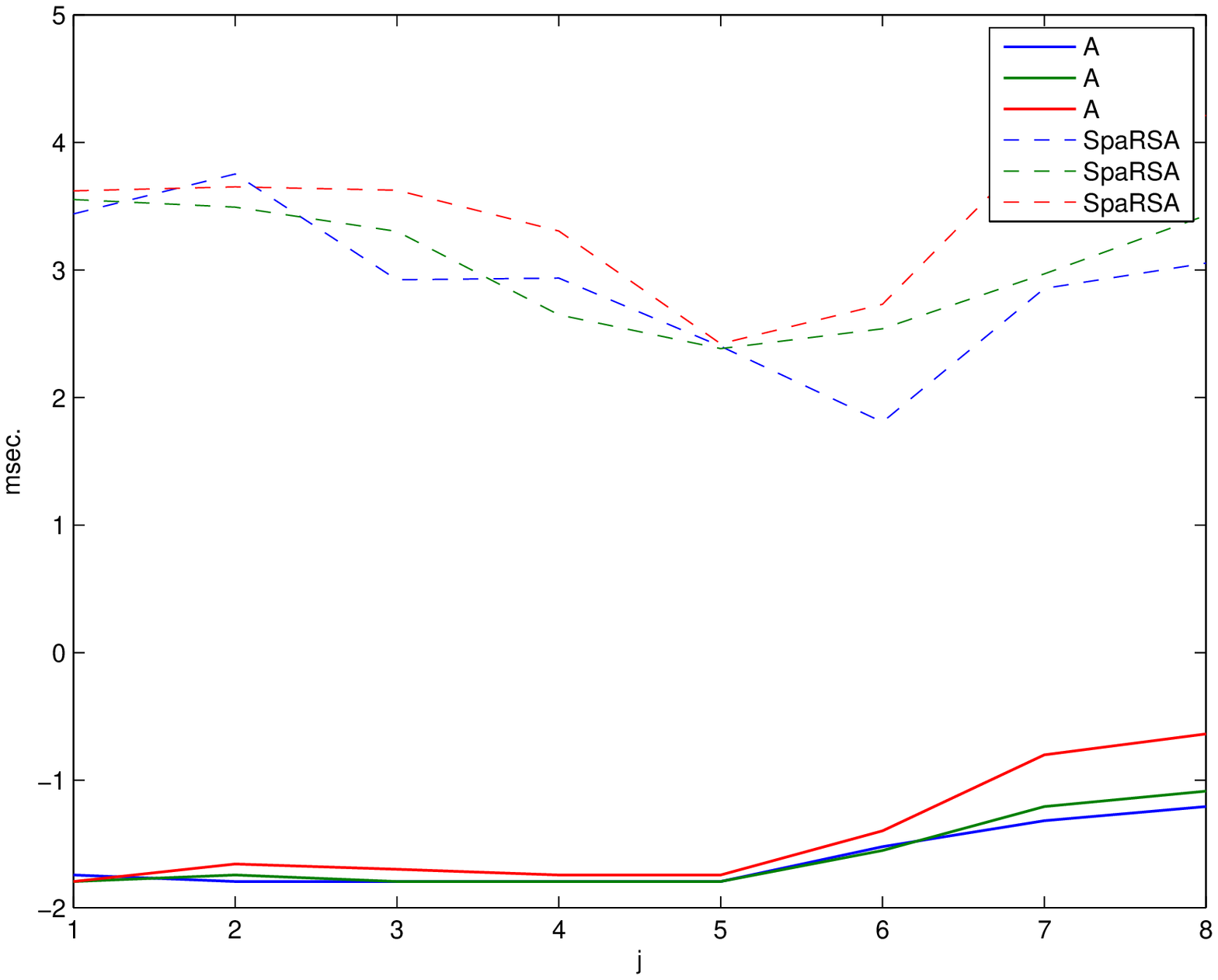}}
\subcaptionbox{$\mathcal{M}_4$}
{\includegraphics[width=0.19\textwidth]{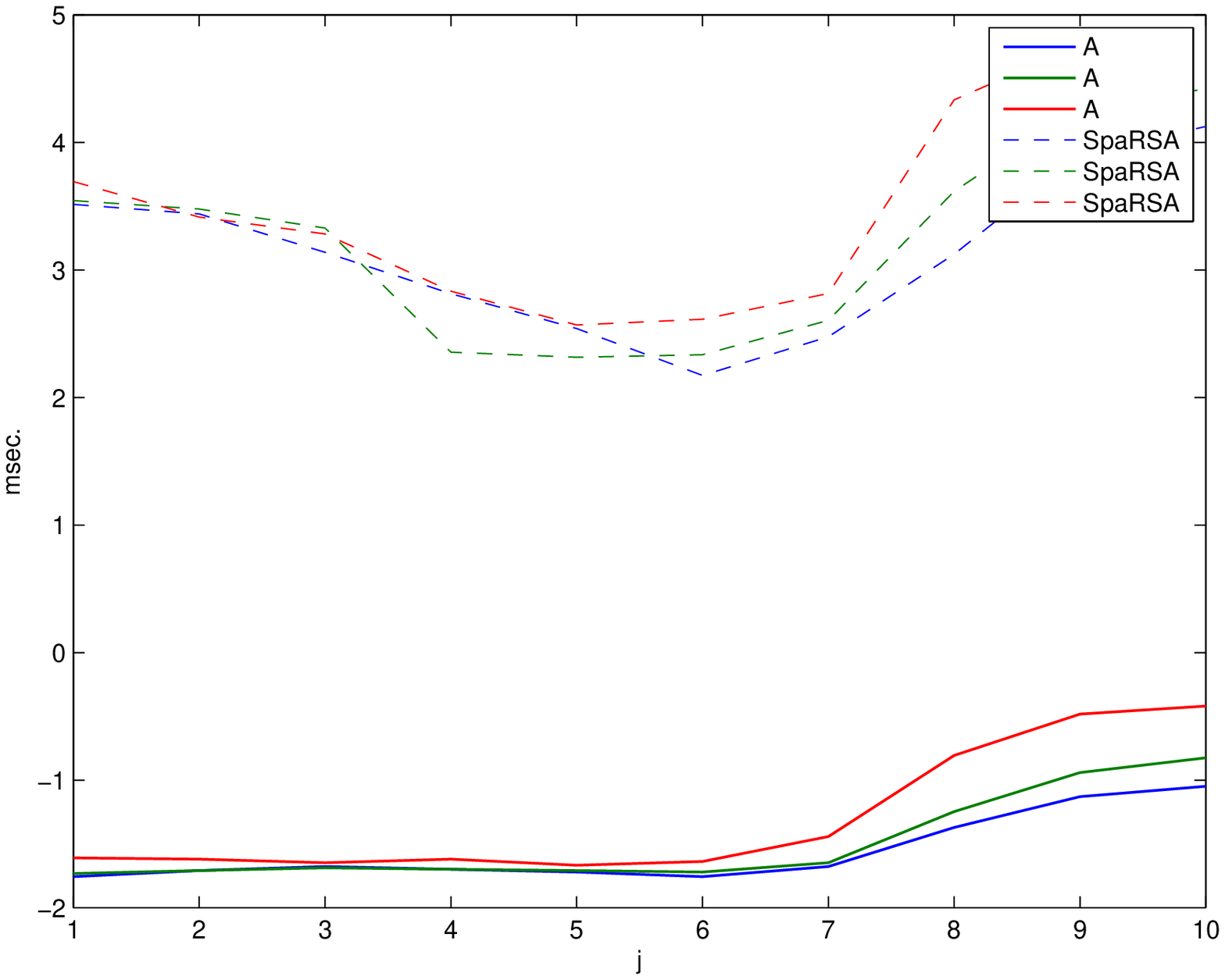}}
\subcaptionbox{$\mathcal{M}_5$}
{\includegraphics[width=0.19\textwidth]{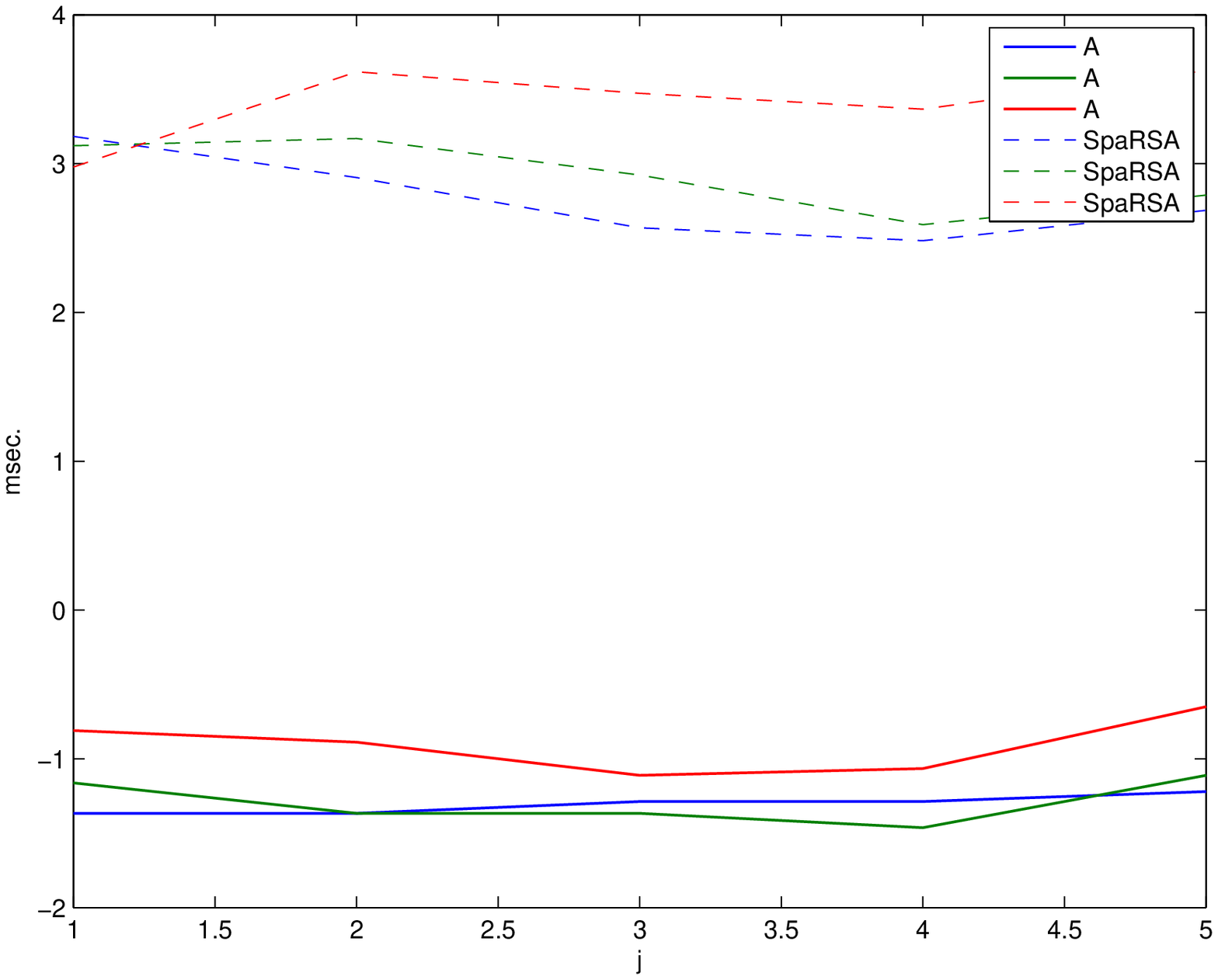}}
\subcaptionbox{$\mathcal{M}_1+\mathcal{N}(0,\frac{0.05^2}{D}I_D)$}
{\includegraphics[width=0.19\textwidth]{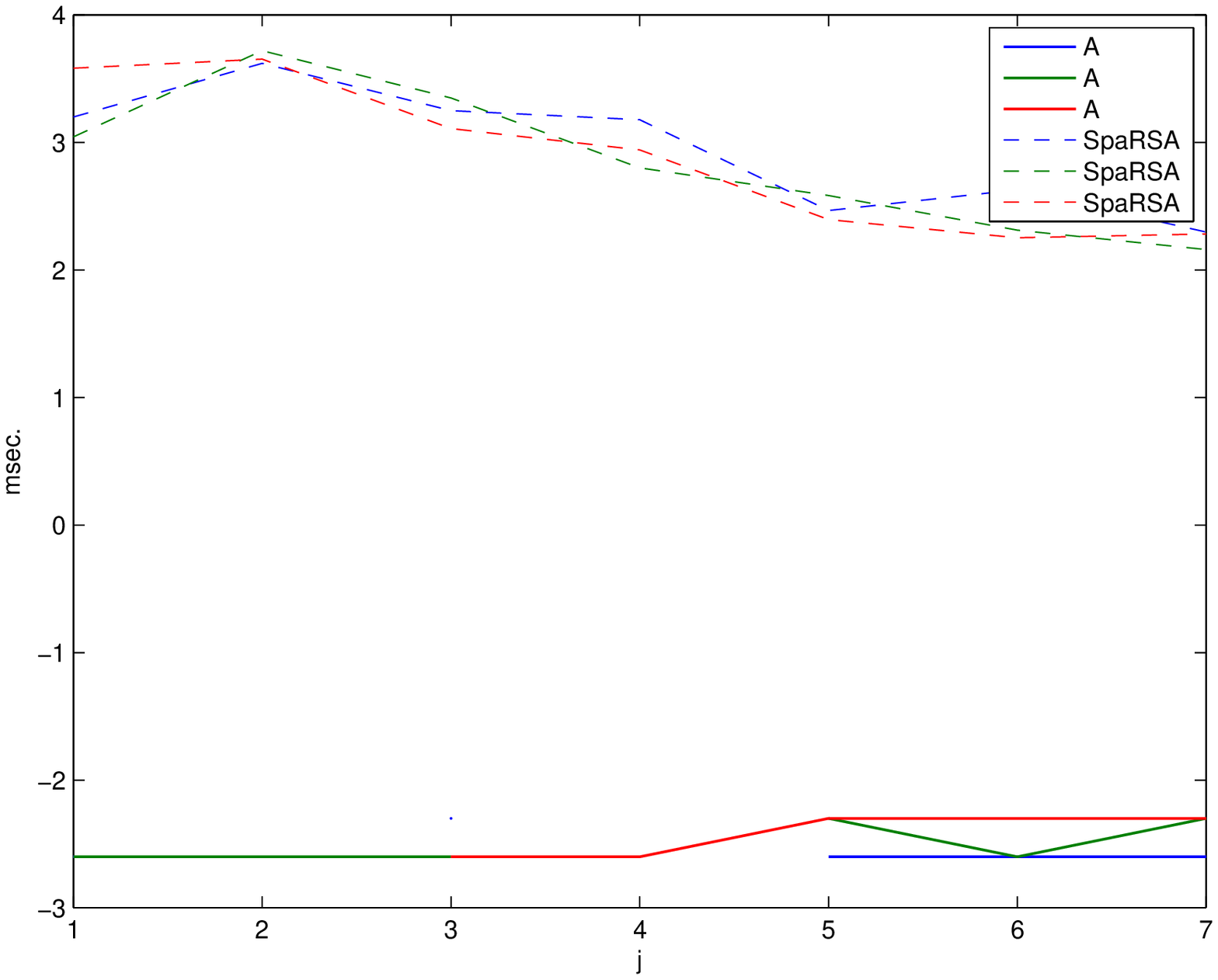}}
\subcaptionbox{$\mathcal{M}_2+\mathcal{N}(0,\frac{0.05^2}{D}I_D)$}
{\includegraphics[width=0.19\textwidth]{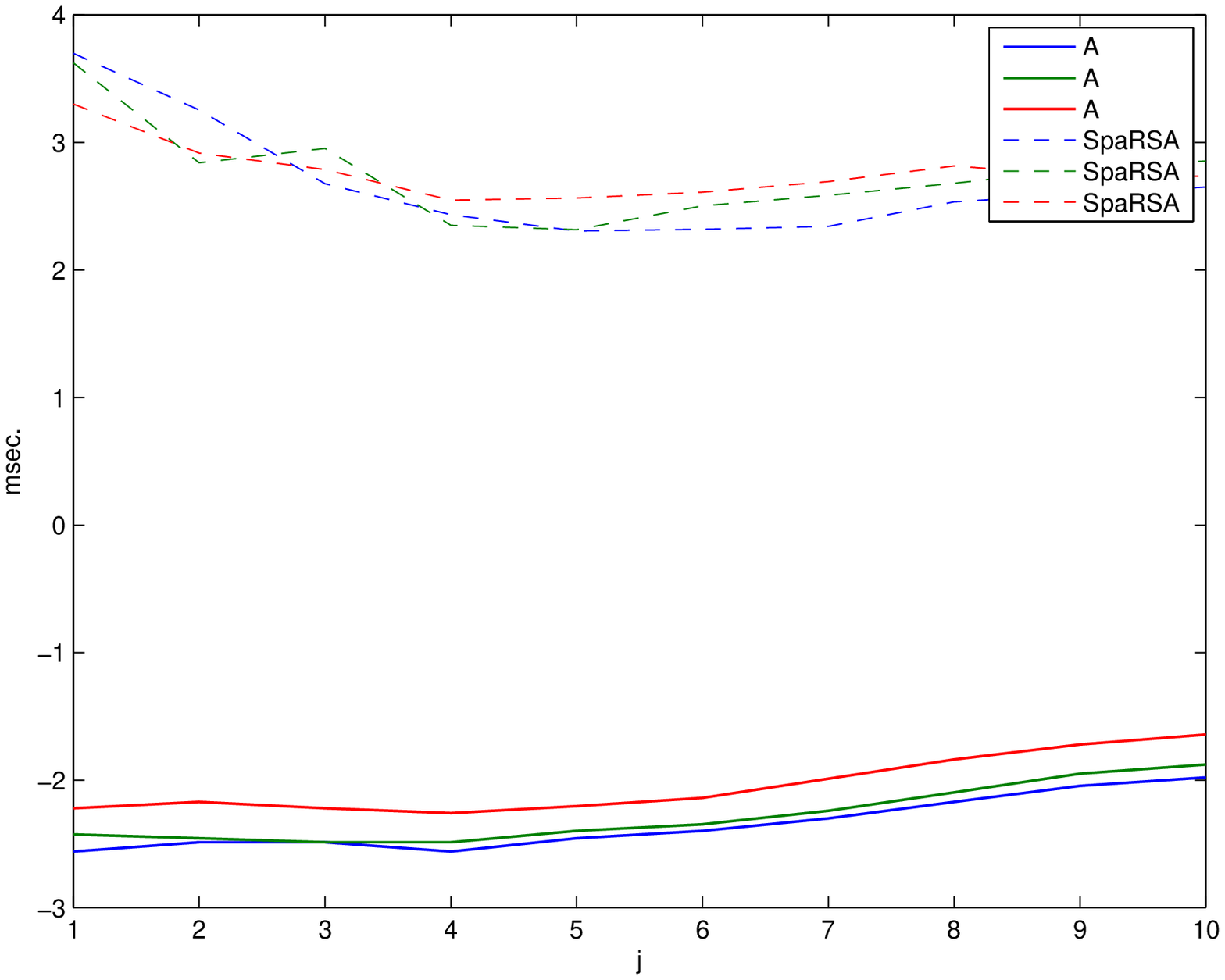}}
\subcaptionbox{$\mathcal{M}_3+\mathcal{N}(0,\frac{0.05^2}{D}I_D)$}
{\includegraphics[width=0.19\textwidth]{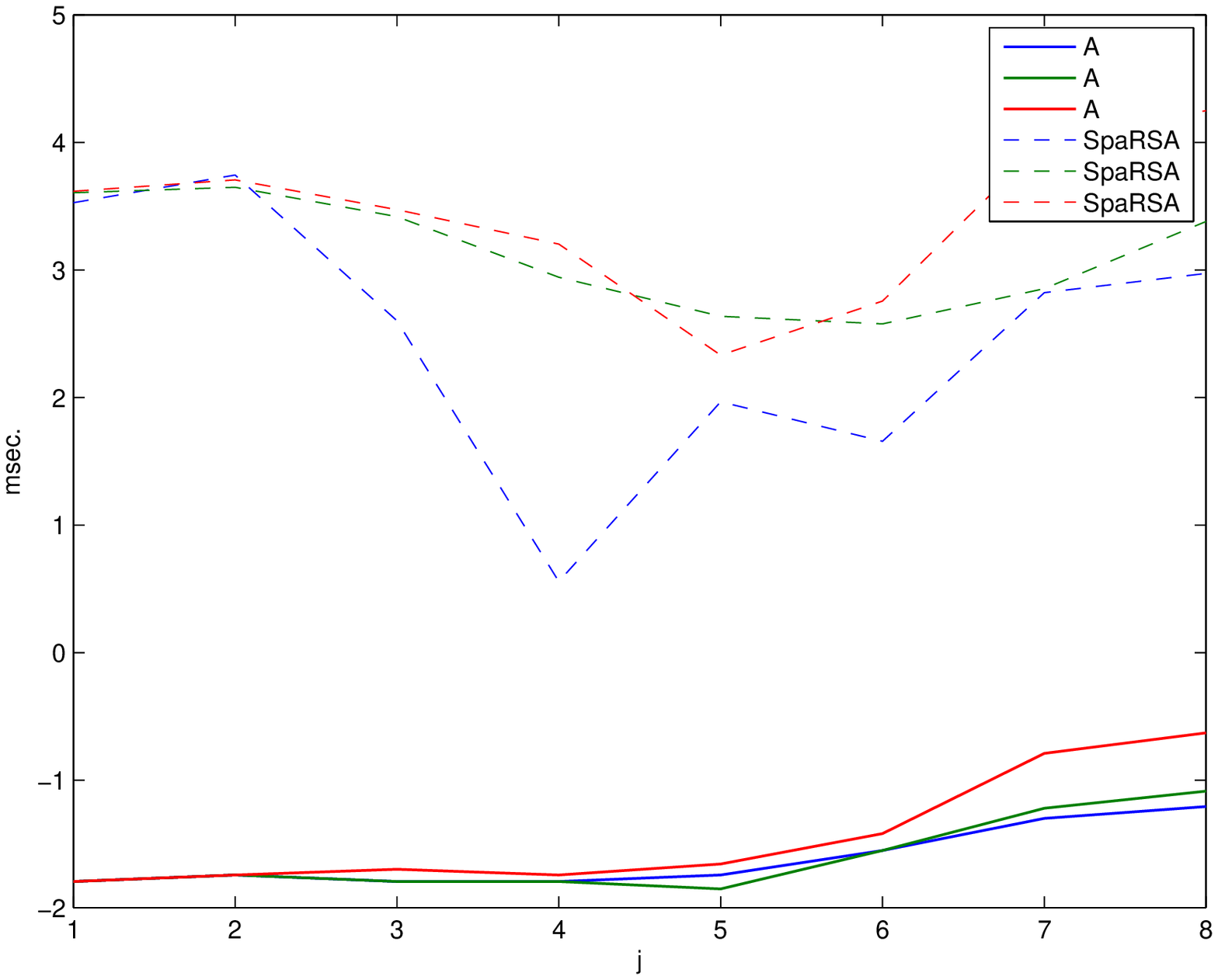}}
\subcaptionbox{$\mathcal{M}_4+\mathcal{N}(0,\frac{0.05^2}{D}I_D)$}
{\includegraphics[width=0.19\textwidth]{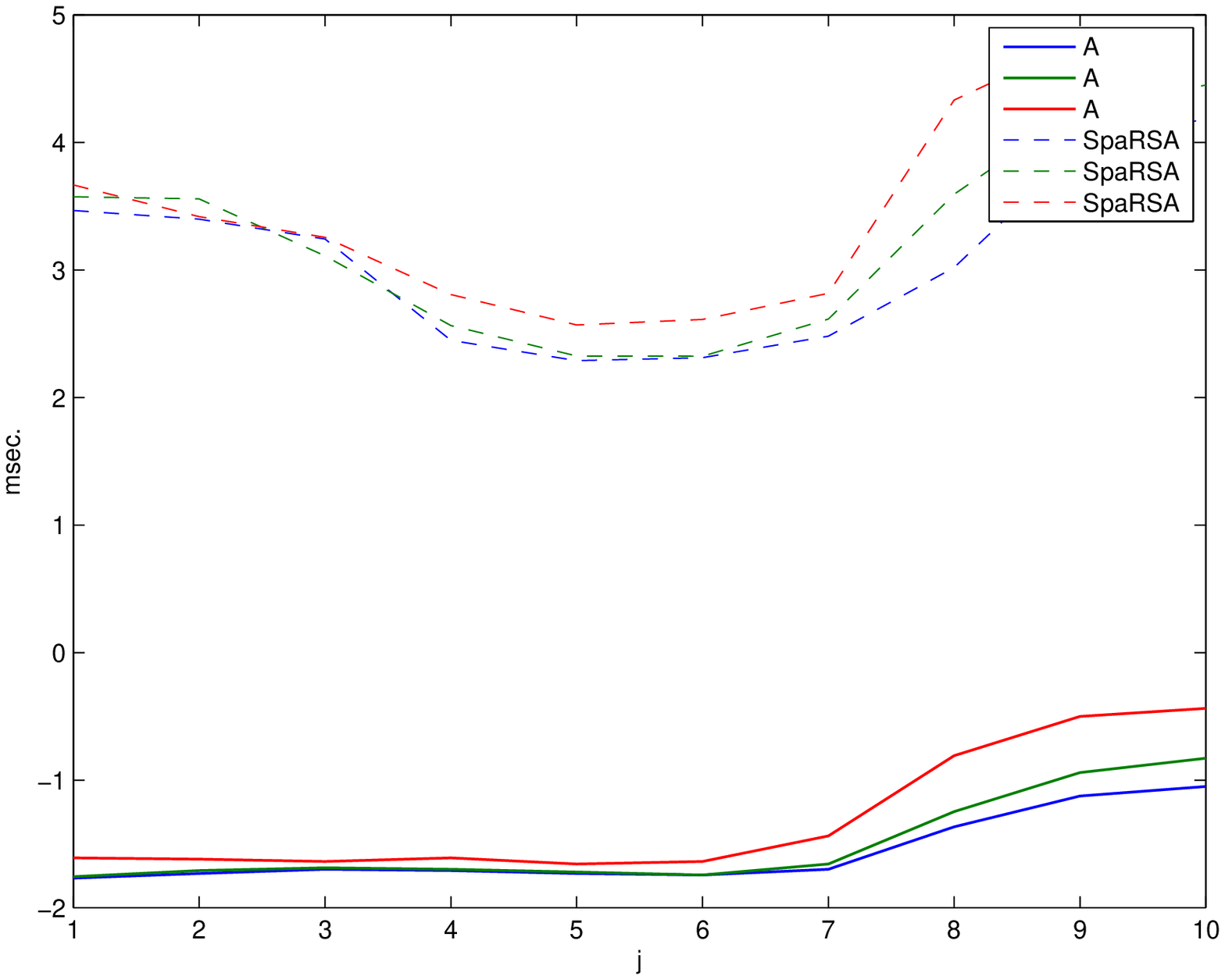}}
\subcaptionbox{$\mathcal{M}_5+\mathcal{N}(0,\frac{0.05^2}{D}I_D)$}
{\includegraphics[width=0.19\textwidth]{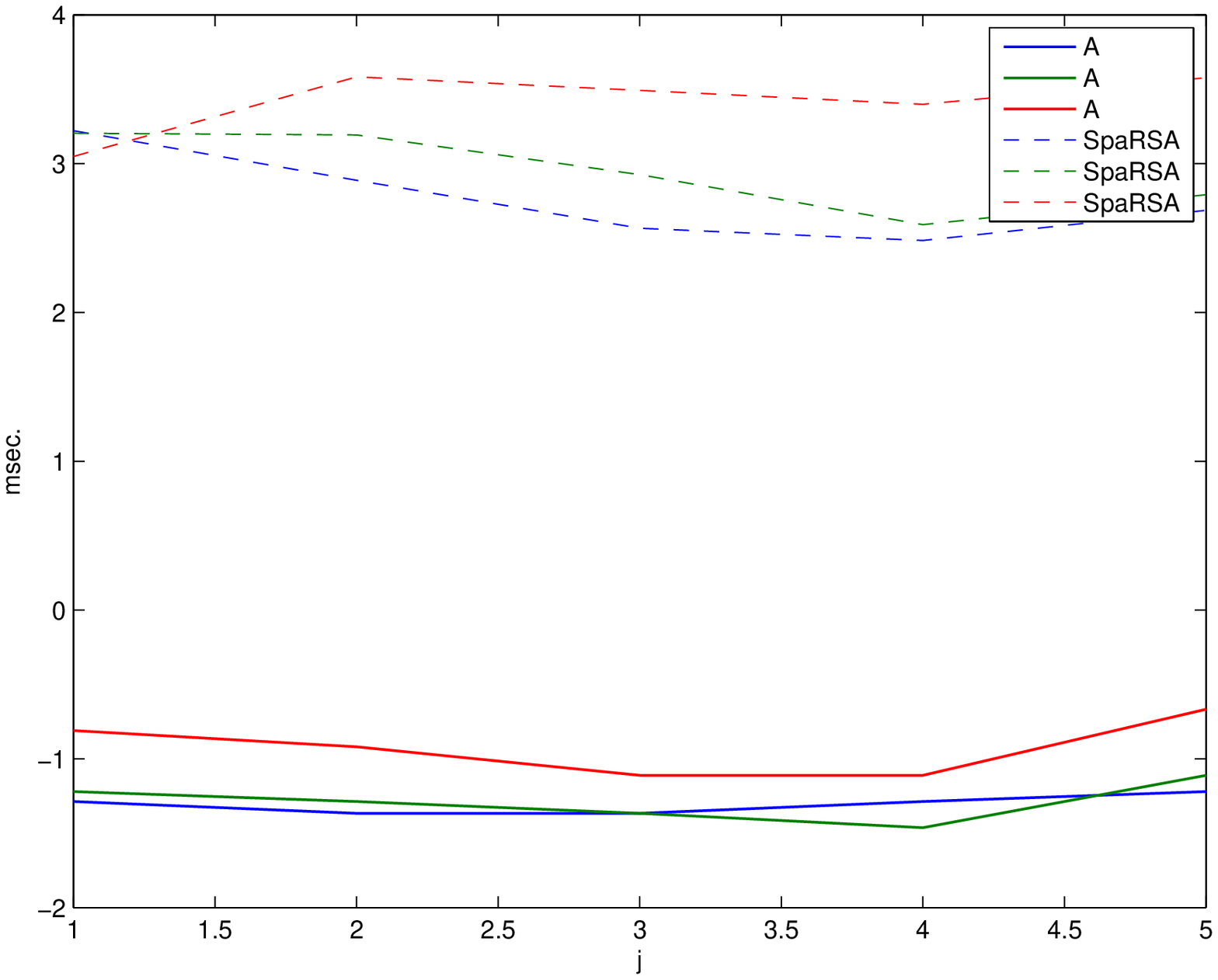}}
\subcaptionbox{$\mathcal{M}_1+\mathcal{N}(0,\frac{0.05^2}{D}I_D)$}
{\includegraphics[width=0.19\textwidth]{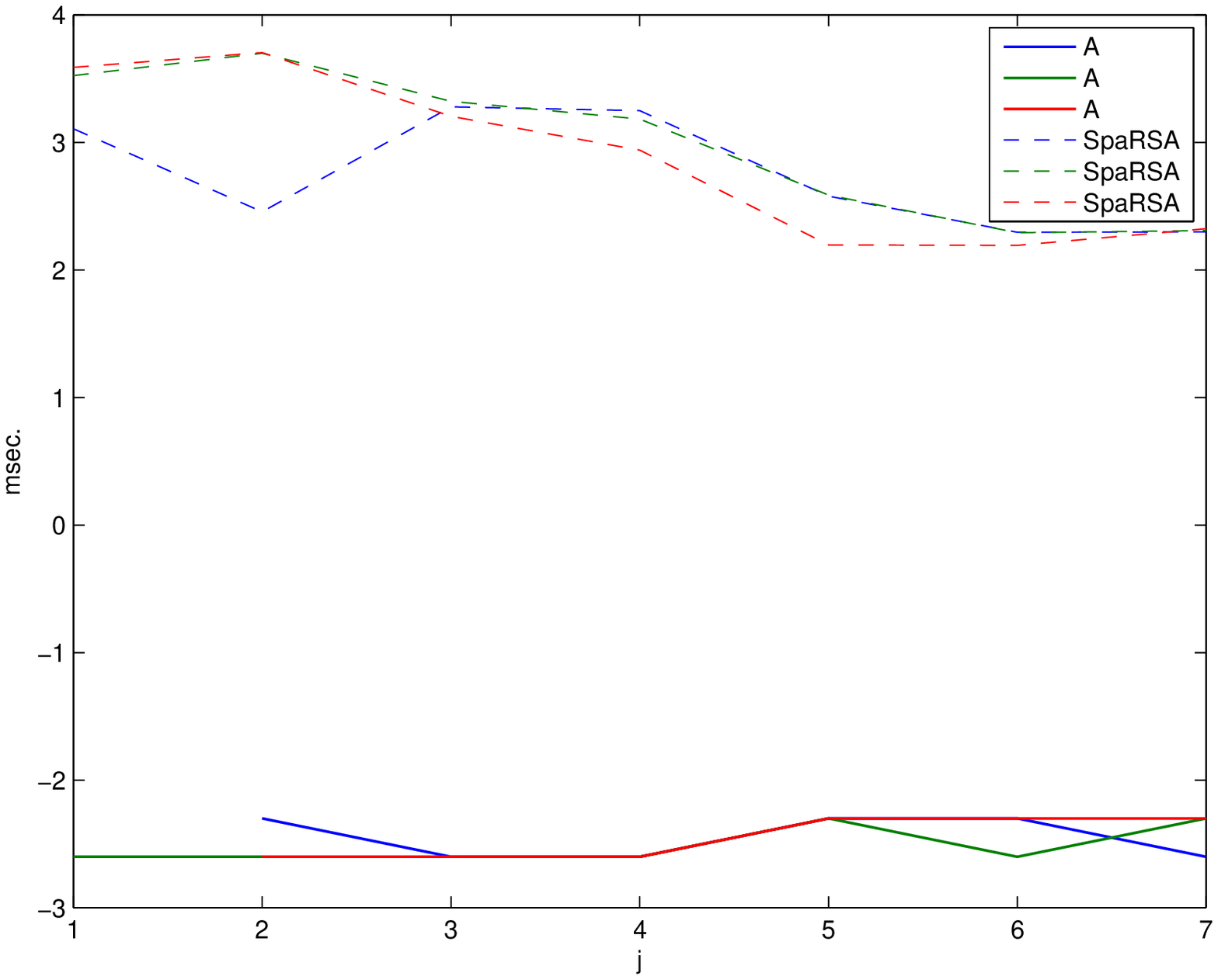}}
\subcaptionbox{$\mathcal{M}_2+\mathcal{N}(0,\frac{0.1^2}{D}I_D)$}
{\includegraphics[width=0.19\textwidth]{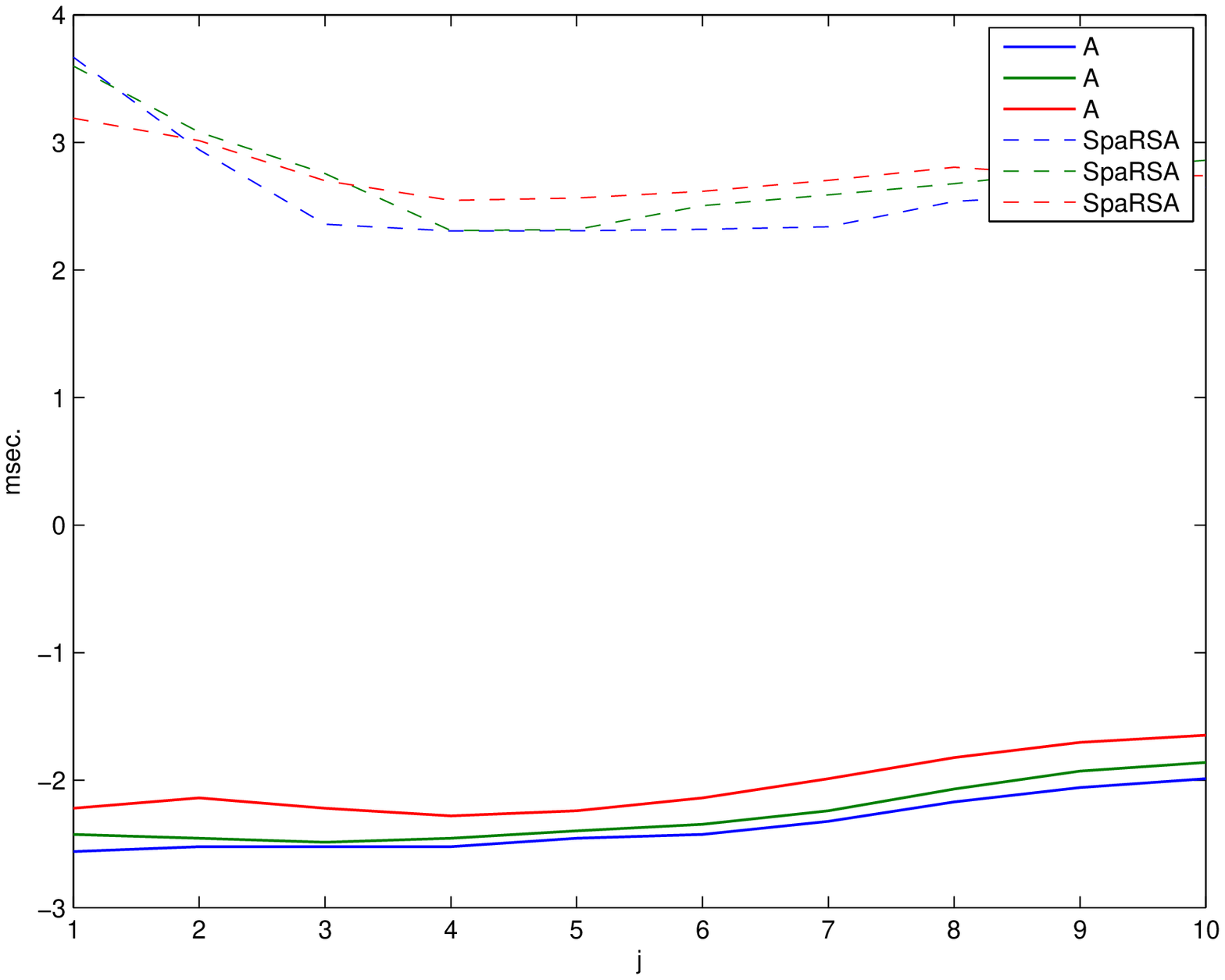}}
\subcaptionbox{$\mathcal{M}_3+\mathcal{N}(0,\frac{0.1^2}{D}I_D)$}
{\includegraphics[width=0.19\textwidth]{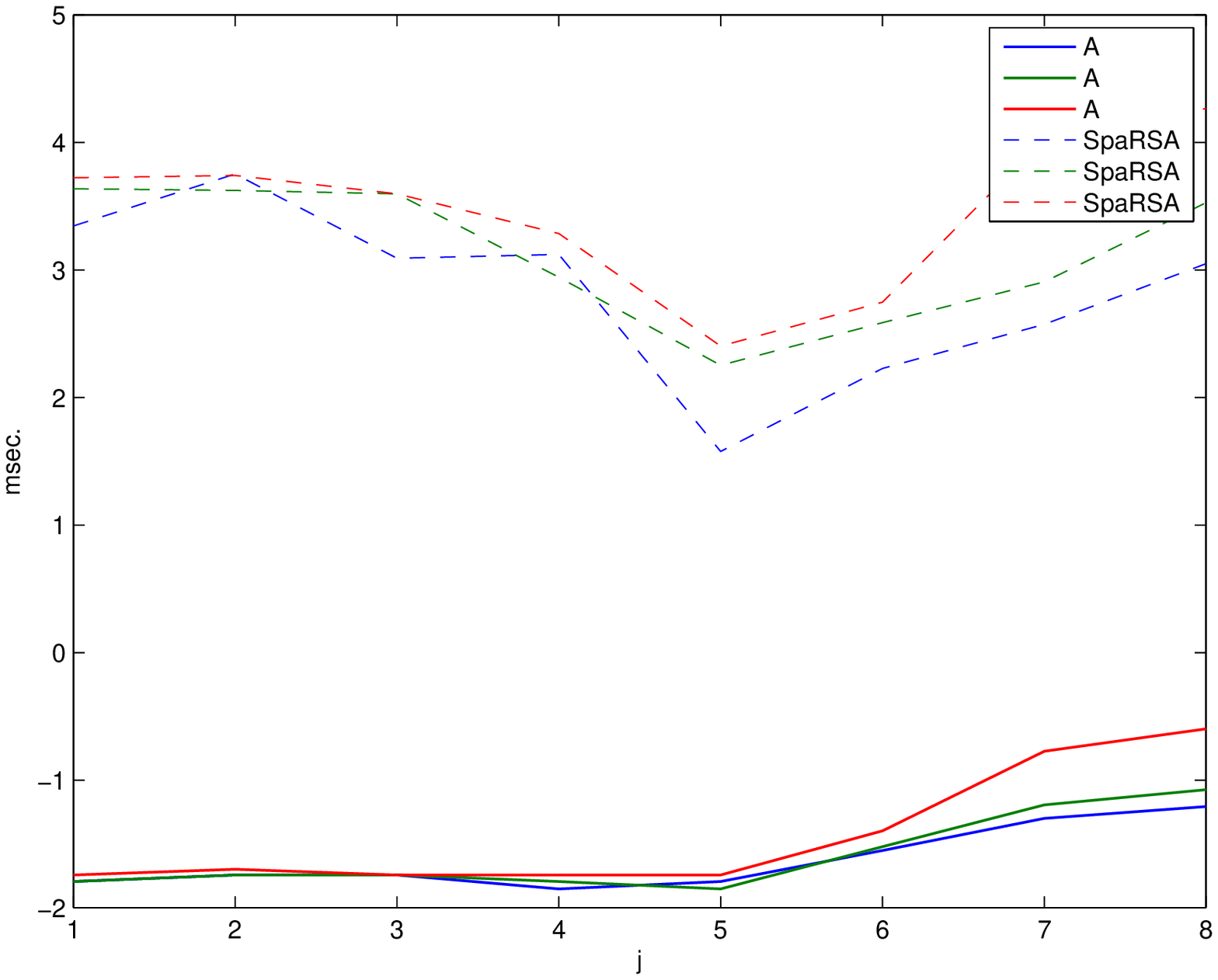}}
\subcaptionbox{$\mathcal{M}_4+\mathcal{N}(0,\frac{0.1^2}{D}I_D)$}
{\includegraphics[width=0.19\textwidth]{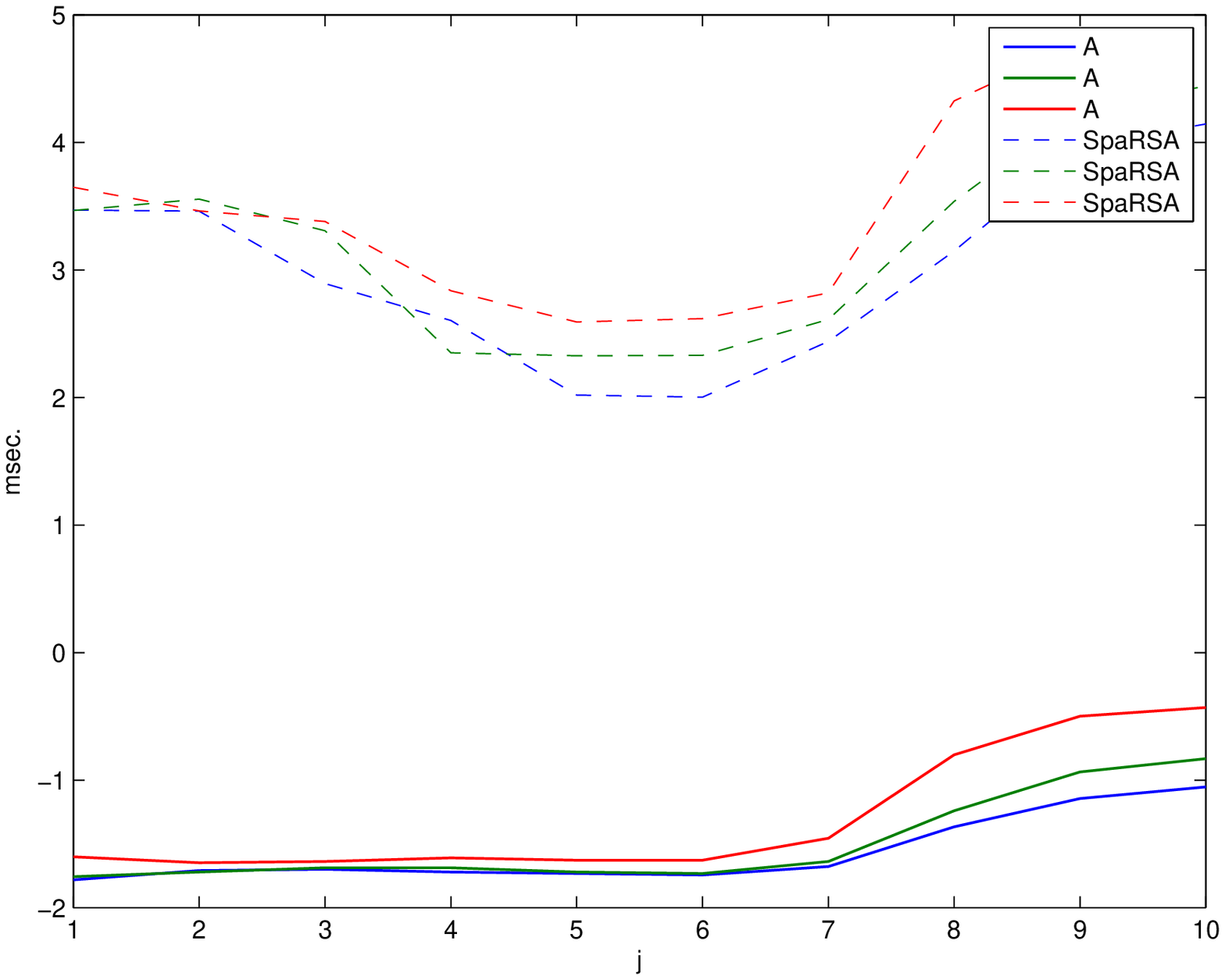}}
\subcaptionbox{$\mathcal{M}_5+\mathcal{N}(0,\frac{0.1^2}{D}I_D)$}
{\includegraphics[width=0.19\textwidth]{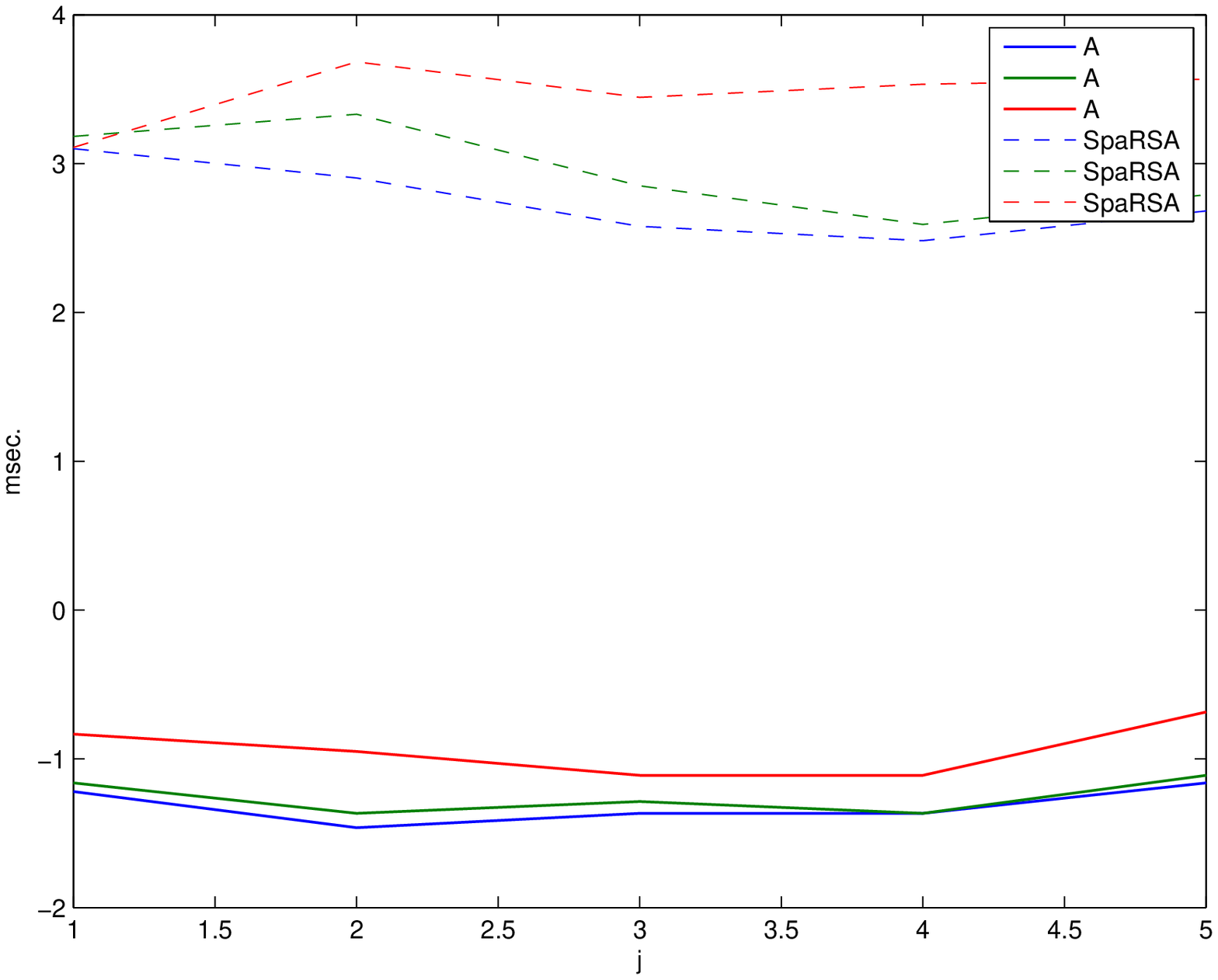}}
\caption{From left to right: data sets $\mathcal{M}_1$ to $\mathcal{M}_5$; from top to bottom: adding noise with increasing variance as above. In each plot, having the horizontal axis representing the scale $j$ and the vertical axis the time (in milliseconds) per point $x$ needed to compute either $\mathcal{A}(x)$ or \textrm{SpaRSA}(x) \cite{DBLP:journals/tsp/WrightNF09} (with the matrices involved in the algorithm, and their transposed, precomputed). Because of time involved, we ran SpaRSA only on $50$ randomly chosen points rather than all the points.}
\label{f:exstiming}
\end{figure}

Finally, in Figure \ref{f:exstiming} we report running times, for the same data sets as in Figure \ref{f:exs}, for our algorithm $\mathcal{A}$ and \textrm{SpaRSA}. These graphs suggest that our algorithm can perform several orders of magnitude faster than \textrm{SpaRSA}.  In the examples shown it took a few seconds to run Algorithm~\ref{alg:RecovAlg} on all the points, with \textrm{SpaRSA} taking a significant fraction of a second to run on a single point.  

\section{Conclusion}

In this paper we discussed the ability of random projection to embed an intrinsically low $d$-dimensional submanifold of $\mathbbm{R}^D$, together with a piecewise linear approximation to the submanifold, into $\mathbbm{R}^{O(d \log d)}$ in a way which (approximately) preserves the fidelity of the embedded piecewise linear approximation to the embedded manifold.  Although any collection of approximating affine spaces suffice, we focussed on the type of multi-scale linear approximations provided by GMRA \cite{CM:MGM2} in particular.  It is worth mentioning that the entire Geometric Wavelet Transform (GWT) \cite{CM:MGM2} of a point near a given manifold can also be preserved by the type of random projections discussed herein.  

Note that the GWT of a point on a given manifold will always be approximated by the sum of at most $Jd$ vectors (where $J$ is the number of scales in the GWT).  So, pessimistically, a random projection needs to preserve all distances in a number of $O(Jd)$-dimensional subspaces which is bounded above by Lemma~\ref{lem:CenterBound} in order to approximately preserve the entire geometric wavelet transform of each point on the manifold.  Thus, the GWT of each point on a given manifold should be preserved in compressed form by a random linear projection onto a subspace whose dimension, $m$, satisfies a variant of Theorem~\ref{thm:Assump1rowbound} with $d$ replaced everywhere by $Jd$.

\appendix

\section{Proof of Lemma~\ref{lem:ManCbound}}
\label{app:ManProof}

To prove this lemma we will modify the proof of Theorem~3.1 in \cite{RandProjSMan}.  The proof of Theorem~3.1 proceeds in two steps.  First, a finite set, $B \subset \mathbbm{R}^D$, of points on/near the given manifold $\mathcal{M}$ is defined.  The main body of the proof then consists of demonstrating that any $m \times D$ matrix, $M'$, which embeds $B$ into $\mathbbm{R}^m$ with $\Theta(\epsilon)$-distortion will also satisfy
$$(1 - \epsilon) \left\| \vec{x} - \vec{y} \right\| ~\leq~ \left\| M' \vec{x} - M' \vec{y} \right\| ~\leq~ (1 + \epsilon) \left\| \vec{x} - \vec{y} \right\|$$
for all $\vec{x}, \vec{y} \in \mathcal{M}$.  Our proof will proceed along a similar path.  We will begin by first defining a modified version of the set, $B$, considered in \cite{RandProjSMan}.  We will call this set $\tilde{B}$.  Then, we will prove that any $m \times D$ matrix which which embeds $\tilde{B}$ into $\mathbbm{R}^m$ with $\Theta(\epsilon)$-distortion will also satisfy item (a) of Assumption Set 2 in Section~\ref{sec:MeasMatrix}.

Let $d_{\mathcal{M}}\left( \vec{x}, \vec{y} \right)$ denote the geodesic distance between $\vec{x},\vec{y} \in \mathcal{M}$.  Furthermore, let \textbf{Tan}$_{\vec{x}}$ denote the $d$-dimensional tangent space to $\mathcal{M}$ at each $\vec{x} \in \mathcal{M}$.  Finally, let 
$$\mathcal{B}_{\mathcal{M},\delta} \left( \vec{x} \right) = \left\{ \vec{y} \in \mathcal{M} ~\big|~ d_{\mathcal{M}}\left( \vec{x}, \vec{y} \right) ~\leq~ \delta \right\}$$ 
for each $\delta \in \mathbbm{R}^{+}$ and $\vec{x} \in \mathcal{M}$.

We are now ready to construct $B \subset \mathbbm{R}^D$ as per \cite{RandProjSMan} as follows:  Set $T = O \left( \frac{\epsilon^2}{D}\cdot\min \left\{ 1, {\rm reach}\left( \mathcal{M} \right) \right\} \right)$ and, for each 
$\vec{x} \in \mathcal{M}$, let $Q_2 \left( \vec{x} \right) \subset \textbf{Tan}_{\vec{x}}$ denote a minimal $\Theta \left( \epsilon \cdot T / \sqrt{D} \right)$-cover of the $d$-dimensional Euclidean ball of radius $T$ centered at 
$\vec{0} \in \textbf{Tan}_{\vec{x}}$.  Next, choose $A \subset \mathcal{M}$ to be a minimal finite cover of $\mathcal{M}$ satisfying 
$$\min_{\vec{a} \in A}~ d_{\mathcal{M}} \left( \vec{a}, \vec{x} \right) ~\leq~ T,$$
for all $\vec{x} \in \mathcal{M}$.    Then,
$$B := \bigcup_{\vec{a} \in A} \left\{ \vec{a} \right\} \cup \left( \vec{a} + Q_2 \left( \vec{a} \right) \right).$$
In the next paragraph we will define our modified set, $\tilde{B} \subset \mathbbm{R}^D$, which is a superset of the set $B$ defined above.

Fix $j \in [J]$ and $k \in [K_j]$.  For each $\vec{a} \in A$ above, let $\vec{a}_{j,k} \in \mathcal{B}_{\mathcal{M},T} \left( \vec{a} \right)$ be such that 
$$\left\| \vec{a}_{j,k} - \vec{c}_{j,k} \right\| ~\leq~ \left\| \vec{y} - \vec{c}_{j,k} \right\| ~\forall \vec{y} \in \mathcal{B}_{\mathcal{M},T} \left( \vec{a} \right).$$
Let $A_{j,k} = \left\{ \vec{a}_{j,k} ~\big|~ \vec{a} \in A \right\}$.  Furthermore, denote the $\left( d+1 \right)$-dimensional vector space spanned by 
$\textbf{Tan}_{\vec{a}_{j,k}} \bigcup \left\{ \vec{c}_{j,k} - \vec{a}_{j,k} \right\}$ by $\textbf{Tan}_{\vec{a},j,k}$, and then let $Q_{j,k} \left( \vec{a} \right) \subset \textbf{Tan}_{\vec{a},j,k}$ be a 
minimal $\Theta \left( \epsilon \cdot T / \sqrt{D} \right)$-cover of the $\left( d+1 \right)$-dimensional Euclidean ball of radius $T$ centered at $\vec{0}$.  To finish, define
$$B_{j,k} := \bigcup_{\vec{a} \in A} \left\{ \vec{a}_{j,k} \right\} \cup \left( \vec{a}_{j,k} + Q_{j,k} \left( \vec{a} \right)  \right)$$
and then set 
$$\tilde{B} := \left( \bigcup_{j \in [J],~k \in [K_j]} B_{j,k} \cup \left\{ \vec{c}_{j,k} \right\} \right) \cup B \cup Q,$$
where $Q \subset \mathbbm{R}^D$ is as defined in Lemma~\ref{lem:RIPBound}.

Note that $\left| \tilde{B} \right|$ will be bounded above by 
$$(J + 1) \cdot K_{J} \left( 1 + \max_{j \in [J],~k \in [K_j]} \left| B_{j,k} \right| \right) + \left| B \right| + \left| Q \right|.$$  
Applying Lemma~\ref{lem:RIPBound} to bound $\left| Q \right|$, Lemma~\ref{lem:CenterBound} to bound $K_{J}$, and appealing to Section~3.2.5 of \cite{RandProjSMan} to bound $\left| B \right|$, the previous line reveals that
\begin{equation}
\left| \tilde{B} \right| ~\ll~ 2^{O \left(J \cdot d \right)} \cdot V \cdot \left( \frac{d}{ \min \left\{ 1, C_1 \right\}} \right)^{O \left( d \right)} \left( \max_{j \in [J],~k \in [K_j]} \left| B_{j,k} \right| \right) + V \left( \frac{D}{\epsilon \cdot \min \left\{ 1, {\rm reach}\left( \mathcal{M} \right) \right\} }  \right)^{O \left( d \right)}.
\label{eqn:BtildeBound}
\end{equation}  
We now finish bounding the cardinality of $\tilde{B}$ by noting that $\left| B_{j,k} \right|$ will always be bounded above by the upper bounds for $\left| B \right|$ in Section~3.2.5 of \cite{RandProjSMan} after every occurrence 
of $K = d$ is replaced with $d+1$.\footnote{Intuitively, we are increasing the effective intrinsic dimensionality of $\mathcal{M}$ from $d$ to $d+1$ in the process of creating our $B_{j,k}$-subsets.}  The stated upper bound on 
$\left| \tilde{B} \right|$ follows.

We will now complete the second portion of our proof by demonstrating that a sufficiently precise linear embedding of $\tilde{B}$ will satisfy item (a) of Assumption Set 2.  First, since $B \subset \tilde{B}$, 
Theorem~3.1 in \cite{RandProjSMan} guarantees that a low-distortion embedding of $\tilde{B}$ will preserve all pairwise distances between points on the manifold $\mathcal{M}$.  Furthermore, any embedding of $\tilde{B}$ will also 
embed all $\vec{c}_{j,k}$-vectors since they form a proper subset of $\tilde{B}$.  Hence, if suffices for us to show that a sufficiently precise linear embedding of $\tilde{B}$ will (approximately) preserve the distance from 
each $\vec{c}_{j,k}$-vector to all points on the manifold $\mathcal{M}$.

Fix $j \in [J]$, $k \in [K_j]$, and $\vec{x} \in \mathcal{M}$.  Let $\vec{a}~' \in A$ be the closest element of $A$ to $\vec{x}$,
$$\vec{a}~' ~=~ \argmin_{\vec{a} \in A}~ d_{\mathcal{M}} \left( \vec{a}, \vec{x} \right).$$
Finally, let $\vec{x}~'_{j,k}$ denote the projection of $\vec{x}$ onto the $(d+1)$-dimensional affine subspace $\vec{a}~'_{j,k} + \textbf{Tan}_{\vec{a}~',j,k}$.  By considering the Taylor series expansion of the unit speed 
parameterization of the geodesic path from $\vec{a}~'_{j,k}$ to $\vec{x}$ on $\mathcal{M}$, we find that
$$\vec{x} ~=~ \vec{x}~'_{j,k} + \vec{r},~\textrm{where}~\left\| \vec{r} \right\| = O \left( \frac{d^2_{\mathcal{M}} \left( \vec{x}, \vec{a}~'_{j,k} \right)}{{\rm reach}\left( \mathcal{M} \right)} \right).$$
In fact, the magnitude of the remainder, $\vec{r}$, is also $O \left( \left\| \vec{x} - \vec{a}~'_{j,k} \right\|^2 \right)$ since $T < {\rm reach}\left( \mathcal{M} \right) / 2$ (see Corollary 2.1 in \cite{RandProjSMan}).  
Furthermore, the definition of $\vec{a}~'_{j,k} \in \mathcal{M}$ implies that $\left\| \vec{x} - \vec{a}~'_{j,k} \right\| = O \left( \left\| \vec{x} - \vec{c}_{j,k} \right\| \right)$.

Continuing with the proof, suppose that an $m \times D$ matrix, $M'$, embeds $\tilde{B}$ into $\mathbbm{R}^m$ with $\Theta(\epsilon)$-distortion.  A trivial variant of Lemma~\ref{lem:SubSpaceBound} then implies that
\begin{align}
\left\| M' \vec{x} - M' \vec{c}_{j,k} \right\| &~\leq~ \left\| M' \vec{x} - M' \vec{x}~'_{j,k} \right\| + \left\| M' \vec{x}~'_{j,k} - M' \vec{c}_{j,k} \right\| ~\leq~ \left\| M' \vec{r} \right\| + \left(1 + \Theta(\epsilon) \right) \left\| \vec{x}~'_{j,k} - \vec{c}_{j,k} \right\| \nonumber \\ &~\leq~ \left(1 + \Theta(\epsilon) \right) \left( \left\| \vec{x} - \vec{c}_{j,k} \right\| + \left\| \vec{r} \right\| \right) + \left\| M' \vec{r} \right\| ~\leq~ \left(1 + \Theta(\epsilon) \right) \left\| \vec{x} - \vec{c}_{j,k} \right\| + \left\| M' \vec{r} \right\| + O \left(\left\| \vec{x} - \vec{a}~'_{j,k} \right\|^2 \right) \nonumber
\end{align}
since $Q_{j,k} \left( \vec{a}~' \right) \subset \textbf{Tan}_{\vec{a}~',j,k}$ is a proper subset of $\tilde{B}$, and $\left( \vec{x}~'_{j,k} - \vec{c}_{j,k} \right) \in \textbf{Tan}_{\vec{a}~',j,k}$.
In addition, the fact that $Q \subset \tilde{B}$ together with Lemma~\ref{lem:RIPBound} guarantees that $M'$ will have the RIP($D$,$d$,$\Theta(\epsilon)$).  This fact combined with the H\"{o}lder inequality finally reveals that 
\begin{align}
\left\| M' \vec{x} - M' \vec{c}_{j,k} \right\| &~\leq~ \left(1 + \Theta(\epsilon) \right) \left\| \vec{x} - \vec{c}_{j,k} \right\| + O \left(\sqrt{\frac{D}{d}} \cdot \left\| \vec{x} - \vec{a}~'_{j,k} \right\|^2 \right) ~\leq~ \left(1 + \Theta(\epsilon) + O \left(\sqrt{\frac{D}{d}} \cdot T \right) \right) \left\| \vec{x} - \vec{c}_{j,k} \right\| \nonumber \\ &~\leq~ \left(1 + O \left( \epsilon \right) \right) \left\| \vec{x} - \vec{c}_{j,k} \right\| \nonumber.
\end{align}
The lower bound for $\left\| M' \vec{x} - M' \vec{c}_{j,k} \right\|$ is established in an analogous fashion.  We have the stated theorem.

\section{Proof of Lemma~\ref{lem:ManApproxbound}}
\label{app:ManApproxProof}

The proof of this Lemma borrows heavilly from the proof of Lemma~\ref{lem:ManCbound}.  Set $T = O \left( \frac{2^{-J} \epsilon^2}{D}\cdot\min \left\{ 1, {\rm reach}\left( \mathcal{M} \right) \right\} \right)$.  We will begin 
by defining the set $B' \subset \mathbbm{R}^D$.  Let $A \subset \mathcal{M}$, $B \subset \mathbbm{R}^D$, and $A_{j,k} = \left\{ \vec{a}_{j,k} ~\big|~ \vec{a} \in A \right\} \subset \mathcal{M}$ for each $j \in [J], k \in [K_j]$
be defined as in Appendix~\ref{app:ManProof} above (except now using the smaller value of $T$ from the second sentence of this appendix).  Let $\widetilde{\textbf{Tan}}_{\vec{a},j,k}$ denote the $\left( 2d+1 \right)$-dimensional 
vector space spanned by 
$$\textbf{Tan}_{\vec{a}_{j,k}} \bigcup \left\{ \vec{c}_{j,k} - \vec{a}_{j,k} \right\} \bigcup \left\{ \Phi^{\rm T}_{j,k}\Phi_{j,k} \vec{y} ~\big|~ \vec{y} \in \mathbbm{R}^D \right\}$$
for each $\vec{a}_{j,k} \in A_{j,k}$.  Furthermore, for each $\vec{a}_{j,k} \in A_{j,k}$, let ${Q}'_{j,k} \left( \vec{a} \right) \subset \widetilde{\textbf{Tan}}_{\vec{a},j,k}$ be a minimal 
$\Theta \left( \epsilon \cdot T / \sqrt{D} \right)$-cover of the $\left( 2d+1 \right)$-dimensional Euclidean ball of radius $T$ centered at $\vec{0} \in \widetilde{\textbf{Tan}}_{\vec{a},j,k}$.  To finish, define
$${B}'_{j,k} := \bigcup_{\vec{a} \in A} \left\{ \vec{a}_{j,k} \right\} \cup \left( \vec{a}_{j,k} + {Q}'_{j,k} \left( \vec{a} \right)  \right)$$
for each $j \in [J], k \in [K_j]$, and then set 
$${B}' := \left( \bigcup_{j \in [J],~k \in [K_j]} B'_{j,k} \cup \left\{ \vec{c}_{j,k} \right\} \right) \cup B \cup Q,$$
where $Q \subset \mathbbm{R}^D$ is as defined in Lemma~\ref{lem:RIPBound}.  It is not difficult to see that $\left| B' \right|$ will be bounded above as per Equation~\ref{eqn:BtildeBound} after $\epsilon$ 
is replaced everywhere by $2^{-J} \epsilon$.  Simplifying yields the stated upper bound.

We will now complete our proof by demonstrating that a sufficiently precise linear embedding of $B'$ will satisfy item (d) of Assumption Set 2.  Fix $j \in [J]$, $k \in [K_j]$, and $\vec{x} \in \mathcal{M}$.  Let $\vec{a}~' \in A$ be the closest element of $A$ to $\vec{x}$,
$$\vec{a}~' ~=~ \argmin_{\vec{a} \in A}~ d_{\mathcal{M}} \left( \vec{a}, \vec{x} \right).$$
Finally, let $\vec{x}~'_{j,k}$ denote the projection of $\vec{x}$ onto the $(2d+1)$-dimensional affine subspace $\vec{a}~'_{j,k} + \widetilde{\textbf{Tan}}_{\vec{a},j,k}$.  By considering the Taylor series expansion of the unit 
speed parameterization of the geodesic path from $\vec{a}~'_{j,k}$ to $\vec{x}$ on $\mathcal{M}$, we find that
$$\vec{x} ~=~ \vec{x}~'_{j,k} + \vec{r},~\textrm{where}~\left\| \vec{r} \right\| = O \left( \frac{d^2_{\mathcal{M}} \left( \vec{x}, \vec{a}~'_{j,k} \right)}{{\rm reach}\left( \mathcal{M} \right)} \right).$$
Furthermore, we recall that the magnitude of the remainder, $\vec{r}$, is also $O \left( \left\| \vec{x} - \vec{a}~'_{j,k} \right\|^2 \right)$ since $T$ is sufficiently small.  

To finish, suppose that an $m \times D$ matrix, $M'$, embeds $B'$ into $\mathbbm{R}^m$ with $\Theta(\epsilon)$-distortion.  A trivial variant of Lemma~\ref{lem:SubSpaceBound} implies that
\begin{align}
\left\| M' \vec{x} - M' \mathbbm{P}_{j,k} \left( \vec{x} \right) \right\| &~\leq~ \left\| M' \vec{x} - M' \vec{x}~'_{j,k} \right\| + \left\| M' \vec{x}~'_{j,k} - M' \mathbbm{P}_{j,k} \left( \vec{x} \right) \right\| ~\leq~ 
\left\| M' \vec{r} \right\| + \left(1 + \Theta(\epsilon) \right) \left\| \vec{x}~'_{j,k} - \mathbbm{P}_{j,k} \left( \vec{x} \right) \right\| \nonumber \\ &~\leq~ 
\left(1 + \Theta(\epsilon) \right) \left( \left\| \vec{x} - \mathbbm{P}_{j,k} \left( \vec{x} \right) \right\| + \left\| \vec{r} \right\| \right) + \left\| M' \vec{r} \right\| \nonumber \\ &~\leq~ 
\left(1 + \Theta(\epsilon) \right) \left\| \vec{x} - \mathbbm{P}_{j,k} \left( \vec{x} \right) \right\| + \left\| M' \vec{r} \right\| + O \left(\left\| \vec{x} - \vec{a}~'_{j,k} \right\|^2 \right) \nonumber
\end{align}
since $Q'_{j,k} \left( \vec{a}~' \right) \subset \widetilde{\textbf{Tan}}_{\vec{a},j,k}$ is a subset of $B'$, and $\left( \vec{x}~'_{j,k} - \mathbbm{P}_{j,k} \left( \vec{x} \right) \right) \in \widetilde{\textbf{Tan}}_{\vec{a},j,k}$.
In addition, the fact that $Q \subset B'$ together with Lemma~\ref{lem:RIPBound} guarantees that $M'$ will have the RIP($D$,$d$,$\Theta(\epsilon)$).  This fact combined with the H\"{o}lder inequality reveals that 
\begin{align}
\left\| M' \vec{x} - M' \mathbbm{P}_{j,k} \left( \vec{x} \right) \right\| &~\leq~ \left(1 + \Theta(\epsilon) \right) \left\| \vec{x} - \mathbbm{P}_{j,k} \left( \vec{x} \right) \right\| + O \left(\sqrt{\frac{D}{d}} \cdot \left\| \vec{x} - \vec{a}~'_{j,k} \right\|^2 \right) 
~\leq~ \left(1 + \Theta(\epsilon) \right) \left\| \vec{x} - \mathbbm{P}_{j,k} \left( \vec{x} \right) \right\| + O \left(\sqrt{\frac{D}{d}} \cdot T^2 \right)  \nonumber \\ 
&~\leq~ \left(1 + \Theta(\epsilon) \right) \left\| \vec{x} - \mathbbm{P}_{j,k} \left( \vec{x} \right) \right\| + 2^{-J}  \nonumber
\end{align}
whenever $T$ is weighted by a sufficiently small (universal) constant.
The lower bound for $\left\| M' \vec{x} - M' \mathbbm{P}_{j,k} \left( \vec{x} \right) \right\|$ is established in an analogous fashion.

\end{document}